\documentclass[%
 reprint,
 amsmath,amssymb,
 aps,
prb,
]{revtex4-2}

\usepackage{graphicx}
\usepackage{dcolumn}
\usepackage{bm}
\usepackage{hyperref}
\usepackage{old-arrows}
\usepackage{xcolor}
\usepackage{float}
\usepackage{ulem}

\begin{document}

\preprint{APS/123-QED}

\title{Equal contribution of even and odd frequency pairing to transport across normal metal-superconductor junctions}

\author{Shun Tamura$^1$, Viktoriia Kornich$^1$, and Bj\"{o}rn Trauzettel$^{1,2}$}
\affiliation{
    $^1$Institute for Theoretical Physics and Astrophysics, University of W\"{u}rzburg, D-97074 W\"{u}rzburg, Germany\\
    $^2$W\"{u}rzburg-Dresden Cluster of Excellence ct.qmat, D-97074 W\"{u}rzburg, Germany
}




\date{\today}

\begin{abstract}
Odd-frequency pairing is an unconventional type of Cooper pairing in superconductors related to the frequency dependence of the corresponding anomalous Green function. We show by a combination of analytical and numerical methods that odd-frequency pairing is ubiquitously present in the current of Andreev-scattered particles across a junction formed by a normal metal (N) and a superconductor (S), even if the superconducting pairing is of conventional $s$-wave, spin singlet type. We carefully analyze the conductance of NS junctions with different pairing symmetries ($s$-wave, $p$-wave, $d$-wave). In all cases, we identify a generic equal balance of even and odd frequency pairing to the contributions related to Andreev reflection. This analysis shows in retrospect that the presence of odd-frequency pairing in electric currents across NS junctions is rather the rule, not the exception. This insight stems from an alternative approach of analyzing the transport problem of hybrid structures. It is based on the Kubo-Greenwood formula with direct access to symmetries of the anomalous Green functions characterizing the superconducting pairing. 
We expect that our predictions substantially enrich the interpretation of transport data across NS junctions in many material combinations.
\end{abstract}

\maketitle



\emph{Introduction.}---
The symmetry of the superconducting pairing potential (SPP) has been the central topic since its discovery. One important (but less investigated) aspect thereof is its frequency dependence. The frequency dependence of the SPP is classified in two distinct ways: even- and odd-frequency pairing. Even-frequency pairing (EFP) applies to all known bulk superconductors to date, no matter whether their pairing is of conventional $s$-wave, spin singlet type or unconventional. Odd-frequency pairing (OFP) is considered to be rather exotic. It refers to the property that the anomalous Green function (related to a particular type of pairing amplitude) is odd under the exchange of time or frequency~\cite{Berezinskii74,doi:10.1143/JPSJ.81.011013,RevModPhys.91.045005,Cayao2020,https://doi.org/10.1002/andp.201900298}. Bulk OFP has not yet been discovered experimentally. In fact, its stability is an interesting research topic by itself~\cite{PhysRevLett.66.1533,PhysRevB.45.13125,PhysRevB.46.10812,P_Coleman_1997,PhysRevLett.107.247202}. In hybrid structures, such as normal metal (N) / superconductor (S) junctions or Josephson junctions, translation symmetry is broken. It has been soon realized that this broken symmetry gives rise to the emergence of odd-frequency pairing in superconducting hybrids~\cite{PhysRevLett.86.4096,PhysRevB.70.012507,RevModPhys.77.1321,PhysRevB.76.054522,PhysRevB.76.054522,PhysRevB.76.054522,PhysRevLett.98.037003,PhysRevB.70.012507,PhysRevB.87.104513,PhysRevB.104.094503}. In Josephson junctions, supplemented with magnetic materials in the weak link, a long-range proximity effect has been considered as an indirect evidence of OFP~\cite{PhysRevLett.104.137002}. More distinct features of OFP (as compared to EFP) have also been predicted, for instance, the paramagnetic Meissner effect, which should appear under certain conditions~\cite{PhysRevB.52.1271,doi:10.1143/JPSJ.66.2556,PhysRevB.89.184508,PhysRevB.95.184506,PhysRevB.104.054507}. Indirect evidence of this particular type of attraction of magnetic flux by superconductivity has been reported in experiments based on low-energy muon spectroscopy~\cite{PhysRevX.5.041021,PhysRevLett.125.026802,PhysRevMaterials.5.114801} and on scanning tunneling spectroscopy (STS)~\cite{PhysRevLett.125.117003,doi:10.1126/sciadv.adf5500}. 

However, it is fair to say that the present-day understanding is that it is difficult to observe evidence for OFP in any type of experiment involving superconductors or hybrid junctions thereof. In this Letter, we argue that the opposite is true for standard transport measurements across NS junctions. In such junctions, it is impossible to observe genuine fingerprints of conventional EFP. In fact, we show below that the transport features related to superconductivity, i.~e.~Andreev reflection in the context of NS junctions, are always equally balanced by EFP and OFP contributions. This observation is deeply connected to the underlying symmetries of retarded and advanced Green functions that enter into linear response expressions for the conductance. It has been overlooked so far, because common methods of calculating these transport properties do not give insight on the impact of EFP or OFP on the conductance. We benchmark our discovery by a number of examples, where the N side is either a one-dimensional (1D) system or a 1D ladder and the S side is either a 1D or a 2D superconductor with different pairing symmetries such as $s$-wave, $p$-wave, and $d$-wave. 
We expect that our predictions substantially enrich the interpretation of transport data across NS junctions in many material combinations

\emph{Conductance across NS junction.}---
We evaluate the conductance $G$ by linear response theory employing
\begin{align}
    &
    G
    =
    -\int dE\frac{df(E)}{dE}\bar{\Gamma}_\mathrm{e}(E),\:\:
    \Gamma_\mathrm{e}(x,x',E)
    =
    \gamma(\tilde{G},\tilde{G}),
    \label{eq:def_conductance}
    \\
    &
    \gamma(g_1,g_2)
    =
    \alpha
    \mathrm{Tr}
    \left[
        P_\mathrm{e}
        g_1(x,x',E)
        \overleftrightarrow{\nabla}
        \overleftrightarrow{\nabla}'
        g_2(x',x,E)
    \right].
    \label{eq:def_Gamma_e}
\end{align}
The spatial average is depicted by using the symbol of the over bar in this Letter: 
$\bar{\beta}=\frac{1}{{(L_2-L_1)}^2}\int_{L_1}^{L_2}dx dx'\beta(x,x')$.
Here, $\alpha=\frac{-e^2\hbar^3\pi}{4m^2}$, $f(E)$ is the Fermi-Dirac distribution function, $g(x)\overleftrightarrow{\nabla}h(x)=[\partial_x g(x)]h(x)-g(x)\partial_x h(x)$, $\overleftrightarrow{\nabla}'$ acts on $x'$,  $P_\mathrm{e}=(\hat{\tau}_0+\hat{\tau}_3)/2$ with Pauli matrices $\hat{\tau}_{j=0,1,2,3}$ in particle-hole space, $m$ is an electron mass, $e$ is an elementary charge, and the trace is taken for particle-hole and spin space.
$\tilde{G}$ is given by
$\tilde{G}(x,x',E) = \frac{1}{2\pi i}[\check{G}^\mathrm{A}(x,x',E)-\check{G}^\mathrm{R}(x,x',E)]$
with the advanced (retarded) Green function (GF) $\check{G}^\mathrm{A(R)}$. 
The symbol of the over tilde is used in this way throughout this Letter.
Equation~\eqref{eq:def_conductance} in combination with Eq.~\eqref{eq:def_Gamma_e} is known as the Kubo-Greenwood formula~\cite{D_A_Greenwood_1958,C_Caroli_1971,PhysRevB.40.8169}.
We evaluate $\bar{\Gamma}_\mathrm{e}(E)$ in the N region, i.e., $x$ and $x'$ are chosen in the N region.

Dividing the GFs into normal and anomalous GFs described by $\check{G}_\mathrm{N}$ and $\check{F}$, respectively, Andreev reflection is described by the anomalous part.
Then, the retarded GF can be expressed as
$\check{G}^\mathrm{R}(x,x',E)
    =
    \begin{pmatrix}
        G_\mathrm{N}^{\mathrm{R},11}(x,x',E) & F^{\mathrm{R},12}(x,x',E)
        \\
        F^{\mathrm{R},21}(x,x',E) & G_\mathrm{N}^{\mathrm{R},22}(x,x',E)
    \end{pmatrix}$ with
\begin{align}
&
    \check{G}^\mathrm{R}(x,x',E)
    =
    -i
    \int d(t-t')e^{i(E+i\eta)(t-t')}
    \Theta(t-t')
    \nonumber\\
    &\times
    \begin{pmatrix}
        \langle\{\Psi_{\sigma}(x,t),\Psi_{\sigma'}^\dagger(x',t')\}\rangle
        &
        \langle\{\Psi_{\sigma}(x,t),\Psi_{\sigma'}(x',t')\}\rangle
        \\
        \langle\{\Psi_{\sigma}^\dagger(x,t),\Psi_{\sigma'}^\dagger(x',t')\}\rangle
        &
        \langle\{\Psi_{\sigma}^\dagger(x,t),\Psi_{\sigma'}(x',t')\}\rangle
    \end{pmatrix},
\end{align}
where $\Theta(t)$ is the Heaviside step function, and $\check{G}_\mathrm{N}^\mathrm{R}$ and $\check{F}^\mathrm{R}$ are normal and anomalous GFs:
$\check{G}_\mathrm{N}^\mathrm{R}
    =
    \begin{pmatrix}
        G_\mathrm{N}^\mathrm{R,11} & 0
        \\
        0 & G_\mathrm{N}^\mathrm{R,22}
    \end{pmatrix}$ and 
    $\check{F}^\mathrm{R}
    =
    \begin{pmatrix}
        0 & F^\mathrm{R,12}
        \\
        F^\mathrm{R,21} & 0
    \end{pmatrix}$.
The advanced GF is defined similarly.
Here, $\Psi_\sigma(x,t)$ is the Heisenberg representation of an annihilation operator with spin $\sigma$, spatial position $x$, and time $t$. 
$\eta$ is a positive infinitesimal number.
$\bar{\Gamma}_\mathrm{e}(E)$ can be divided into normal transmission $\bar{\Gamma}_\mathrm{N}(E)$ and Andreev reflection $\bar{\Gamma}_F(E)$ terms,
\begin{align}
    \bar{\Gamma}_\mathrm{e}(E)=\bar{\Gamma}_\mathrm{N}(E)+\bar{\Gamma}_F(E),
    \label{eq:decompose1}
\end{align}
with
$\Gamma_\mathrm{N}(x,x',E) = \gamma(\tilde{G}_{\mathrm{N}},\tilde{G}_{\mathrm{N}})$ and 
$\Gamma_F(x,x',E) = \gamma(\tilde{F},\tilde{F})$.
There are no cross terms between normal and anomalous GFs.


\emph{Even and odd-frequency pairing contributions.}---
\begin{figure}[t]
    \centering
    \includegraphics[width=8.0cm]{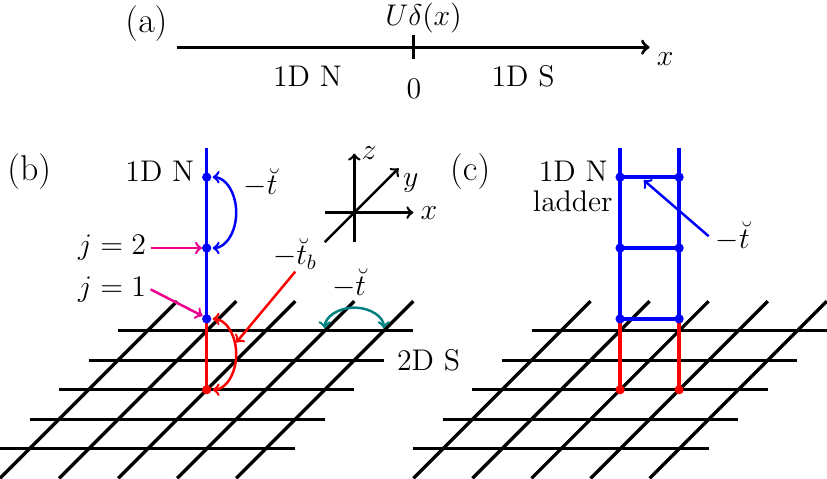}
    \caption{
    Schematic illustration of three types of junctions.
    (a) Continuum 1D N/1D S junction,
    (b) 1D N/2D S lattice model, and 
    (c) 1D N ladder/2D S lattice model.
    }
    \label{fig:schematic}
\end{figure}
In NS junctions, OFP induced at the interface can penetrate into the N region and contribute to Andreev reflection.
We decompose $\Gamma_F(x,x',E)$ into EFP and OFP components.
The advanced (retarded) GF can be written as the sum of even and odd components
$\check{F}^{\mathrm{A(R)}}=\check{F}^{\mathrm{A(R),even}}+\check{F}^{\mathrm{A(R),odd}}$.
Then, $\Gamma_F$ is decomposed as $\Gamma_F=\Gamma_F^\mathrm{ee}+\Gamma_F^\mathrm{oo}+\Gamma_F^\mathrm{eo}$
with
\begin{align}
    &
    \Gamma_F^{\mathrm{ee(oo)}}(x,x',E)
    =
    \gamma(\tilde{F}^\mathrm{even(odd)},\tilde{F}^\mathrm{even(odd)}),
    \\
    &
    \Gamma_F^{\mathrm{eo}}(x,x',E)
    =
    \gamma(\tilde{F}^\mathrm{even},\tilde{F}^\mathrm{odd})
    +
    \gamma(\tilde{F}^\mathrm{odd},\tilde{F}^\mathrm{even}).
\end{align}
We analyze the odd-frequency contribution to $\bar{\Gamma}_F(E)$ for three distinct systems illustrated in Fig.~\ref{fig:schematic}. 
Remarkably, we demonstrate that $\bar{\Gamma}_F^\mathrm{ee}(E)=\bar{\Gamma}_F^\mathrm{oo}(E)$~\cite{current_conserve}.
$\bar{\Gamma}_F^\mathrm{eo}(E)$ is zero due to particle-hole symmetry [proof is given in the Supplemental Material (SM)~\cite{SM}]. 
Hence, we do not discuss it.
Figure~\ref{fig:schematic}(a) shows the continuum 1D NS junction, where we analytically prove the equal contribution of EFP and OFP to $\bar{\Gamma}_F(E)$. 
Figure~\ref{fig:schematic}(b) shows the 1D N/2D S junction inspired by scanning tunneling spectroscopy.
In this setup, we analyze  $s$-, $p_x$-, and $d$-wave SPPs. 
We demonstrate that only $s$-wave junctions exhibit Andreev reflection since both EFP and OFP vanish at the interface between 1D N and 2D S for $p_x$- and $d$-wave junctions. 
Hence, they cannot penetrate into the 1D N. 
These cancellations do not occur for the setup shown in Fig.~\ref{fig:schematic}(c), where the normal metal has more spatial structure.

\emph{1D N/1D S continuum model.}---
\begin{figure}[t]
    \centering
    \includegraphics[width=8.6cm]{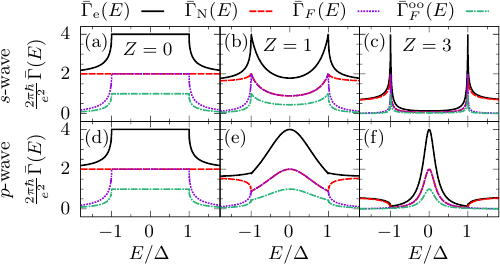}
    \caption{
    $\bar{\Gamma}_\mathrm{e}(E)$ and its components are plotted as a function of $E$ for several values of $Z=2mU/(k_\mathrm{F}\hbar^2)$.
    (a)--(c) $s$-wave and (d)--(f) $p$-wave junctions.
    $Z=0$ for (a) and (d), $1$ for (b) and (e), and $3$ for (c) and (f).
    $\Delta/\mu=0.01$ for all plots.
    }
    \label{fig:Gamma_continuum}
\end{figure}
We now present our analytical results for the 1D N/1D S continuum model.
The Bogoliubov-de Gennes (BdG) Hamiltonian is
$H(x,x')=\delta(x-x')\hat{\sigma}_0\hat{\tau}_3\varepsilon(x)+\Theta(x)\Theta(x')\Delta(x,x')$ with $\varepsilon(x)=-\frac{\hbar^2}{2m}\frac{d^2}{dx^2}-\mu+U\delta(x)$, $\mu$ the chemical potential, $U$ the barrier potential at the interface, and $\hat{\sigma}_{j=0,1,2,3}$ Pauli matrices in spin space.
As the SPP $\Delta(x,x')$, we study $s$-wave and $p$-wave cases: $\Delta(x,x')=\Delta\delta(x-x')i\hat{\sigma}_2i\hat{\tau}_2$ for $s$-wave, $\Delta(x,x')=\Delta\hat{\sigma}_{1}\int dk  \begin{pmatrix}0&e^{ik(x-x')}\\-e^{-ik(x-x')}&0\end{pmatrix}\mathrm{sgn}(k)$ for $p$-wave [see also Fig.~\ref{fig:schematic}(a)].
We define the dimensionless parameter $Z=\frac{2mU}{k_\mathrm{F}\hbar^2}$ with $k_\mathrm{F}=\sqrt{\frac{2m\mu}{\hbar^2}}$.
We derive the GFs along the lines of Ref.~\cite{PhysRev.175.559,PhysRevB.41.4017}.
Explicit expressions are given in the SM~\cite{SM}.
Employing Eqs.~\eqref{eq:def_conductance} and \eqref{eq:def_Gamma_e}, we reproduce the differential conductance of Blonder, Tinkham, and Klapwijk (BTK) theory~\cite{PhysRevB.25.4515,doi:10.1143/JPSJ.61.1685}: $\bar{\Gamma}_\mathrm{N}(E)=\frac{e^2}{\pi\hbar}[1-{|b(E)|}^2]$ and $\bar{\Gamma}_F(E)=\frac{e^2}{\pi\hbar}\frac{k_\mathrm{h}}{k_\mathrm{e}}{|a(E)|}^2$, where the electron (hole) wave number is given by $k_\mathrm{e(h)}=\sqrt{\frac{2m}{\hbar^2}[\mu+(-) E]}$, and 
$a(E)$ and $b(E)$ are hole (Andreev) and electron reflection coefficients, respectively. 
We choose $L_1=-\infty$ and $L_2=0$.

The EFP and OFP contributions are 
\begin{align}
    &
    \frac{4\pi\hbar}{e^2}
    \Gamma_F^\mathrm{ee(oo)}(x,x',E)
    =
    \frac{k_\mathrm{h}}{k_\mathrm{e}}{|a(E)|}^2
    +
    \frac{k_\mathrm{e}}{k_\mathrm{h}}{|a(-E)|}^2
    \nonumber\\
    &
    -(+)
    \frac{{(k_\mathrm{e}+k_\mathrm{h})}^2}{2k_\mathrm{e}k_\mathrm{h}}
    \mathrm{Re}
    \left[
        a(E)a^*(-E)
        e^{-i(k_\mathrm{e}-k_\mathrm{h})(x+x')}
    \right]
    \nonumber\\
    &
    +(-)
    \frac{{(k_\mathrm{e}-k_\mathrm{h})}^2}{2k_\mathrm{e}k_\mathrm{h}}
    \mathrm{Re}
    \left[
        a(E)a(-E)
        e^{-i(k_\mathrm{e}+k_\mathrm{h})(x+x')}
    \right].
    \label{eq:Gamma_F_ee_oo_x_continuum}
\end{align}
After averaging over $x$ and $x'$, the last two terms in Eq.~\eqref{eq:Gamma_F_ee_oo_x_continuum} vanish.
Then, we obtain $\bar{\Gamma}_F^\mathrm{ee}(E)=\bar{\Gamma}_F^\mathrm{oo}(E)$
for $E\neq0$~\cite{k_particle_hole} (see the SM~\cite{SM} for further details).
For the $s$-wave junction with a fully transparent barrier [$Z=0$ shown Fig.~\ref{fig:Gamma_continuum}(a)],
perfect Andreev reflection occurs, and $\frac{2\pi\hbar}{e^2}\bar{\Gamma}_\mathrm{e}(E)\sim4$ holds for $|E|<|\Delta|$~\cite{Gamma_almost_two}.
As the value of $Z$ increases [$Z=1$ and $Z=3$ shown in Figs.~\ref{fig:Gamma_continuum}(b) and (c), respectively], the shape of $\bar{\Gamma}_\mathrm{e}(E)$ approaches the U-shaped density of states reflecting the $s$-wave SPP\@.
Accordingly, the amplitude of Andreev reflection is suppressed.
For any values of $Z$, $\bar{\Gamma}_\mathrm{N}(E)=\bar{\Gamma}_F(E)$ holds for $|E|<|\Delta|$ due to the normalization of the coefficients: $\frac{k_\mathrm{h}}{k_\mathrm{e}}{|a(E)|}^2+{|b(E)|}^2=1$ for $|E|<|\Delta|$~\cite{PhysRevB.25.4515}.
The presence of Andreev reflection [$\bar{\Gamma}_F(E)\neq0$] is, thus, inherently connected to the presence of
OFP~\cite{Lee2019_nature,Parab2019_scientific_reports,Soulen_science,Zareapour_2017,PhysRevB.99.014510,C4NR07262F}.
For $p$-wave junctions [Figs.~\ref{fig:Gamma_continuum}(d)--(f)], $\bar{\Gamma}_\mathrm{e}(E=0)$ takes a constant value due to the presence of a Majorana state~\cite{A_Yu_Kitaev_2001,PhysRevB.101.214507,PhysRevB.101.094506}. 
Half of it stems from Andreev reflection $\bar{\Gamma}_F(E=0)$.
Experimental conductance exhibiting a zero energy peak larger than the value of the normal state signifies the existence of Andreev reflection, a distinct indicator of the presence of OFP. 
Hence, in Refs.~\cite{doi:10.1126/science.1222360,PhysRevLett.114.017001,PhysRevLett.116.257003,Sun2017,Heedt2021}, signatures of OFP have been observed in retrospect.

\emph{1D N/2D S lattice model.}---
\begin{figure}[t]
    \centering
    \includegraphics[width=8.6cm]{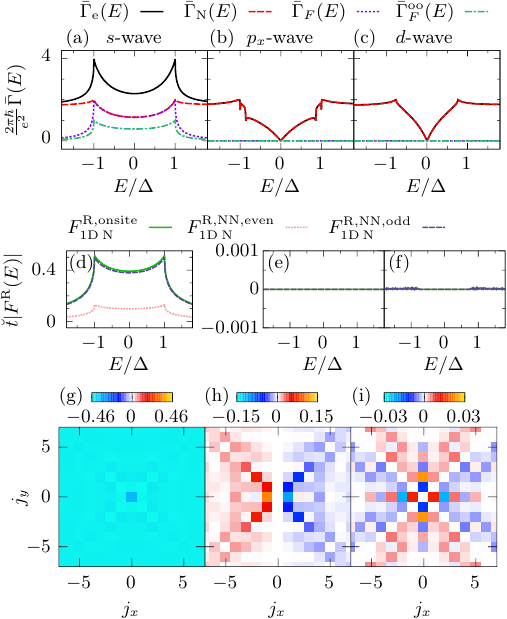}
    \caption{
    (a)--(c) $\bar{\Gamma}_\mathrm{e}(E)$ and its components are plotted as a function of $E$.
    $\bar{\Gamma}^\mathrm{eo}_F(E)=0$ numerically and is not plotted.
    (d)--(f) The absolute value of onsite and NN retarded GF in 1D N is plotted as a function of $E$.
    (a)--(f) Averaging length $L=500$, $L_x=10^7$, and $\eta/\breve{t}=10^{-7}$.
    (g)--(i) Onsite component of the anomalous GF in Matsubara frequency representation in the 2D S close to the 1D N is plotted as functions of $j_x$ and $j_y$ with Matsubara frequency $\omega_n/\Delta=0.1$ and $L_x=2000$. 
    (a), (d) and (g) $s$-wave, (b), (e) and (h) $p_x$-wave, and (c), (f) and (i) $d$-wave S junctions.
    (g) $\breve{t}\mathrm{Re}F_{\mathrm{2D,SS}}^\mathrm{onsite,even}$,
    (h) $\breve{t}\mathrm{Re}F_{\mathrm{2D,ST}}^\mathrm{onsite,odd}$, and
    (i) $\breve{t}\mathrm{Re}F_{\mathrm{2D,SS}}^\mathrm{onsite,even}$.
    The imaginary part for (g)--(i) is zero.
    }
    \label{fig:1d_2d_v5}
\end{figure}
Let us now consider the model illustrated in Fig.~\ref{fig:schematic}(b).
The Hamiltonian is given by
\begin{align}
    H
    =&
    -\breve{t}\sum_{j>0,\sigma}
    \left(
        c_{j,\sigma}^\dagger c_{j+1,\sigma}+\mathrm{H.c.}
    \right)
    -\mu_\mathrm{N}\sum_{j>0,\sigma}c_{j,\sigma}^\dagger c_{j,\sigma}
    \nonumber\\
    &
    -\breve{t}_\mathrm{b}
    \left(
        c_{1,\sigma}^\dagger b_{\mathbf{j}_0,\sigma}+\mathrm{H.c.}
    \right)
    + H_\Delta
    \nonumber\\
    &
    -\breve{t}\sum_{\langle\mathbf{i},\mathbf{j}\rangle,\sigma}
    \left(
        b_{\mathbf{i},\sigma}^\dagger b_{\mathbf{j},\sigma}+\mathrm{H.c.}
    \right)
    -\mu_\mathrm{S}\sum_{\mathbf{j},\sigma}b_{\mathbf{j},\sigma}^\dagger b_{\mathbf{j},\sigma},
\end{align}
where $c_{j,\sigma}$ ($b_{\mathbf{j},\sigma}$) is an annihilation operator in 1D N (2D S) with the $j$-th ($\mathbf{j}$)-th site and spin $\sigma$.
Here, $\breve{t}$ is a hopping integral within 1D N and 2D S, $\breve{t}_\mathrm{b}$ is a hopping integral between 1D N and 2D S, $\mu_\mathrm{N(S)}$ is a chemical potential in 1D N (2D S), and $\mathbf{j}_0=(0,0)$.
We utilize $\Delta/\breve{t}=0.1$, $\breve{t}_\mathrm{b}/\breve{t}=1$, $\mu_\mathrm{N}/\breve{t}=-0.5$, and $\mu_\mathrm{S}/\breve{t}=-1$.
We impose periodic boundary conditions in the $x$-direction with $L_x$ sites and an infinite system in the $y$-direction~\cite{bondary_condition}.
We consider $s$-, $p_x$-, and $d$-wave SPPs for $H_\Delta=\sum_{k,\sigma,\sigma'} b_{k,\sigma}^\dagger\hat{\Delta}_{\sigma,\sigma'}(k)b_{-k,\sigma'}^\dagger +\mathrm{H.c.}$ with momentum $k$, where $\hat{\Delta}_{\sigma,\sigma'}(k)$ is given by 
$\Delta i\hat{\sigma}_2$,
$\Delta\sin k_x \hat{\sigma}_3 i\hat{\sigma}_2$, and 
$\frac{\Delta}{2}(\cos k_x-\cos k_y)i\hat{\sigma}_2$, respectively.
Without loss of generality, we assume that $\Delta$ is real and positive.

For the lattice model, we use a discretized version of Eq.~\eqref{eq:def_Gamma_e}:
$\Gamma_\mathrm{e}(j,j',E)
    =
    \pi\hbar
    \mathrm{Tr}
    \left[
        \hat{P}_\mathrm{e}
        \hat{J}_j
        \hat{\tilde{G}}_{j,j'}(E)
        \hat{J}_{j'}
        \hat{\tilde{G}}_{j',j}(E)
    \right]$ with $\hat{P}_\mathrm{e}=\mathrm{diag}(P_\mathrm{e},P_\mathrm{e})$, $\hat{J}_j
    =
    \begin{pmatrix}
        0 & J_{j,j+1}
        \\
        J_{j+1,j} & 0
    \end{pmatrix}
    $, $J_{j,j+1}=J_{j+1,j}^*=\frac{e\breve{t}}{i\hbar}\hat{\sigma}_0\hat{\tau}_0$, and 
    $\hat{\tilde{G}}_{j,j'}
    =
    \begin{pmatrix}
        \tilde{G}_{j,j'} & \tilde{G}_{j,j'+1}
        \\
        \tilde{G}_{j+1,j'} & \tilde{G}_{j+1,j'+1}
    \end{pmatrix}$~\cite{PhysRevB.23.6851,PhysRevLett.47.882}. 
Here, the trace in $\Gamma_\mathrm{e}(j,j',E)$ is taken for the spin, particle-hole, and neighboring two spatial lattice sites spanned from $j$ to $j+1$.
The spatial average is defined by $\bar{\Gamma}_\mathrm{e}(E) = \frac{1}{L^2}\sum_{j,j'=1}^{L} \Gamma_\mathrm{e}(j,j',E)$ with $j$ and $j'$ chosen in the 1D N region.
As shown in Figs.~\ref{fig:1d_2d_v5}(a)--(c), only the $s$-wave junction has a non-zero Andreev reflection [$\bar{\Gamma}_F(E)\neq 0$]. 
For $p$-wave and $d$-wave junctions, $\bar{\Gamma}_\mathrm{e}(E)$ exhibits a V-shaped structure reflecting the density of states~\cite{Pan2001,RevModPhys.79.353,doi:10.1143/JPSJ.81.011005}.
Numerical equivalence of $\bar{\Gamma}_F^\mathrm{ee}$ and $\bar{\Gamma}_F^\mathrm{oo}$ is shown in the SM~\cite{SM}.
The onsite ($j=1$) and nearest neighbor (NN) between $j=1$ and $j=2$ components [see Fig.~\ref{fig:schematic}(b)] of the retarded anomalous GF in the 1D N are plotted in Figs.~\ref{fig:1d_2d_v5}(d)--(f)~\cite{def_F_ret}.
In Figs.~\ref{fig:1d_2d_v5}(d) and (f), $s$- and $d$-wave junctions, respectively, the spin-singlet (SS) EFP and OFP components are shown, and in Figs.~\ref{fig:1d_2d_v5}(e), the $p_x$-wave junction, and the spin-triplet (ST) EFP and OFP components are shown~\cite{spin_rotate_symmetry}.
In Fig.~\ref{fig:1d_2d_v5}(d), we confirm that EFP and OFP penetrate into 1D N~\cite{GRE}.
For $p_x$- and $d$-wave cases, both EFP and OFP do not penetrate into 1D N [Figs.\ref{fig:1d_2d_v5}(e) and (f)]~\cite{Numerial_error_d}.

Let us explain why EFP and OFP can (cannot) penetrate into 1D N for the $s$-wave ($p_x$- and $d$-wave) junction.
As an example, the onsite components of the anomalous GF in 2D S close to 1D N are illustrated in Figs.~\ref{fig:1d_2d_v5}(g)--(i) (NN pairings are shown in the SM~\cite{SM}).
We define the onsite SS EFP (ST OFP) component of the anomalous GF with Matsubara frequency ($\omega_n$) in 2D S as follows
\begin{align}
    &
    F_\mathrm{2D,SS(ST)}^\mathrm{onsite,even(odd)}(\mathbf{j},i\omega_n)
    \nonumber\\
    =&
    \frac{1}{4}\sum_{\zeta=\pm1}
    g(\zeta)
    \left[
        F_{\mathbf{j},\mathbf{j},\uparrow,\downarrow}^{12}(\zeta i\omega_n)
        -(+)
        F_{\mathbf{j},\mathbf{j},\downarrow,\uparrow}^{12}(\zeta i\omega_n)
    \right]
\end{align}
with $g(\pm1)=1$ for SS EFP and $g(\pm1)=\pm1$ for ST OFP~\cite{why_matsubara}.
When  $F_\mathrm{2D,SS(ST)}^\mathrm{onsite,even(odd)}(\mathbf{j}_0,i\omega_n)$ is non-zero, the onsite pairing can penetrate into 1D N\@.

For the $s$-wave junction, the onsite anomalous GF (SS EFP) does not exhibit a sign change due to the isotropy of the $s$-wave SPP.
Hence, this onsite pairing can penetrate into 1D N [Fig.~\ref{fig:1d_2d_v5}(g)].
There are no cancellations for NN EFP and OFP. 
Thus, they can also penetrate into 1D N [Fig.~\ref{fig:1d_2d_v5}(d)].
For the $p_x$-wave junction, the onsite anomalous GF (ST OFP)  [Fig.~\ref{fig:1d_2d_v5}(h)] exhibits a sign change at $j_x=0$ since the $p_x$-wave SPP changes its sign in the $\pm x$ direction. 
Then, OFPs cancel each other at $j_x=0$ and cannot penetrate into 1D N\@.
For the $d$-wave S junction [Fig.~\ref{fig:1d_2d_v5}(i)], the onsite anomalous GF (SS EFP) also exhibits a sign change at $j_x=\pm j_y$ reflecting $d$-wave symmetry.
Then, the EFP contributions cancel each other and cannot penetrate into 1D N\@.
For $p_x$-wave and $d$-wave junctions, NN EFP contributions also cancel each other and cannot penetrate into 1D N~\cite{SM}.
The same argument applies to NN OFP contributions.

\emph{1D N ladder/2D S model.}---
\begin{figure}[t]
    \centering
    \includegraphics[width=8.6cm]{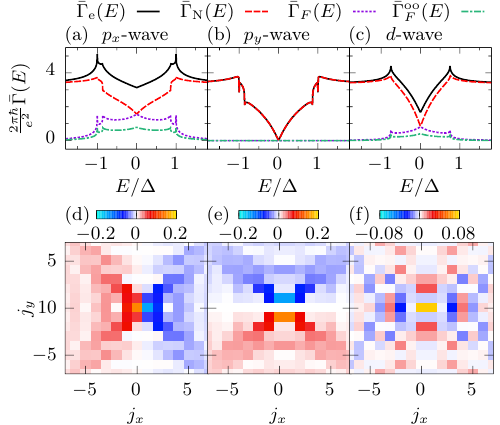}
    \caption{
    (a)--(c) $\bar{\Gamma}_\mathrm{e}(E)$ and its components are plotted as a function of $E$ with $L=500$, $L_x=2\times10^6$, and $\eta/\breve{t}=10^{-7}$.
    $\bar{\Gamma}^\mathrm{eo}_F(E)=0$ numerically and is not plotted.
    (d)--(f) Onsite component of the anomalous GF close to the 1D N ladder is plotted as functions of $j_x$ and $j_y$ at $\omega_n/\Delta=0.1$ with $L_x=2000$.
    (a) and (d) $p_x$-wave, (b) and (e) $p_y$-wave, and (c) and (f) $d$-wave S junction.
    (d) and (e) $\breve{t}$Re$F_\mathrm{2D,ST}^\mathrm{onsite,odd}$, 
    (f) $\breve{t}$Re$F_\mathrm{2D,SS}^\mathrm{onsite,even}$.
    The imaginary part for (d)--(f) is zero.
    }
    \label{fig:Gamma_ladder_v6}
\end{figure}
From the results of the 1D N/2D S junctions, we expect that EFP and OFP can penetrate into the N lead 
if we replace the 1D N lead with a 1D N ladder [Fig.~\ref{fig:schematic}(c)].
Note that this setup mimics a double tip in STS experiments.
The 1D N ladder is connected to $(j_x,j_y)=(0,0)$ and $(1,0)$.
We plot $\bar{\Gamma}_\mathrm{e}(E)$ and its components in Figs.~\ref{fig:Gamma_ladder_v6}(a)--(c) accompanied with the onsite pairing of anomalous GFs in Figs.~\ref{fig:Gamma_ladder_v6}(d)--(f) for $p_x$-, $p_y$-, and $d$-wave junctions.
NN pairings are shown in the SM~\cite{SM}.
The SPP for the $p_y$-wave case is given by $\hat{\Delta}(k)=\Delta\sin k_y \hat{\sigma}_3i\hat{\sigma}_2$.
For $p$-wave junctions, depending on the orientation of SPPs ($p_x$- or $p_y$-wave), EFP and OFP can penetrate into the 1D N ladder.
For the $p_x$-wave junction [Fig.~\ref{fig:Gamma_ladder_v6}(a)], we observe that EFP and OFP contribute to $\bar{\Gamma}_\mathrm{e}(E)$ since onsite OFPs in the $x$-direction do not cancel each other as shown in Fig.~\ref{fig:Gamma_ladder_v6}(d).
However, for the $p_y$-wave junction [Fig.~\ref{fig:Gamma_ladder_v6}(b)], $\bar{\Gamma}_\mathrm{e}(E)$ and its components are qualitatively the same as the ones in Fig.~\ref{fig:1d_2d_v5}(b).
Then, the OFP contributions cancel each other [Fig.~\ref{fig:Gamma_ladder_v6}(e)] (NN EFPs and NN OFPs also cancel and cannot penetrate into the 1D N ladder~\cite{SM}).
For the $d$-wave junction, shown in Fig.~\ref{fig:Gamma_ladder_v6}(c), EFP and OFP contribute to $\bar{\Gamma}_\mathrm{e}(E)$,
where the EFP contributions do not cancel each other [Fig.~\ref{fig:Gamma_ladder_v6}(f)].

\emph{Conclusions.}---
We have analyzed the impact of even- and odd-frequency pairing on the conductance across generic NS junctions based on linear response theory. We have identified an equal balance of even- and odd-frequency pairing contributions to the conductance related to Andreev reflection. The larger the transparency across the junction, the more pronounced are these contributions typically. 
Hence, we prove that the presence of Andreev reflection in transport across NS junctions manifests the existence of odd-frequency pairing in a variety of hybrid structures.

\begin{acknowledgments}
We thank Y. Tanaka for helpful discussions. This work was supported by the W{\"u}rzburg-Dresden Cluster of Excellence ct.qmat, EXC2147, project-id 390858490, the DFG (SFB 1170), and the Bavarian Ministry of Economic Affairs, Regional Development and Energy within the High-Tech Agenda Project ``Bausteine f{\"u}r das Quanten Computing auf Basis topologischer Materialen''.
\end{acknowledgments}

\let\oldaddcontentsline\addcontentsline
\renewcommand{\addcontentsline}[3]{}
\bibliography{bibliography}

\begin{thebibliography}{65}
\bibitem [{cur()}]{current_conserve}%
  \BibitemOpen
  \href@noop {} {}\bibinfo {note} {$\Gamma_\mathrm{e}(x,x',E)$, $\Gamma_\mathrm{N}(x,x',E)$, and $\Gamma_F(x,x',E)$ do not depend on $x$, and $x'$ due to current conservation. However, $\Gamma_F^\mathrm{ee}(x,x',E)$, and $\Gamma_F^\mathrm{oo}(x,x',E)$ depend on $x$ and $x'$, and spatial averaging is needed.}\BibitemShut {Stop}%
\bibitem [{k_p()}]{k_particle_hole}%
  \BibitemOpen
  \href@noop {} {}\bibinfo {note} {At $E=0$, the particle and hole momentum coincide ($k_\mathrm{e}=k_\mathrm{h}=k_\mathrm{F}$), and the decompositon into EFP and OFP is not unique. On the other hand, for $E\neq0$, the decomposition is unique.}\BibitemShut {Stop}%
\bibitem [{bon()}]{bondary_condition}%
  \BibitemOpen
  \href@noop {} {}\bibinfo {note} {We utilize the recursive Green function method~\cite{PhysRevB.55.5266} to calculate the Green function. Then, the system can be infinite in one direction, but we must impose periodic or open boundary conditions in other directions.}\BibitemShut {Stop}%
\bibitem [{def()}]{def_F_ret}%
  \BibitemOpen
  \href@noop {} {}\bibinfo {note} {In Figs.~\ref{fig:1d_2d_v5}(d)--(f), $F^{\mathrm{R,onsite}}_\mathrm{1D\:N}$ and $F^{\mathrm{R,NN,even(odd)}}_\mathrm{1D\:N}$ are defined as follows. For $s$- and $d$-wave junctions, the onsite component is SS $s$-wave EFP: \begin{align} F^\mathrm{R,onsite}_\mathrm{1D\:N,SS}(E) = \frac{1}{2} \left[ F^\mathrm{R,12}_{1,1,\uparrow,\downarrow}(E) - F^\mathrm{R,12}_{1,1,\downarrow,\uparrow}(E) \right], \end{align} and the NN components are SS $s$-wave EFP and SS $p$-wave OFP: \begin{align} & F^\mathrm{R,NN,even(odd)}_\mathrm{1D\:N,SS}(E) = \frac{1}{4} \left[ F^\mathrm{R,12}_{1,2,\uparrow,\downarrow}(E) - F^\mathrm{R,12}_{1,2,\downarrow,\uparrow}(E) \right] \nonumber\\ & +(-) \frac{1}{4} \left[ F^\mathrm{A,12}_{1,2,\uparrow,\downarrow}(-E) - F^\mathrm{A,12}_{1,2,\downarrow,\uparrow}(-E) \right]. \end{align} For $p$--wave junctions, the onsite component is ST $s$-wave OFP: \begin{align} F^\mathrm{R,onsite}_\mathrm{1D\:N,ST}(E) = \frac{1}{2} \left[
  F^\mathrm{R,12}_{1,1,\uparrow,\downarrow}(E) + F^\mathrm{R,12}_{1,1,\downarrow,\uparrow}(E) \right], \end{align} and the NN components are ST $p$-wave EFP and ST $s$-wave OFP: \begin{align} & F^\mathrm{R,NN,even(odd)}_\mathrm{1D\:N,ST}(E) = \frac{1}{4} \left[ F^\mathrm{R,12}_{1,2,\uparrow,\downarrow}(E) + F^\mathrm{R,12}_{1,2,\downarrow,\uparrow}(E) \right] \nonumber\\ & +(-) \frac{1}{4} \left[ F^\mathrm{A,12}_{1,2,\uparrow,\downarrow}(-E) + F^\mathrm{A,12}_{1,2,\downarrow,\uparrow}(-E) \right]. \end{align}}\BibitemShut {NoStop}%
\bibitem [{spi()}]{spin_rotate_symmetry}%
  \BibitemOpen
  \href@noop {} {}\bibinfo {note} {Due to spin rotational symmetry, $s$-wave and $d$-wave junctions can only have SS components, and $p$-wave S junctions can only have ST components. For onsite pairings, the SS OFP is forbidden by Fermi-Dirac statistics.}\BibitemShut {Stop}%
\bibitem [{GRE()}]{GRE}%
  \BibitemOpen
  \href@noop {} {}\bibinfo {note} {The EFP and OFP components of retarded GFs are not even and odd function of $E$, respectively. EFP and OFP satisfy $F^\mathrm{R,even}(E)=F^\mathrm{A,even}(-E)$ and $F^\mathrm{R,odd}(E)=-F^\mathrm{A,odd}(-E)$, respectively.}\BibitemShut {Stop}%
\bibitem [{Num()}]{Numerial_error_d}%
  \BibitemOpen
  \href@noop {} {}\bibinfo {note} {The numerical error for $d$-wave junctions is larger than that for $p_x$-wave junctions since EFP and OFP contributions between the $x$- and $y$-direction cancel each other for $d$-wave junctions. We adopt periodic boundary conditions in the $x$-direction and infinite length in the $y$-direction. When the system size in the $x$-direction is finite, there is no perfect cancelation in the $x$ and $y$-direction [Fig.~\ref{fig:1d_2d_v5}(f)]. However, for $p_x$-wave junctions, EFP and OFP in the $+x$- and $-x$-direction cancel. Hence, in this case, there is almost no finite size effect [Fig.~\ref{fig:1d_2d_v5}(e)].}\BibitemShut {Stop}%
\bibitem [{why()}]{why_matsubara}%
  \BibitemOpen
  \href@noop {} {}\bibinfo {note} {We use the Matsubara frequency representation to reduce finite-size effects. To calculate the GF in real space, we cannot access large $L_x$ due to the limitation of numerical resources. For finite $L_x$, the finite size effects of the GF reduce when we adopt the Matsubara frequency representation for not too small Matsubara frequency.}\BibitemShut {Stop}%
\end{thebibliography}
\let\addcontentsline\oldaddcontentsline

\clearpage
\onecolumngrid
\makeatletter
\widetext
\setcounter{secnumdepth}{3}
\setcounter{equation}{0}
\setcounter{figure}{0}
\setcounter{section}{0}
\renewcommand{\thesection}{S\arabic{section}}
\renewcommand{\theequation}{S\arabic{equation}}
\renewcommand{\thefigure}{S\arabic{figure}}

\begin{center}
\textbf{\large Supplemental Material: Equal contribution of even and odd frequency pairing to transport across normal metal-superconductor junctions}
\end{center}

\tableofcontents
\section{\label{sub:continuum_model}Continuum models}

\subsection{\label{sub:conductance_linear}Conductance in 1D continuum model}
In linear response theory, the conductance $G$ given between two points in space $L_1$ and $L_2$ is written in the form
\begin{align}
    G
    =&
    -
    \int dE\frac{df(E)}{dE}
    \bar{\Gamma}_\mathrm{e}(E),
    \label{eq:conductance}
    \\
    \bar{\Gamma}_\mathrm{e}(E)
    =&
    \frac{1}{{(L_2-L_1)}^2}
    \int_{L_1}^{L_2} dx dx'
    \Gamma_\mathrm{e}(x,x',E),
    \label{eq:spatial_average_app}
\end{align}
with
\begin{align}
    \Gamma_\mathrm{e}(x,x',E)
    =&
    \frac{-e^2\hbar^3\pi}{4m^2}
    \mathrm{Tr}
    \left[
        P_\mathrm{e}
        \tilde{G}_{1}(x,x',E)\overleftrightarrow{\nabla} \overleftrightarrow{\nabla}' \tilde{G}_{2}(x',x,E)
    \right],
    \label{eq:def_Gamma_e_app}
    \\
    g(x)\overleftrightarrow{\nabla} h(x)
    =&
    g(x)\partial_x h(x) - [\partial_x g(x)] h(x),
    \\
    \tilde{G}_1(x,x',E)
    =&
    \begin{cases}
        \frac{1}{2\pi i}
        \left[
            \check{G}^\mathrm{A}(x,x',E) - \check{G}^\mathrm{R}(x,x',E)
        \right],
        & x=x',
        \\
        \frac{1}{\sqrt{2}\pi i}
        \left[
            \sin\theta\check{G}^\mathrm{A}(x,x',E) - \cos\theta\check{G}^\mathrm{R}(x,x',E)
        \right],
        & x\neq x',
    \end{cases}
    \label{eq:def_tilde1}
    \\
    \tilde{G}_2(x,x',E)
    =&
    \begin{cases}
        \frac{1}{2\pi i}
        \left[
            \check{G}^\mathrm{A}(x,x',E) - \check{G}^\mathrm{R}(x,x',E)
        \right],
        & x=x',
        \\
        \frac{1}{\sqrt{2}\pi i}
        \left[
            \cos\theta\check{G}^\mathrm{A}(x,x',E) - \sin\theta\check{G}^\mathrm{R}(x,x',E)
        \right],
        & x\neq x',
    \end{cases}
    \label{eq:def_tilde2}
    \\
    P_\mathrm{e}=&\frac{1}{2}(\hat{\tau}_0+\hat{\tau}_3),
\end{align}
for continuum systems with Pauli matrices $\hat{\tau}_{j=0,1,2,3}$ acting on particle-hole space.
Here, $x$ and $x'$ are chosen in the normal metal region since the charge current is not conserved in the superconducting region.
$\check{G}^{\mathrm{A(R)}}(x,x',E)$ is the advanced (retarded) Green function, $\check{G}^{\mathrm{A(R)}}(x,x',E)=\check{G}(x,x',E-(+)i\eta)$ with a positive infinitesimal number $\eta$, 
$\overleftrightarrow{\nabla}'$ acts on $x'$, $\mathrm{Tr}$ in Eq.~\eqref{eq:def_Gamma_e_app} is taken for inner degrees of freedom e.g., spin and orbital, and $f(E)$ is the Fermi-Dirac distribution function.
$\theta\in\mathbb{R}$ is an inner degree of freedom. Any choice of $\theta$ results in the same $\bar{\Gamma}_\mathrm{e}(E)$ due to the following relation:
\begin{align}
    0=&
    \mathrm{Tr}
    \left[
        P_\mathrm{e}
        \check{G}^\mathrm{A}(x,x',E)\overleftrightarrow{\nabla} \overleftrightarrow{\nabla}' \check{G}^\mathrm{A}(x',x,E)
    \right]
    =
    \mathrm{Tr}
    \left[
        P_\mathrm{e}
        \check{G}^\mathrm{R}(x,x',E)\overleftrightarrow{\nabla} \overleftrightarrow{\nabla}' \check{G}^\mathrm{R}(x',x,E)
    \right]
\end{align}
for $x\neq x'$.
Equation~\eqref{eq:conductance} in combination with Eqs.~\eqref{eq:spatial_average_app}, \eqref{eq:def_Gamma_e_app}, \eqref{eq:def_tilde1}, and \eqref{eq:def_tilde2} with $\theta=\pi/4$ is called the Kubo-Greenwood formula~\cite{D_A_Greenwood_1958,PhysRevB.40.8169}.
For $x\neq x'$, Eqs.~\eqref{eq:def_tilde1} and \eqref{eq:def_tilde2} with $\theta=n\pi/2$ reduces to
\begin{align}
    \Gamma_\mathrm{e}(x,x',E)
    =&
    \frac{-e^2\hbar^3\pi}{4m^2}
    \frac{1}{2\pi^2}
    \mathrm{Tr}
    \left[
        P_\mathrm{e}
        \check{G}^\mathrm{A}(x,x',E)\overleftrightarrow{\nabla} \overleftrightarrow{\nabla}' \check{G}^\mathrm{R}(x',x,E)
    \right],
    \label{eq:Caroli_app}
\end{align}
which is called Caroli formula~\cite{C_Caroli_1971}.
Due to the charge current conservation, Eq.~\eqref{eq:spatial_average_app} can be recast into the form
\begin{align}
   \bar{\Gamma}_\mathrm{e}(E)
    =&
    \Gamma_\mathrm{e}(x,x',E),
    \label{eq:current_conservation}
\end{align}
where the right hand side of Eq.~\eqref{eq:current_conservation} does not depend on the positions $x$ and $x'$~\cite{PhysRevB.40.8169,PhysRevB.23.6851,PhysRevLett.47.882}.
We assume one dimensional conductors in Eq.~\eqref{eq:spatial_average_app}, but extensions to higher dimensions are straightforward.

We set $\theta=\pi/2$ unless it is specified otherwise since numerical errors become smallest at $\theta=n\pi/2$ with $n\in\mathbb{Z}$.

\subsection{\label{sub:odd_freq}Odd frequency pairing}
Odd-frequency pairing is described by the anomalous Green function in superconductor junctions.
The full Green function is given by 
\begin{align}
&
    \check{G}((x,\tau),(x',0))
    =
    -
    \langle\boldsymbol{\psi}(x,\tau)\boldsymbol{\psi}^\dagger(x',0)\rangle,
    \label{eq:def_G_time}
\end{align}
where ${[\boldsymbol{\psi}(x,\tau)]}^\mathrm{T}=(\psi_\uparrow(x,\tau),\psi_\downarrow(x,\tau),\psi_\uparrow^\dagger(x,\tau),\psi_\downarrow^\dagger(x,\tau))$
with T indicating the transpose of a vector, positive imaginary time $\tau$, and the expectation value taken at temperature $1/\beta$.
The Matsubara frequency representation of this Green function is given by
\begin{align}
    \check{G}(x,x',i\omega_n)
    =&
    \mathcal{F}(\check{G}((x,\tau),(x',0))),
    \\
    \mathcal{F}(f(\tau))
    =&
    \frac{1}{2}\left(1-e^{-i\omega_n\beta\hbar}\right)
    \int_0^{\beta\hbar} d\tau e^{i\omega_n \tau}f(\tau).
    \label{eq:app_def_green_function}
\end{align}
$\check{G}(x,x',i\omega_n)$ has four components, which we denote as
\begin{align}
    G_{\mathrm{N},\sigma,\sigma'}^{11}(x,x',i\omega_n)
    =&
    -
    \mathcal{F}(\langle \psi_{\sigma}(x,\tau)\psi_{\sigma'}^\dagger(x',0)\rangle),
    \label{eq:def_GN}
    \\
    G^{22}_{\mathrm{N},\sigma,\sigma'}(x,x',i\omega_n)
    =&
    -
    \mathcal{F}(\langle \psi_{\sigma}^\dagger(x,\tau)\psi_{\sigma'}(x',0)\rangle),
    \label{eq:def_GN_bar}
    \\
    F^{12}_{\sigma,\sigma'}(x,x',i\omega_n)
    =&
    -
    \mathcal{F}(\langle \psi_{\sigma}(x,\tau)\psi_{\sigma'}(x',0)\rangle),
    \label{eq:app_def_F}
    \\
    F^{21}_{\sigma,\sigma'}(x,x',i\omega_n)
    =&
    -
    \mathcal{F}(\langle \psi_{\sigma}^\dagger(x,\tau)\psi_{\sigma'}^\dagger(x',0)\rangle).
    \label{eq:app_def_F_bar}
\end{align}
Equations~\eqref{eq:app_def_F} and \eqref{eq:app_def_F_bar} are called anomalous Green functions.
Let $\check{G}_\mathrm{N}(x,x',i\omega_n)$ and $\check{F}(x,x',i\omega_n)$ be
\begin{align}
    \check{G}_\mathrm{N}(x,x',i\omega_n)
    =&
    \begin{pmatrix}
        G_{\mathrm{N},\sigma,\sigma}^{11}(x,x',i\omega_n) & 0
        \\
        0&G_{\mathrm{N},\sigma,\sigma'}^{22}(x,x',i\omega_n)
    \end{pmatrix},
    \label{eq:def_GN_app}
    \\
    \check{F}(x,x',i\omega_n)
    =&
    \begin{pmatrix}
        0&F_{\sigma,\sigma'}^{12}(x,x',i\omega_n) 
        \\
        F_{\sigma,\sigma'}^{21}(x,x',i\omega_n) & 0
    \end{pmatrix}.
    \label{eq:def_F_app}
\end{align}
We define even (odd) frequency components thereof by
\begin{align}
    \check{F}^{\mathrm{even(odd)}}(x,x',i\omega_n)
    =&
    \frac{1}{2}
    \left[
        \check{F}(x,x',i\omega_n)
        +(-)
        \check{F}(x,x',-i\omega_n)
    \right].
\end{align}
Evidently,  $\check{F}^{\mathrm{even(odd)}}(x,x',i\omega_n)$ is an even (odd) function of $\omega_n$. 

Making use of an analytic continuation, $i\omega_n\rightarrow z$ with $z\in\mathbb{C}$, we can extend the definition of the Green function to the complex plain: $\check{G}(x,x',z)$.
From Eqs.~\eqref{eq:def_GN_app} and \eqref{eq:def_F_app}, the Green function with complex frequency $z$ can be written as
\begin{align}
    \check{G}(x,x',z)
    =&
    \check{G}_\mathrm{N}(x,x',z)
    +
    \check{F}(x,x',z).
    \label{eq:def_G_z}
\end{align}
Then, advanced (retarded) Green function can be obtained by $\check{G}^\mathrm{A(R)}(x,x',E)=\check{G}(x,x',E-(+)i\eta)$ with positive infinitesimal number $\eta$.
Likewise, we obtain advanced even (odd) frequency anomalous Green function $\check{F}^{\mathrm{A},\mathrm{even(odd)}}(x,x',E)$ and retarded one $\check{F}^{\mathrm{R},\mathrm{even(odd)}}(x,x',E)$.
\subsection{\label{sub:decomp_cond}Decomposition of $\bar{\Gamma}_\mathrm{e}(E)$}
From Eq.~\eqref{eq:def_G_z}, we can write the advanced (retarded) Green function as
\begin{align}
    \check{G}^\mathrm{A(R)}(x,x',E)
    =&
    \check{G}_\mathrm{N}^\mathrm{A(R)}(x,x',E)
    +
    \check{F}^\mathrm{A(R)}(x,x',E),
    \\
    \check{G}_\mathrm{N}^\mathrm{A(R)}(x,x',E)
    =&
    \begin{pmatrix}
        G_{\mathrm{N},\sigma,\sigma'}^{\mathrm{A(R)},11}(x,x',E) & 0
        \\
        0 & G_{\mathrm{N},\sigma,\sigma'}^{\mathrm{A(R)},22}(x,x',E)
    \end{pmatrix},
    \\
    \check{F}^\mathrm{A(R)}(x,x',E)
    =&
    \begin{pmatrix}
        0 & F_{\sigma,\sigma'}^{\mathrm{A(R)},12}(x,x',E)
        \\
        F_{\sigma,\sigma'}^{\mathrm{A(R)},21}(x,x',E) & 0
    \end{pmatrix},
\end{align}
where $\check{G}_\mathrm{N}^\mathrm{A(R)}(x,x',E)$ denotes the normal Green function and $\check{F}^\mathrm{A(R)}(x,x',E)$ the anomalous Green function.
We can also decompose $\tilde{G}_{1}(x,x',E)$ and $\tilde{G}_{2}(x,x',E)$ given by Eqs.~\eqref{eq:def_tilde1} and \eqref{eq:def_tilde2}, respectively, as $\tilde{G}_{1(2)}(x,x',E)=\tilde{G}_{\mathrm{N},1(2)}(x,x',E)+\tilde{F}_{1(2)}(x,x',E)$ in the same manner.
Then, Eq.~\eqref{eq:spatial_average_app} can be written as
\begin{align}
    \bar{\Gamma}_\mathrm{e}(E)
    =&
    \bar{\Gamma}_\mathrm{N}(E) + \bar{\Gamma}_F(E),
    \\
    \Gamma_\mathrm{N}(x,x',E)
    =&
    \frac{-e^2\hbar^3\pi}{4m^2}
    \mathrm{Tr}
    \left[
        P_\mathrm{e}
        \tilde{G}_{\mathrm{N},1}(x,x',E)
        \overleftrightarrow{\nabla}
        \overleftrightarrow{\nabla}'
        \tilde{G}_{\mathrm{N},2}(x',x,E)
    \right],
    \\
    \Gamma_F(x,x',E)
    =&
    \frac{-e^2\hbar^3\pi}{4m^2}
    \mathrm{Tr}
    \left[
        P_\mathrm{e}
        \tilde{F}_{1}(x,x',E)
        \overleftrightarrow{\nabla}
        \overleftrightarrow{\nabla}'
        \tilde{F}_{2}(x',x,E)
    \right].
    \label{eq:Gamma_F}
\end{align}

Next, we further decompose $\check{F}^{\mathrm{A(R)}}(x,x',E)$ into two parts $\check{F}^{\mathrm{A(R)}}=\check{F}^\mathrm{A(R),even}+\check{F}^\mathrm{A(R),odd}$ with
\begin{align}
    \check{F}^\mathrm{A,even(odd)}(x,x',E)
    =&
    \frac{1}{2}
    \left[
        \check{F}(x,x',E-i\eta)
        +(-)
        \check{F}(x,x',-E+i\eta)
    \right],
    \label{eq:F_decompose1}
    \\
    \check{F}^\mathrm{R,even(odd)}(x,x',E)
    =&
    \frac{1}{2}
    \left[
        \check{F}(x,x',E+i\eta)
        +(-)
        \check{F}(x,x',-E-i\eta)
    \right],
    \label{eq:F_decompose2}
\end{align}
and define $\tilde{F}_{1}^\mathrm{even(odd)}(x,x',E)$ and $\tilde{F}_{2}^\mathrm{even(odd)}(x,x',E)$ in a similar manner as Eqs.~\eqref{eq:def_tilde1} and \eqref{eq:def_tilde2}, respectively, by using Eqs.~\eqref{eq:F_decompose1} and \eqref{eq:F_decompose2}.
Making use of Eqs.~\eqref{eq:F_decompose1} and \eqref{eq:F_decompose2}, Eq.~\eqref{eq:Gamma_F} can be rewritten as
\begin{align}
    \Gamma_F(x,x',E)
    =&
    \Gamma_F^{\mathrm{ee}}(x,x',E)
    +
    \Gamma_F^{\mathrm{oo}}(x,x',E)
    +
    \Gamma_F^{\mathrm{eo}}(x,x',E)
    \label{eq:F_decompose3}
\end{align}
with
\begin{align}
    \Gamma_F^\mathrm{ee(oo)}(x,x',E)
    =&
    \frac{-e^2\hbar^3\pi}{4m^2}
    \mathrm{Tr}
    \left[
        P_\mathrm{e}
        \tilde{F}_{1}^\mathrm{even(odd)}(x,x',E)
        \overleftrightarrow{\nabla}
        \overleftrightarrow{\nabla}'
        \tilde{F}_{2}^\mathrm{even(odd)}(x',x,E)
    \right],
    \label{eq:def_Gamma_ee_oo}
    \\
    \Gamma_F^\mathrm{eo}(x,x',E)
    =&
    \frac{-e^2\hbar^3\pi}{4m^2}
    \left\{
    \mathrm{Tr}
    \left[
        P_\mathrm{e}
        \tilde{F}_{1}^\mathrm{even}(x,x',E)
        \overleftrightarrow{\nabla}
        \overleftrightarrow{\nabla}'
        \tilde{F}_{2}^\mathrm{odd}(x',x,E)
    \right]
    \right.
    \nonumber\\
    &\hspace{13mm}\left.
    +
    \mathrm{Tr}
    \left[
        P_\mathrm{e}
        \tilde{F}_{1}^\mathrm{odd}(x,x',E)
        \overleftrightarrow{\nabla}
        \overleftrightarrow{\nabla}'
        \tilde{F}_{2}^\mathrm{even}(x',x,E)
    \right]
    \right\}.
    \label{eq:def_Gamma_eo}
\end{align}
Generally, each term on the right hand side of Eq.~\eqref{eq:F_decompose2} depends on $x$ and $x'$ although the left hand side of Eq.~\eqref{eq:F_decompose3} is independent of $x$ and $x'$.

It is noted that there are two types current conservation in the N region: charge and particle current conservation.
Similar to Eq.~\eqref{eq:def_Gamma_e_app}, the differential conductance for the particle current is given by $\Gamma_\mathrm{p}(x,x',E)=\Gamma_\mathrm{N}(x,x',E)-\Gamma_F(x,x',E)$.
Both $\Gamma_\mathrm{e}(x,x',E)$ and $\Gamma_\mathrm{p}(x,x',E)$ do not depend on the positions $x$ and $x'$. Therefore, $\Gamma_\mathrm{N}(x,x',E)$ and $\Gamma_F(x,x',E)$ also do not depend on $x$ and $x'$.
However, $\Gamma_F^\mathrm{ee,oo,eo}(x,x',E)$ generally depends on $x$ and $x'$.

The cross term between even and odd-frequency, $\bar{\Gamma}_F^\mathrm{eo}(E)$, is zero due to particle-hole symmetry.
The anomalous Green function satisfies the following relations according to the definitions Eqs.~\eqref{eq:app_def_green_function}, \eqref{eq:app_def_F}, and \eqref{eq:app_def_F_bar}.
\begin{align}
    {[
        F_{\sigma',\sigma}^{12}(x',x,z)
    ]}^*
    =&
    F_{\sigma,\sigma'}^{21}(x,x',z^*),
    \label{eq:app_F_star}
\end{align}
with $z\in\mathbb{C}$ and $Z=\mathrm{Tr}e^{-\beta H}$.
Also, $ {[ F_{\sigma,\sigma'}^{12}(x,x',z) ]}^*$ is given by
\begin{align}
    {[
        F_{\sigma,\sigma'}^{12}(x,x',z)
    ]}^*
    =&
    - F_{\sigma,\sigma'}^{21}(x,x',-z^*).
\end{align}
Let us define $\tilde{\check{F}}_{1}(x,x',E)$ in the following equation.
\begin{align}
    &
    \tilde{\check{F}}_{1}(x,x',E)
    \\
    =&
    \begin{pmatrix}
        0&\tilde{F}_{1}^{12}(x,x',E)
        \\
        \tilde{F}_{1}^{21}(x,x',E)&0
    \end{pmatrix},
    \nonumber\\
    =&
    \frac{1}{\sqrt{2}\pi i}
    \begin{pmatrix}
        0&\sin\theta F^{12}(x,x',E-i\eta) - \cos\theta F^{12}(x,x',E+i\eta)
        \\
        \sin\theta F^{21}(x,x',E-i\eta) - \cos\theta F^{21}(x,x',E+i\eta) &0 
    \end{pmatrix},
\end{align}
for $x\neq x'$, and
\begin{align}
    \tilde{\check{F}}_{1}(x,x',E)
    =&
    \frac{1}{2\pi i}
    \begin{pmatrix}
        0&F^{12}(x,x',E-i\eta) - F^{12}(x,x',E+i\eta)
        \\
        F^{21}(x,x',E-i\eta) - F^{21}(x,x',E+i\eta) &0 
    \end{pmatrix},
\end{align}
for $x=x'$.
$\tilde{\check{F}}_{2}(x,x',E)$ is defined in the same way.
$\tilde{\check{F}}_1(x,x',E)$ satisfies
\begin{align}
    {\left[\tilde{\check{F}}_{1}(x',x,E)\right]}^\dagger
    =&
    \tilde{\check{F}}_{2}(x,x',E).
    \label{eq:app_tilde_F_dagger}
\end{align}
From Eq.~\eqref{eq:app_tilde_F_dagger}, we obtain ${\left[\tilde{\check{F}}_{1(2)}^\mathrm{even(odd)}(x',x,E)\right] }^\dagger=\tilde{\check{F}}_{2(1)}^\mathrm{even(odd)}(x,x',E)$. 

The component of the differential conductance for the cross term between even and odd-frequency pairing satisfies
\begin{align}
    {[\Gamma_F^\mathrm{eo}(x,x',E)]}^*
    =&
    \frac{-e^2\hbar^3\pi}{4m^2}
    \mathrm{Tr}
    { \left[
        P_\mathrm{e}
        \tilde{\check{F}}^\mathrm{even}_1(x,x',E)
        \overleftrightarrow{\nabla}
        \overleftrightarrow{\nabla}'
        \tilde{\check{F}}^\mathrm{odd}_2(x',x,E)
    \right]}^\dagger
    +
    (\mathrm{even}\leftrightarrow\mathrm{odd})
    \nonumber\\
    =&
    \frac{-e^2\hbar^3\pi}{4m^2}
    \mathrm{Tr}
    { \left\{
        P_\mathrm{e}
        {\left[
            \tilde{\check{F}}^\mathrm{odd}_2(x',x,E)
        \right]}^\dagger
        \overleftrightarrow{\nabla}
        \overleftrightarrow{\nabla}'
        {\left[
            \tilde{\check{F}}^\mathrm{even}_1(x,x',E)
        \right]}^\dagger
    \right\}}
    +
    (\mathrm{even}\leftrightarrow\mathrm{odd})
    \nonumber\\
    =&
    \Gamma_F^\mathrm{eo}(x,x',E).
    \label{eq:app_Gamma_F_star1}
\end{align}

Particle-hole symmetry of the Hamiltonian gives
\begin{align}
    C^{-1}H(x,x')C
    =&
    -H(x,x'),
    \\
    H(x,x')
    =&
    \begin{pmatrix}
        \varepsilon_{\sigma,\sigma'}(x,x') & \Delta_{\sigma,\sigma'}(x,x')
        \\
        -\Delta_{\sigma,\sigma'}^*(x,x') & -\varepsilon_{\sigma',\sigma}(x',x)
    \end{pmatrix},
\end{align}
where the charge conjugation operator is given by $C=\hat{\tau}_1 K$, $K$ is a complex conjugation operator, $\varepsilon_{\sigma,\sigma'}(x,x')$ is a normal part of the Hamiltonian, and $\Delta$ is a superconducting pair potential ($\Delta_{\sigma,\sigma'}(x,x')=-\Delta_{\sigma',\sigma}(x',x)$).
The Green function satisfies
\begin{align}
    C^{-1}\check{G}(x,x',z)C
    =&
    -G(x,x',-z^*).
\end{align}
Then, $\Gamma_F^\mathrm{eo}(x,x',E)$ satisfies
\begin{align}
    {[\Gamma_F^\mathrm{eo}(x,x',E)]}^*
    =&
    \frac{-e^2\hbar^3\pi}{4m^2}
    \mathrm{Tr}
    \left[
        C^{-1}
        P_\mathrm{e}
        C C^{-1}
        \tilde{\check{F}}_1^\mathrm{even}(x,x',E)
        C
        \overleftrightarrow{\nabla}
        \overleftrightarrow{\nabla}'
        C^{-1}
        \tilde{\check{F}}_2^\mathrm{odd}(x',x,E)
        C
    \right]
    +
    (\mathrm{even}\leftrightarrow\mathrm{odd})
    \nonumber\\
    =&
    -
    \frac{-e^2\hbar^3\pi}{4m^2}
    \mathrm{Tr}
    \left[
        P_\mathrm{h}
        \tilde{\check{F}}_2^\mathrm{even}(x,x',E)
        \overleftrightarrow{\nabla}
        \overleftrightarrow{\nabla}'
        \tilde{\check{F}}_1^\mathrm{odd}(x',x,E)
    \right]
    +
    (\mathrm{even}\leftrightarrow\mathrm{odd})
    \nonumber\\
    =&
    -\Gamma_F^\mathrm{eo}(x',x,E),
\end{align}
where we used the relation $C^{-1} \tilde{\check{F}}^\mathrm{even(odd)}(x,x',E) C=-(+)\tilde{\check{F}}^\mathrm{even(odd)}(x,x',E)$ and $P_\mathrm{h}=(\hat{\tau}_0-\hat{\tau}_3)/2$.
Hence, after averaging over $x$ and $x'$, we obtain
\begin{align}
    {[\bar{\Gamma}^\mathrm{eo}_F(E)]}^*=-\bar{\Gamma}^\mathrm{eo}_F(E).
    \label{eq:app_Gamma_F_star2}
\end{align}
Equations~\eqref{eq:app_Gamma_F_star1} and \eqref{eq:app_Gamma_F_star2} yield $\bar{\Gamma}_F^\mathrm{eo}(E)=0$.

\subsection{\label{sub:BTK}Blonder-Tinkham-Klapwijk (BTK) approach}
We can calculate the conductance for NS junctions by utilizing the Blonder, Tinkham, and Klapwijk (BTK) approach~\cite{PhysRevB.25.4515}.
In this subsection, we explain the BTK approach for the calculation of the conductance across NS junctions.
For the S side, we take two prominent examples: $s$-wave spin-singlet and $p$-wave spin-triplet. 
In general, Bogoliubov-de Gennes (BdG) Hamiltonian is given by
\begin{align}
    H=&\frac{1}{2}\int dxdx'\Psi^\dagger(x)H(x,x')\Psi(x'),
    \\
    H(x,x')
    =&
    \begin{pmatrix}
        \varepsilon_{\sigma,\sigma'}(x,x') & \Delta_{\sigma,\sigma'}(x,x')
        \\
        -\Delta^*_{\sigma,\sigma'}(x,x') & -\varepsilon_{\sigma',\sigma}(x',x)
    \end{pmatrix}
\end{align}
with $\Psi(x)={\left(\psi_\uparrow(x),\psi_\downarrow(x),\psi_\uparrow^\dagger(x),\psi_\downarrow^\dagger(x)\right)}^\mathrm{T}$.
Let us consider a 1D normal metal ($x<0$) superconductor ($x>0$) junction.
As a normal part of the Hamiltonian, $\varepsilon_{\sigma,\sigma'}(x,x')$, we consider following function:
\begin{align}
    \varepsilon_{\sigma,\sigma'}(x,x')
    =&
    \delta(x-x')
    \hat{\sigma}_0\hat{\tau}_3
    \left[
        -\frac{\hbar^2}{2m}\frac{d^2}{dx^2}-\mu+U\delta(x)
    \right],
\end{align}
where $m$ is a mass of electron, $\mu$ is a chemical potential, and $U$ is a barrier potential at the interface.
As a pair potential, we consider an $s$-wave spin-singlet and $p$-wave spin-triplet.
For the $s$-wave spin-singlet case, the pair potential is given by $\Delta_{\sigma,\sigma'}(x,x')=\Delta\Theta(x)\delta(x-x') i\hat{\sigma}_2$ with the Heaviside step function $\Theta(x)$ and $\Delta\in\mathbb{R}$.
For the $p$-wave spin-singlet case, we adopt $\Delta_{\sigma,\sigma'}(x,x')=\Delta\Theta(x)\Theta(x')\int dk e^{ik(x-x')}\mathrm{sgn}(k)$.
In momentum space, the pair potential for the $p$-wave S is given by $\Delta(k)=\Delta\mathrm{sgn}(k)$ with the signum function $\mathrm{sgn}(k)$.
We can divide the BdG Hamiltonian into two disconnected parts with basis ${(\psi_\uparrow(x),\psi_\downarrow^\dagger(x))}^\mathrm{T}$ and ${(\psi_\downarrow(x),\psi_\uparrow^\dagger(x))}^\mathrm{T}$, respectively.
Then, if we choose the basis ${(\psi_\uparrow(x),\psi_\downarrow^\dagger(x))}^\mathrm{T}$, the Hamiltonian can be reduced to a $2\times2$ matrix:
\begin{align}
    H_{2\times2}(x,x')
    =&
    \delta(x-x')\left[-\frac{\hbar^2}{2m}\frac{d^2}{dx^2}-\mu+U\delta(x)\right]\hat{\tau}_3
    +
    \Theta(x)\Delta_{2\times2}(x,x'),
    \label{eq:app_Hamiltonian_continuum}
    \\
    \Delta_{2\times2}(x,x')
    =&
    \begin{cases}
        \Delta\delta(x-x')\hat{\tau}_1
        & s\mathrm{-wave},
        \\
        \Delta\Theta(x')\int dk 
        \begin{pmatrix}
            0&e^{ik(x-x')}
            \\
            -e^{-ik(x-x')}&0
        \end{pmatrix}
        \mathrm{sgn}(k)
        & p\mathrm{-wave}
    \end{cases}
    \label{eq:app_gap_continuum}
\end{align}
with Pauli matrices $\hat{\tau}_{j=0,1,2,3}$ acting on particle-hole space.

In this approach, we first write down the scattering states in N and S regions at a given energy $E$:
\begin{align}
    \Psi^\mathrm{N}(x,E)
    =&
    e^{ik_\mathrm{F}x}\Psi_\mathrm{e}^\mathrm{N}+ae^{ik_\mathrm{F}x}\Psi_\mathrm{h}^\mathrm{N}+be^{-ik_\mathrm{F}x}\Psi^\mathrm{N}_\mathrm{e},
    \\
    \Psi^\mathrm{S}(x,E)
    =&
    ce^{ik_\mathrm{F}x}\Psi^\mathrm{S}_\mathrm{e}+de^{-ik_\mathrm{F}x}\Psi^\mathrm{S}_\mathrm{h}
\end{align}
with
\begin{align}
&
    \Psi_\mathrm{e}^\mathrm{N}
    =
    {(1,0)}^\mathrm{T},\:
    \Psi_\mathrm{h}^\mathrm{N}
    =
    {(0,1)}^\mathrm{T},
    \\
    &
    \Psi_\mathrm{e}^\mathrm{S}
    =
    {(u_\mathrm{e},v_\mathrm{e})}^\mathrm{T},\:
    \Psi_\mathrm{h}^\mathrm{S}
    =
    {(v_\mathrm{h},u_\mathrm{h})}^\mathrm{T},
    \\
    &
    \begin{pmatrix}
        E_+ &\Delta(k)
        \\
        \Delta(k) & -E_+
    \end{pmatrix}
    \begin{pmatrix}
        u_\mathrm{e}\\v_\mathrm{e}
    \end{pmatrix}
    =
    E
    \begin{pmatrix}
        u_\mathrm{e}\\v_\mathrm{e}
    \end{pmatrix},
    \label{eq:eigen_BTK_e}
    \\
    &
    \begin{pmatrix}
        -E_- &\Delta(-k)
        \\
        \Delta(-k) & E_-
    \end{pmatrix}
    \begin{pmatrix}
        v_\mathrm{h}\\u_\mathrm{h}
    \end{pmatrix}
    =
    E
    \begin{pmatrix}
        v_\mathrm{h}\\u_\mathrm{h}
    \end{pmatrix},
    \label{eq:eigen_BTK_h}
    \\
    &
    E_\pm
    =
    \begin{cases}
    \sqrt{E^2-\Delta^2(\pm k)} & E>|\Delta(\pm k)|,
    \\
    i\sqrt{\Delta^2(\pm k)-E^2} & |E|<|\Delta(\pm k)|,
    \\
    -\sqrt{E^2-\Delta^2(\pm k)} & E<|\Delta(\pm k)|,
    \end{cases}
    \\
    &
    k_\mathrm{F}=\sqrt{\frac{2m}{\hbar^2}\mu}
\end{align}
with $\Delta(k)=\Delta$ for the $s$-wave pair potential, and $\Delta(k)=\Delta\mathrm{sgn}(k)$ for the $p$-wave pair potential.
We approximate the wave vector with the Fermi momentum $k_\mathrm{F}$ assuming that the Fermi energy satisfies $\mu\gg|\Delta|$ and $\mu\gg |E|$.
From Eqs.~\eqref{eq:eigen_BTK_e} and \eqref{eq:eigen_BTK_h}, we obtain
\begin{align}
    \frac{v_\mathrm{e}}{u_\mathrm{e}}
    =&
    \begin{cases}
    \frac{\Delta(k)}{E+\mathrm{sgn}(E)\sqrt{E^2-\Delta^2(k)}} & |E|>|\Delta(k)|,
    \\
    \frac{\Delta(k)}{E+i\sqrt{\Delta^2(k)-E^2}} & |E|<|\Delta(k)|,
    \end{cases}
    \\
    \frac{v_\mathrm{h}}{u_\mathrm{h}}
    =&
    \begin{cases}
    \frac{\Delta(-k)}{E+\mathrm{sgn}(E)\sqrt{E^2-\Delta^2(-k)}} & |E|>|\Delta(-k)|,
    \\
    \frac{\Delta(-k)}{E+i\sqrt{\Delta^2(-k)-E^2}} & |E|<|\Delta(-k)|.
    \end{cases}
\end{align}
Next, we solve the boundary conditions at $x=0$:
\begin{align}
    &
    \Psi^\mathrm{N}(x=0)=\Psi^\mathrm{S}(x=0),
    \\ 
    &
    \frac{d}{dx}\Psi^\mathrm{S}(x)\big|_{x=0}
    -
    \frac{d}{dx}\Psi^\mathrm{N}(x)\big|_{x=0}
    =
    \frac{2mU}{\hbar^2}\Psi^\mathrm{N}(x=0).
\end{align}
These boundary conditions imply
\begin{align}
    a(E)
    =&
    \frac{4\frac{v_{\mathrm{e}}}{u_{\mathrm{e}}}}{4+Z^2\left(1-\frac{v_{\mathrm{e}}}{u_{\mathrm{e}}}\frac{v_{\mathrm{h}}}{u_{\mathrm{h}}}\right)},
    \label{eq:a_1}
    \\
    b(E)
    =&
    iZ
    \frac{(2+iZ)\left(1-\frac{v_{\mathrm{e}}}{u_{\mathrm{e}}}\frac{v_{\mathrm{h}}}{u_{\mathrm{h}}}\right)}{4+Z^2\left(1-\frac{v_{\mathrm{e}}}{u_{\mathrm{e}}}\frac{v_{\mathrm{h}}}{u_{\mathrm{h}}}\right)},
    \label{eq:a_2}
    \\
    c(E)
    =&
    \frac{2(2+iZ)\frac{1}{u_\mathrm{e}}}{4+Z^2\left(1-\frac{v_{\mathrm{e}}}{u_{\mathrm{e}}}\frac{v_{\mathrm{h}}}{u_{\mathrm{h}}}\right)},
    \\
    d(E)
    =&
    \frac{-2iZ\frac{v_\mathrm{e}}{u_\mathrm{e}}\frac{1}{u_\mathrm{h}}}{4+Z^2\left(1-\frac{v_{\mathrm{e}}}{u_{\mathrm{e}}}\frac{v_{\mathrm{h}}}{u_{\mathrm{h}}}\right)},
    \\
    Z=&\frac{2mU}{k_\mathrm{F}\hbar^2}.
    \label{eq:def_Z}
\end{align}

Finally, the conductance is given by
\begin{align}
   G_\mathrm{BTK}=&-\int dE\frac{df(E)}{dE}\Gamma_\mathrm{BTK}(E)
   \\
   \Gamma_\mathrm{BTK}(E)
   =&
   \frac{e^2}{2\pi \hbar}
   \left[
      1+{|a(E)|}^2-{|b(E)|}^2
      \right].
   \label{eq:BTK_s_wave}
\end{align}

When $|E|<|\Delta|$, coefficients $a(E)$ and $b(E)$ satisfy ${|a(E)|}^2+{|b(E)|}^2=1$, and Eq.~\eqref{eq:BTK_s_wave} can be recast as
\begin{align}
    \Gamma_\mathrm{BTK}(|E|<|\Delta|)
    =&
    \frac{e^2}{2\pi \hbar} 2{|a(E)|}^2.
\end{align}
We obtain the same results if we choose the basis ${(\psi_\downarrow(x),\psi_\uparrow^\dagger(x))}^\mathrm{T}$. 
The total differential conductance is given by the sum of the two channels.
\subsection{\label{sub:Green_func_McMillan1}Green function in continuum system}
We explain how to derive the Green function in the continuum NS junction (McMillan's method~\cite{PhysRev.175.559}).
The BdG Hamiltonian is given by Eqs.~\eqref{eq:app_Hamiltonian_continuum} and \eqref{eq:app_gap_continuum}
with basis ${(\psi_\uparrow(x),\psi_\downarrow^\dagger(x))}^\mathrm{T}$.
We can repeat the same procedure for the basis ${(\psi_\downarrow(x),\psi_\uparrow^\dagger(x))}^\mathrm{T}$ and we obtain the same differential conductance.
We suppose $\Delta\in\mathbb{R}$.
To calculate the retarded Green function, we consider the following wave functions.
The wave functions for $x<0$ at complex energy $\tilde{E}=E+i\eta$ ($E\in\mathbb{R}$) can be written as
\begin{align}
    \Psi_{\mathrm{e}}^{\rightarrow}(x)
    =&
    \Psi^\mathrm{N}_\mathrm{e}
    e^{ik_\mathrm{e} x}
    +
    a_1
    \Psi^\mathrm{N}_\mathrm{h}
    e^{ik_\mathrm{h} x}
    +
    a_2
    \Psi^\mathrm{N}_\mathrm{e}
    e^{-ik_\mathrm{e} x},
    \label{eq:psi_e_right}
   \\
   \Psi_{\mathrm{h}}^{\rightarrow}(x)
   =&
      \Psi^\mathrm{N}_\mathrm{h}
      e^{-ik_\mathrm{h} x}
      +
      a_3
      \Psi^\mathrm{N}_\mathrm{e}
      e^{-ik_\mathrm{e} x}
      +
      a_4
      \Psi^\mathrm{N}_\mathrm{h}
      e^{ik_\mathrm{h} x},
   \\
   \Psi_{\mathrm{e}}^{\leftarrow}(x)
   =&
      b_5
      \Psi^\mathrm{N}_\mathrm{e}
      e^{-ik_\mathrm{e} x}
      +
      b_6
      \Psi^\mathrm{N}_\mathrm{h}
      e^{ik_\mathrm{h} x},
   \\
   \Psi_{\mathrm{h}}^{\leftarrow}(x)
   =&
      b_7
      \Psi^\mathrm{N}_\mathrm{h}
      e^{ik_\mathrm{h} x}
      +
      b_8
      \Psi^\mathrm{N}_\mathrm{e}
      e^{-ik_\mathrm{e} x},
\end{align}
and likewise, for $x>0$,
\begin{align}
   \Psi_{\mathrm{e}}^{\rightarrow}(x)
   =&
      b_1
      \Psi^\mathrm{S}_\mathrm{e+}
      e^{ik_+ x}
      +
      b_2
      \Psi^\mathrm{S}_\mathrm{h-}
      e^{-ik_- x},
   \\
   \Psi_{\mathrm{h}}^{\rightarrow}(x)
   =&
      b_3
      \Psi^\mathrm{S}_\mathrm{h-}
      e^{-ik_- x}
      +
      b_4
      \Psi^\mathrm{S}_\mathrm{e+}
      e^{ik_+ x},
   \\
   \Psi_{\mathrm{e}}^{\leftarrow}(x)
   =&
      \Psi^\mathrm{S}_\mathrm{e-}
      e^{-ik_+ x}
      +
      a_5
      \Psi^\mathrm{S}_\mathrm{h-}
      e^{-ik_- x}
      +
      a_6
      \Psi^\mathrm{S}_\mathrm{e+}
      e^{ik_+ x},
   \\
   \Psi_{\mathrm{h}}^{\leftarrow}(x)
   =&
      \Psi^\mathrm{S}_\mathrm{h+}
      e^{ik_- x}
      +
      a_7
      \Psi^\mathrm{S}_\mathrm{e+}
      e^{ik_+ x}
      +
      a_8
      \Psi^\mathrm{S}_\mathrm{h-}
      e^{-ik_- x}
    \label{eq:psi_h_left}
\end{align}
with
\begin{align}
    k_\mathrm{e(h)}
    =&
   \sqrt{\frac{2m}{\hbar^2}[\mu+(-)\tilde{E}]},\:\:
   k_{\pm}
   =
   \sqrt{\frac{2m}{\hbar^2}[\mu\pm\Omega(k_\pm)]},
   \label{eq:app_def_k_eh_pm}
   \\
   \Omega(k_\pm)
   =&
   \begin{cases}
   \sqrt{\tilde{E}^2-{|\Delta(k_\pm)|}^2} & \mathrm{Re}\tilde{E}>|\Delta(k_\pm)|,
   \\
   i\sqrt{{|\Delta(k_\pm)|}^2-\tilde{E}^2} & |\tilde{E}|<|\Delta(k_\pm)|,
   \\
   -\sqrt{\tilde{E}^2-{|\Delta(k_\pm)|}^2} & \mathrm{Re}\tilde{E}<-|\Delta(k_\pm)|.
   \end{cases}
\end{align}
Here, 
$\Psi^\mathrm{N}_\mathrm{e}={(1,0)}^\mathrm{T}$, and $\Psi^\mathrm{N}_\mathrm{h}={(0,1)}^\mathrm{T}$.
$\Psi^\mathrm{S}_{\mathrm{e}\pm}={(u_{\mathrm{e}\pm},v_{\mathrm{e}\pm})}^\mathrm{T}$ and
$\Psi^\mathrm{S}_{\mathrm{h}\pm}={(v_{\mathrm{h}\pm},u_{\mathrm{h}\pm})}^\mathrm{T}$
satisfy
\begin{align}
&
   \begin{pmatrix}
      \omega_{\mathrm{e},\pm} & \Delta(\pm k_+)
      \\
      \Delta(\pm k_+) &-\omega_{\mathrm{e},\pm}
   \end{pmatrix}
   \begin{pmatrix}
      u_{\mathrm{e},\pm}\\
      v_{\mathrm{e},\pm}
   \end{pmatrix}
   =
   \tilde{E}
   \begin{pmatrix}
      u_{\mathrm{e},\pm}\\
      v_{\mathrm{e},\pm}
   \end{pmatrix},
   \\
   &
   \begin{pmatrix}
      -\omega_{\mathrm{h},\pm} & \Delta(\pm k_-)
      \\
      \Delta(\pm k_-) &\omega_{\mathrm{h},\pm} 
   \end{pmatrix}
   \begin{pmatrix}
      v_{\mathrm{h},\pm}\\
      u_{\mathrm{h},\pm}
   \end{pmatrix}
   =
   \tilde{E}
   \begin{pmatrix}
      v_{\mathrm{h},\pm}\\
      u_{\mathrm{h},\pm}
   \end{pmatrix},
   \\
   &
   \omega_{\mathrm{e},\pm}
   =
   \begin{cases}
       \sqrt{\tilde{E}^2-{|\Delta(\pm k_+)|}^2} & \mathrm{Re}\tilde{E}>|\Delta(\pm k_+)|,
       \\
       i\sqrt{{|\Delta(\pm k_+)|}^2-\tilde{E}^2} & |\tilde{E}|<|\Delta(\pm k_+)|,
       \\
       -\sqrt{\tilde{E}^2-{|\Delta(\pm k_+)|}^2} & \mathrm{Re}\tilde{E}<-|\Delta(\pm k_+)|,
   \end{cases}
   \\
   &
   \omega_{\mathrm{h},\pm}
   =
   \begin{cases}
       \sqrt{\tilde{E}^2-{|\Delta(\pm k_-)|}^2} & \mathrm{Re}\tilde{E}>|\Delta(\pm k_-)|,
       \\
       i\sqrt{{|\Delta(\pm k_-)|}^2-\tilde{E}^2} & |\tilde{E}|<|\Delta(\pm k_-)|,
       \\
       -\sqrt{\tilde{E}^2-{|\Delta(\pm k_-)|}^2} & \mathrm{Re}\tilde{E}<-|\Delta(\pm k_-)|,
   \end{cases}
   \\
   &
   \frac{v_{\mathrm{e},\pm}}{u_{\mathrm{e},\pm}}
   =
   \frac{\Delta(\pm k_+)}{\tilde{E}+\omega_{\mathrm{e},\pm}},
   \label{eq:v_u_ep}
   \\
   &
   \frac{v_{\mathrm{h},\pm}}{u_{\mathrm{h},\pm}}
   =
   \frac{\Delta(\pm k_-)}{\tilde{E}+\omega_{\mathrm{h},\pm}}.
   \label{eq:v_u_hm}
\end{align}
Here, for the spin-triplet $p$-wave, we used the approximation $\int dx'\Delta(x,x')u_{\mathrm{e(h),\pm}}(x')\sim \zeta^{\rightarrow(\leftarrow)}\Delta u_{\mathrm{e(h),\pm}}(x)$ with $\zeta^\rightarrow=1$ for right going wave and $\zeta^\leftarrow=-1$ for left going wave~\cite{PhysRevB.41.4017}.

The coefficients $a_{j=1,\ldots,8}$, and $b_{j=1,\ldots,8}$ are determined by
the boundary conditions at $x=0$:
\begin{align}
    &
   \Psi_{\mathrm{e,h}}^{\rightleftarrows} (x=-\delta)
   =
   \Psi_{\mathrm{e,h}}^{\rightleftarrows} (x=\delta),
   \\
   &
   \frac{\partial \Psi_{\mathrm{e,h}}^{\rightleftarrows}(x)}{\partial x}\Big|_{x=\delta}
   -
   \frac{\partial \Psi_{\mathrm{e,h}}^{\rightleftarrows}(x)}{\partial x}\Big|_{x=-\delta}
   =
   \frac{2mU}{\hbar^2}\Psi_\mathrm{e,h}^{\rightleftarrows}(0),
\end{align}
with a positive infinitesimal number $\delta$.
The solutions for $a_{j=1,\ldots,8}$ and $b_{j=1,\ldots,8}$ are lengthy expressions. 
We only display the results for $a_{1,2,3,4}$ below.

We also derive the eigenvectors: $\int dx \tilde{\Psi}^\mathrm{T}(x)H(x,x')=E\tilde{\Psi}^\mathrm{T}(x)$. 
$\tilde{\Psi}$ is given from Eq.~\eqref{eq:psi_e_right} to Eq.~\eqref{eq:psi_h_left} with replacing $u_{\mathrm{e(h)},\pm}$ and $v_{\mathrm{e(h)},\pm}$ by $\tilde{u}_{\mathrm{e(h)},\pm}$ and $\tilde{v}_{\mathrm{e(h)},\pm}$, respectively, which are given by
\begin{align}
   \begin{pmatrix}
      \tilde{u}_{\mathrm{e},\pm}&
      \tilde{v}_{\mathrm{e},\pm}
   \end{pmatrix}
   \begin{pmatrix}
      \omega_{\mathrm{e},\pm} & \zeta\Delta(\pm k_+)
      \\
      \zeta\Delta(\pm k_+) &-\omega_{\mathrm{e},\pm}
   \end{pmatrix}
   =&
   \tilde{E}
   \begin{pmatrix}
      \tilde{u}_{\mathrm{e},\pm}&
      \tilde{v}_{\mathrm{e},\pm}
   \end{pmatrix},
   \\
   \begin{pmatrix}
      \tilde{v}_{\mathrm{h},\pm}&
      \tilde{u}_{\mathrm{h},\pm}
   \end{pmatrix}
   \begin{pmatrix}
      -\omega_{\mathrm{h},\pm} & \zeta\Delta(\pm k_-)
      \\
      \zeta\Delta(\pm k_-) &\omega_{\mathrm{h},\pm} 
   \end{pmatrix}
   =&
   \tilde{E}
   \begin{pmatrix}
      \tilde{v}_{\mathrm{h},\pm}&
      \tilde{u}_{\mathrm{h},\pm}
   \end{pmatrix},
\end{align}
with $\zeta=1$ for the spin-singlet $s$-wave and $\zeta=-1$ for the spin-triplet $p$-wave.

The retarded Green function is given by
\begin{align}
    \check{G}^\mathrm{R}(x,x',E)
    =&
    \begin{cases}
        \alpha_1\Psi_\mathrm{e}^{\rightarrow}(x){[\tilde{\Psi}_\mathrm{e}^{\leftarrow}(x')]}^\mathrm{T}
        +
        \alpha_2\Psi_\mathrm{h}^{\rightarrow}(x){[\tilde{\Psi}_\mathrm{h}^{\leftarrow}(x')]}^\mathrm{T}
        +
        \alpha_3\Psi_\mathrm{e}^{\rightarrow}(x){[\tilde{\Psi}_\mathrm{h}^{\leftarrow}(x')]}^\mathrm{T}
        +
        \alpha_4\Psi_\mathrm{h}^{\rightarrow}(x){[\tilde{\Psi}_\mathrm{e}^{\leftarrow}(x')]}^\mathrm{T}
        & x>x',
        \\
        \beta_1\Psi_\mathrm{e}^{\leftarrow}(x){[\tilde{\Psi}_\mathrm{e}^{\rightarrow}(x')]}^\mathrm{T}
        +
        \beta_2\Psi_\mathrm{h}^{\leftarrow}(x){[\tilde{\Psi}_\mathrm{h}^{\rightarrow}(x')]}^\mathrm{T}
        +
        \beta_3\Psi_\mathrm{e}^{\leftarrow}(x){[\tilde{\Psi}_\mathrm{h}^{\rightarrow}(x')]}^\mathrm{T}
        +
        \beta_4\Psi_\mathrm{h}^{\leftarrow}(x){[\tilde{\Psi}_\mathrm{e}^{\rightarrow}(x')]}^\mathrm{T}
        & x'>x.
    \end{cases}
    \label{eq:GR_continuum}
\end{align}
In this equation, we take the limit $\eta\rightarrow0$.
The arrows in Eqs.~\eqref{eq:psi_e_right}--\eqref{eq:psi_h_left} stand for asymptotic states 
$\Psi_\mathrm{e,h}^\rightarrow(x\rightarrow\infty)=\Psi_\mathrm{e,h}^\leftarrow(x\rightarrow-\infty)=0$.
In the opposite limit, they diverge. 
Hence, Eq.~\eqref{eq:GR_continuum} converges in both limit $(x,x')\rightarrow(\infty,-\infty)$ and $(x,x')\rightarrow(-\infty,\infty)$.
It is noted that we assume that the S pair potential is real, and the Green function is given by Eq.~\eqref{eq:GR_continuum}. 

The retarded Green function $\check{G}^\mathrm{R}(x,x',\tilde{E})$ satisfies $[\tilde{E}-H(k,x)]\check{G}^\mathrm{R}(x,x',\tilde{E})=\check{G}^\mathrm{R}(x,x',\tilde{E})[\tilde{E}-H(k,x')]=\delta(x-x')$.
The coefficients $\alpha_{j=1,\ldots,4}$ and $\beta_{j=1,\ldots,4}$ are determined by the boundary conditions
\begin{align}
    &
    \check{G}^\mathrm{R}(x+\delta,x,E) = \check{G}^\mathrm{R}(x-\delta,x,E),
    \\
    &
    \frac{\partial \check{G}^\mathrm{R}(x,x',E)}{\partial x}\Big|_{x=x'+\delta}
    -
    \frac{\partial \check{G}^\mathrm{R}(x,x',E)}{\partial x}\Big|_{x=x'-\delta}
    =
    \frac{2m}{\hbar^2}\hat{\tau}_3.
\end{align}
The obtained results are
\begin{align}
    &
    \alpha_1=\frac{m}{ik_\mathrm{e}\hbar^2},\:
    \alpha_2=\frac{a_1}{a_3}\frac{b_5}{b_7}\alpha_1,\:
    \alpha_3=-\frac{b_6}{b_7}\alpha_1,\:
    \alpha_4=-\frac{b_8}{b_5}\alpha_2,
    \nonumber\\
    &
    \beta_1=\alpha_1,\:
    \beta_2=\alpha_2,\:
    \beta_3=-\frac{b_8}{b_5}\beta_2,\:
    \beta_4=-\frac{b_6}{b_7}\beta_1.
\end{align}

Then, the retarded Green function for $x,x'<0$ can be written as
\begin{align}
    \check{G}^\mathrm{R}(x,x',E)
    =&
        \frac{m}{ik_\mathrm{e}\hbar^2}
        \begin{pmatrix}
            e^{ik_\mathrm{e}|x-x'|}+a_2e^{-ik_\mathrm{e}(x+x')} & f(x,x')a_1e^{-i(k_\mathrm{e}x-k_\mathrm{h}x')}
            \\
            f(x,x')a_\mathrm{1}e^{i(k_\mathrm{h}x-k_\mathrm{e}x')}  & \frac{a_1}{a_3}\left[e^{-ik_\mathrm{h}|x-x'|}+a_4e^{ik_\mathrm{h}(x+x')}\right]
        \end{pmatrix},
    \label{eq:app_R_Green_continuum}
\end{align}
where the coefficients $a_1$, $a_2$, $a_3$, and $a_4$ are given by
\begin{align}
    a_1
    =&
     \frac{2 k_\mathrm{e}  \frac{v_{\mathrm{e},+}}{u_{\mathrm{e},+}} (k_{-}+k_{+})}
   {\left[i k_\mathrm{F} Z (k_\mathrm{e}-k_\mathrm{h}-k_{-}+k_{+}) + k_\mathrm{e}k_\mathrm{h} + k_\mathrm{F}^2 Z^2 + k_{-}k_{+} \right] (1-\frac{v_{\mathrm{e},+}}{u_{\mathrm{e},+}} \frac{v_{\mathrm{h},-}}{ u_{\mathrm{h},-}}) + (k_\mathrm{e} k_{+} + k_\mathrm{h}k_{-} ) \frac{v_{\mathrm{e},+}}{u_{\mathrm{e},+}} \frac{v_{\mathrm{h},-}}{u_{\mathrm{h},-}} +k_\mathrm{e} k_{-} + k_\mathrm{h} k_{+}},
   \label{eq:a_1_general}
   \\
   a_2
   =&
   \frac{
       \left[
           k_\mathrm{F} Z (k_\mathrm{e}+k_\mathrm{h}+k_{-}-k_{+}) 
           +i k_\mathrm{F}^2 Z^2
       \right]
       (1 -\frac{v_{\mathrm{e},+}}{u_{\mathrm{e},+}} \frac{v_{\mathrm{h},-}}{u_{\mathrm{h},-}})-i  (k_\mathrm{e}-k_{+}) (k_\mathrm{h}+k_{-})+i \frac{v_{\mathrm{e},+}}{u_{\mathrm{e},+}} \frac{v_{\mathrm{h},-}}{u_{\mathrm{h},-}} (k_\mathrm{e}+k_{-}) (k_\mathrm{h}-k_{+})
   }
   {
       \left[
           k_\mathrm{F} Z (k_\mathrm{e}-k_\mathrm{h}-k_{-}+k_{+}) 
           -
           i k_\mathrm{F}^2 Z^2
       \right]
       ( 1-\frac{v_{\mathrm{e},+}}{u_{\mathrm{e},+}} \frac{v_{\mathrm{h},-}}{u_{\mathrm{h},-}})
       -
       i  (k_\mathrm{e}+k_{+}) (k_\mathrm{h}+k_{-})
       +
       i \frac{v_{\mathrm{e},+}}{u_{\mathrm{e},+} } \frac{v_{\mathrm{h},-}}{u_{\mathrm{h},-}} (k_\mathrm{e}-k_{-}) (k_\mathrm{h}-k_{+})
   },
   \label{eq:a_2_general}
   \\
   a_3
   =&
   a_1\frac{k_\mathrm{h}}{k_\mathrm{e}}
   \frac{v_{\mathrm{h},-}u_{\mathrm{e},+}}{v_{\mathrm{e},+}u_{\mathrm{h},-}},
   \label{eq:a_3_general}
   \\
   a_4
   =&
   \frac{
       \left[
           -k_\mathrm{F} Z (k_\mathrm{e}+k_\mathrm{h}-k_{-}+k_{+}) 
           +
           i k_\mathrm{F}^2 Z^2 
       \right]
       (1-\frac{v_{\mathrm{e},+}}{u_{\mathrm{e},+}}\frac{v_{\mathrm{h},-}}{u_{\mathrm{h},-}}  )
       -i (k_\mathrm{e}+k_{+}) (k_\mathrm{h}-k_{-})
       +
       i \frac{v_{\mathrm{e},+}}{u_{\mathrm{e},+}}\frac{v_{\mathrm{h},-}}{u_{\mathrm{h},-}} (k_\mathrm{e}-k_{-}) (k_\mathrm{h}+k_{+})
   }
   {
       \left[
           k_\mathrm{F} Z (k_\mathrm{e}-k_\mathrm{h}-k_{-}+k_{+}) 
           -
           i k_\mathrm{F}^2 Z^2 
       \right]
       (1-\frac{v_{\mathrm{e},+}}{u_{\mathrm{e},+}}\frac{v_{\mathrm{h},-}}{u_{\mathrm{h},-}}  )
       -
       i  (k_\mathrm{e}+k_{+}) (k_\mathrm{h}+k_{-})
       +
       i  \frac{v_{\mathrm{e},+}}{u_{\mathrm{e},+}}\frac{v_{\mathrm{h},-}}{u_{\mathrm{h},-}}
       (k_\mathrm{e}-k_{-}) (k_\mathrm{h}-k_{+})
   },
   \label{eq:a_4_general}
\end{align}
and $f(x,x')=1$ for the spin-singlet $s$-wave and $f(x,x')=\mathrm{sgn}(x-x')$ for the spin-triplet $p$-wave.
The advanced Green function can be determined by $\check{G}^\mathrm{A}(x,x',E)={[\check{G}^\mathrm{R}(x',x,E)]}^\dagger$.

Eq.~\eqref{eq:app_def_k_eh_pm} satisfies
\begin{align}
    k_\mathrm{e(h)}(-E)
    =&
    k_\mathrm{h(e)}(E),
    \\
    {[k_\pm(-E)]}^*
    =&
    k_\mp(E).
\end{align}
From Eq.~\eqref{eq:a_1_general}, $a_1$ with positive and negative energy satisfy following relation:
\begin{align}
    \frac{a_1^*(-E)}{k_\mathrm{e}(-E)}
    =&
    -
    \frac{a_1(E)}{k_\mathrm{e}(E)}.
\end{align}
The odd-frequency component of the retarded Green function is given by
\begin{align}
    F^{\mathrm{R,odd}}(x,x',E)
    =&
    \frac{mf(x,x')}{2i\hbar^2}
    \frac{a_1(E)}{k_\mathrm{e}(E)}
    \begin{pmatrix}
        0&
        e^{-i[k_\mathrm{e}(E)x-k_\mathrm{h}(E)x']}
        \mp
        e^{-i[k_\mathrm{h}(E)x-k_\mathrm{e}(E)x']}
        \\
        e^{-i[k_\mathrm{h}(E)x-k_\mathrm{e}(E)x']}
        \mp 
        e^{-i[k_\mathrm{e}(E)x-k_\mathrm{h}(E)x']}
        &0
    \end{pmatrix}.
\end{align}
Here, $-$ sign is for the spin-singlet $s$-wave, and $+$ sign is for the spin-triplet $p$-wave.

From Eqs.~\eqref{eq:spatial_average_app} and \eqref{eq:def_Gamma_e_app}, we obtain
\begin{align}
    \bar{\Gamma}_\mathrm{e}(E)
    =&
    \frac{e^2}{2\pi\hbar}
    \left[
        1+\frac{k_\mathrm{h}}{k_\mathrm{e}}{|a_1(E)|}^2-{|a_2(E)|}^2
    \right],
    \label{eq:app_Gamma_e}
    \\
    \bar{\Gamma}_\mathrm{N}(E)
    =&
    \frac{e^2}{2\pi\hbar}
    \left[
        1-{|a_2(E)|}^2
    \right],
    \\
    \bar{\Gamma}_{F}(E)
    =&
    \frac{e^2}{2\pi\hbar}
    \frac{k_\mathrm{h}}{k_\mathrm{e}}
    {|a_1(E)|}^2.
\end{align}
Applying Andreev approximation, i.e., $k_\mathrm{F}=k_\mathrm{e}=k_\mathrm{h}$, we obtain the differential conductance of the BTK approach given by Eq.~\eqref{eq:BTK_s_wave} with the BTK's notation $a_1(E)=a(E)$ and $a_2(E)=b(E)$.
It is noted that when we average over $x$ and $x'$ according to Eq.~\eqref{eq:spatial_average_app}, the contribution from $x=x'$ is not relevant for sufficiently large value of $L_2-L_1$, where the derivative of the Green function is not well defined at $x=x'$.
Therefore, we can neglect the contribution from $x=x'$.
\begin{align}
    \Gamma_F^\mathrm{ee(oo)}(x,x',E)
    =&
    \frac{e^2}{2\pi\hbar}
    \frac{1}{8}
    \left\{
    -(+)
    \frac{{(k_\mathrm{e}+k_\mathrm{h})}^2}{k_\mathrm{e}k_\mathrm{h}}
    2\mathrm{Re}
    \left[
        a^*(-E)a_1(E)
        \sin\theta\cos\theta
        e^{-i(k_\mathrm{e}-k_\mathrm{h})(x+x')}
    \right]
    \right.
    \nonumber\\
    &
    \hspace{1.4cm}
    \left.
    +(-)
    \frac{{(k_\mathrm{e}-k_\mathrm{h})}^2}{k_\mathrm{e}k_\mathrm{h}}
    \mathrm{Re}
    \left[
        a_1(-E)a_1(E)
        e^{-i(k_\mathrm{e}+k_\mathrm{h})(x-x')}
    \right]
    \right.
    \nonumber\\
    &
    \hspace{1.4cm}
    \left.
    +
    \frac{2k_\mathrm{e}}{k_\mathrm{h}}
    {|a_1(-E)|}^2
    +
    \frac{2k_\mathrm{h}}{k_\mathrm{e}}
    {|a_1(E)|}^2
    \right\},
    \label{eq:Gamma_ee_oo_conti_theta_app}
    \\
    \Gamma_{F}^\mathrm{eo}(x,x',E)
    =&
    \frac{e^2}{2\pi\hbar}
    \frac{1}{2}
    \left[
        \frac{k_\mathrm{h}}{k_\mathrm{e}}
        {|a_1(E)|}^2
        -
        \frac{k_\mathrm{e}}{k_\mathrm{h}}
        {|a_1(-E)|}^2
    \right].
\end{align}
After spatial averaging, we obtain 
\begin{align}
    \bar{\Gamma}_F^\mathrm{ee(oo,eo)}(E)
    =&
    \frac{1}{L^2}
    \int_{-L}^0 dx dx'
    \Gamma_F^\mathrm{ee(oo,eo)}(x,x',E).
\end{align}
Note that terms depending on $x$ and $x'$ vanish for $L\rightarrow\infty$.
Then, the final result reads
\begin{align}
    \bar{\Gamma}_F^\mathrm{ee}(E)
    =
    \bar{\Gamma}_F^\mathrm{oo}(E)
    =&
    \frac{e^2}{2\pi\hbar}
    \frac{1}{4}
    \left[
    \frac{k_\mathrm{h}}{k_\mathrm{e}}{|a_1(E)|}^2
    +
    \frac{k_\mathrm{e}}{k_\mathrm{h}}{|a_1(-E)|}^2
    \right],
    \label{eq:bar_Gamma_ee_oo_Kubo}
    \\
    \bar{\Gamma}_F^\mathrm{eo}(E)
    =&
    \frac{e^2}{2\pi\hbar}
    \frac{1}{2}
    \left[
        \frac{k_\mathrm{h}}{k_\mathrm{e}}
        {|a_1(E)|}^2
        -
        \frac{k_\mathrm{e}}{k_\mathrm{h}}
        {|a_1(-E)|}^2
    \right],
    \label{eq:bar_Gamma_eo_Kubo}
\end{align}
for $E\neq0$.
For $E=0$, $k_\mathrm{e}=k_\mathrm{h}=k_\mathrm{F}$ holds, and we obtain $\bar{\Gamma}_F^\mathrm{eo}(E=0)=0$ and
\begin{align}
    \bar{\Gamma}_F^\mathrm{ee(oo)}(E=0)
    =&
    \frac{e^2}{2\pi\hbar}
    \frac{1}{2}
    {|a_1(0)|}^2
    \left[
        1
        -(+)
        2
        \sin\theta\cos\theta
    \right].
    \label{eq:bar_Gamma_theta_app}
\end{align}
Here, the angle $\theta$ appears as introduced in Eqs.~\eqref{eq:def_tilde1} and \eqref{eq:def_tilde2}.
Equation~\eqref{eq:bar_Gamma_theta_app} indicate that the decomposition into even and odd-frequency contributions at $E=0$ is not unique.
Note that this angle $\theta$ does not affect the total conductance measured in the laboratory.

Here, we show the differential conductance and its components for the basis ${(\psi_\uparrow(x),\psi_\downarrow^\dagger(x))}^\mathrm{T}$. By utilizing the basis ${(\psi_\downarrow(x),\psi_\uparrow^\dagger(x))}^\mathrm{T}$, we obtain exactly the same differential conductance. Hence, the total differential conductance and its components are given by the twice of Eqs.~\eqref{eq:app_Gamma_e}-\eqref{eq:bar_Gamma_theta_app}.

\subsection{\label{sub:Green_func_McMillan2}Approximation for momentum $k_\mathrm{e}$ and $k_\mathrm{h}$}
For $|E|\ll\mu$ and $|\Delta|\ll\mu$, we can approximate the wave numbers as $k_\mathrm{F}=k_\mathrm{e}=k_\mathrm{h}=k_+=k_-$.
$a_1$, $a_2$, $a_3$, and $a_4$ can be written as
\begin{align}
    a_1(E)
    =&
    \frac{4\frac{v_{\mathrm{e}+}}{u_{\mathrm{e}+}}}{4+Z^2\left(1-\frac{v_{\mathrm{e}+}}{u_{\mathrm{e}+}}\frac{v_{\mathrm{h}-}}{u_{\mathrm{h}-}}\right)},
    \\
    a_2(E)
    =&
    iZ
    \frac{(2+iZ)(1-\frac{v_{\mathrm{e}+}}{u_{\mathrm{e}+}}\frac{v_{\mathrm{h}-}}{u_{\mathrm{h}-}})}{4+Z^2\left(1-\frac{v_{\mathrm{e}+}}{u_{\mathrm{e}+}}\frac{v_{\mathrm{h}-}}{u_{\mathrm{h}-}}\right)},
    \\
    a_3(E)
    =&
    \frac{4\frac{v_{\mathrm{h}-}}{u_{\mathrm{h}-}}}{4+Z^2\left(1-\frac{v_{\mathrm{e}+}}{u_{\mathrm{e}+}}\frac{v_{\mathrm{h}-}}{u_{\mathrm{h}-}}\right)},
    \\
    a_4(E)
    =&
    iZ
    \frac{(-2+iZ)(1-\frac{v_{\mathrm{e}+}}{u_{\mathrm{e}+}}\frac{v_{\mathrm{h}-}}{u_{\mathrm{h}-}})}{4+Z^2\left(1-\frac{v_{\mathrm{e}+}}{u_{\mathrm{e}+}}\frac{v_{\mathrm{h}-}}{u_{\mathrm{h}-}}\right)}.
\end{align}
The coefficient $a_1(E)$ for the $s$-wave junction is given by
\begin{align}
    a_1(E)
    =&
    \begin{cases}
        \frac{2\Delta}{2E+(2+Z^2)\mathrm{sgn}(E)\sqrt{E^2-{|\Delta|}^2}} & |E|>|\Delta|,
        \\
        \frac{2\Delta}{2E+i(2+Z^2)\sqrt{{|\Delta|}^2-E^2}} & |E|<|\Delta|,
    \end{cases}
    \label{eq:app_a1_s}
\end{align}
and, for the $p$-wave junction, it becomes
\begin{align}
    a_1(E)
    =&
    \begin{cases}
        \frac{2\Delta}{E(2+Z^2)+2\mathrm{sgn}(E)\sqrt{E^2-{|\Delta|}^2}} & |E|>|\Delta|,
        \\
        \frac{2\Delta}{E(2+Z^2)+2i\sqrt{{|\Delta|}^2-E^2}} & |E|<|\Delta|.
    \end{cases}
    \label{eq:app_a1_p}
\end{align}
For both $s$- and $p$-wave junctions, from Eqs.~\eqref{eq:app_a1_s} and \eqref{eq:app_a1_p}, we obtain
\begin{align}
    \bar{\Gamma}_F^\mathrm{eo}(E)=0.
\end{align}

Then, the contribution of the anomalous Green function to $\bar{\Gamma}_F(E)$ for the $s$-wave case is
\begin{align}
    \frac{2\pi\hbar}{e^2}
    \bar{\Gamma}_F^{\mathrm{ee}}(E)
    =&
    \frac{2\pi\hbar}{e^2}
    \bar{\Gamma}_F^{\mathrm{oo}}(E)
    =
    \frac{1}{4}
    \left[
        {|a_1(E)|}^2
        +
        {|a_1(-E)|}^2
    \right]
    \nonumber\\
    =&
    \begin{cases}
        \frac{2\Delta^2{(E+\mathrm{sgn}(E)\sqrt{E^2-\Delta^2})}^2}{{\left\{\Delta^2(2+Z^2)-\mathrm{sgn}(E)E\left[\mathrm{sgn}(E)E+\sqrt{E^2-\Delta^2}\right](4+Z^2)\right\}}^2}
        & |E|>|\Delta|,
        \\
        \frac{2\Delta^2}{\Delta^2{(2+Z^2)}^2-E^2Z^2(4+Z^2)}
        & |E|<|\Delta|,
    \end{cases}
\end{align}
for $E\neq0$.
Likewise, for the $p$-wave case, the components of $\bar{\Gamma}_F(E)$ are
\begin{align}
    \frac{2\pi\hbar}{e^2}
    \bar{\Gamma}_F^{\mathrm{ee}}(E)
    =&
    \frac{2\pi\hbar}{e^2}
    \bar{\Gamma}_F^{\mathrm{oo}}(E)
    =
    \frac{1}{4}
    \left[
        {|a_1(E)|}^2
        +
        {|a_1(-E)|}^2
    \right]
    \nonumber\\
    =&
    \begin{cases}
        \frac{2\Delta^2{(E+\mathrm{sgn}(E)\sqrt{E^2-\Delta^2})}^2}
        {{\left\{-\mathrm{sgn}(E)2\Delta^2+\left[(\mathrm{sgn}(E)E+\sqrt{E^2-\Delta^2}\right](4+Z^2)\right\}}^2}
        & |E|>|\Delta|,
        \\
        \frac{2\Delta^2}{4\Delta^2+E^2Z^2(4+Z^2)}
        & |E|<|\Delta|,
    \end{cases}
    \label{eq:app_Gamma_ee_p}
\end{align}
for $E\neq0$.
For the $p$-wave case, the limit $\lim_{E\rightarrow0}\frac{2\pi\hbar}{e^2}\bar{\Gamma}_F^\mathrm{oo}(E)=1/2$ irrespective of the value of $Z$ due to the presence of a Majorana state.

\section{\label{sec:lattilce_model}Lattice models}

\subsection{\label{sec:1d_lattice}1D N/1D S junction on lattice}
To compare our results between 1D continuum model and 1D lattice model, we first present the results about 1D N/1D S junctions derived on the lattice. 
The mean-field Hamiltonian is given by
\begin{align}
    H
    =&
    H_0+H_\Delta,
    \label{eq:app_1d_N_1d_S_Hamiltonian}
    \\
    H_0
    =&
    -\breve{t}\sum_{j<-1,j\geq0,\sigma}
    \left(
        c_{j,\sigma}^\dagger c_{j+1,\sigma}+\mathrm{H.c.}
    \right)
    -
    \breve{t}_\mathrm{b}\sum_{\sigma}\left(c_{-1,\sigma}^\dagger c_{0,\sigma}+\mathrm{H.c.}\right)
    -\mu\sum_{j,\sigma}c_{j,\sigma}^\dagger c_{j,\sigma},
    \\
    H_\Delta^s
    =&
    \Delta\sum_{j\geq0}\left(c_{j,\uparrow}^\dagger c_{j,\downarrow}^\dagger + \mathrm{H.c.} \right),
    \\
    H_\Delta^p
    =&
    \frac{\Delta}{2i}\sum_{j\geq0}\left(c_{j,\uparrow}^\dagger c_{j+1,\downarrow}^\dagger -c_{j,\downarrow}^\dagger c_{j+1,\uparrow}^\dagger + \mathrm{H.c.} \right),
\end{align}
with $H_\Delta=H_\Delta^{s(p)}$ for the $s(p)$-wave junction.
Here, $\breve{t}$ the hopping integral in 1D N and 1D S, $\breve{t}_\mathrm{b}$ the hopping integral between 1D N and 1D S, $\mu$ the chemical potential, and $\Delta\in\mathbb{R}$ the superconducting pair potential.
\subsection{\label{sec:conductnace_lattilce_model}Conductance in lattice model}
$\Gamma_\mathrm{e}(i,j,E)$ in the lattice model is given by
\begin{align}
    \Gamma_\mathrm{e}(i,j,E)
    =&
    \pi\hbar
    \mathrm{Tr}
    \left[
        P_\mathrm{e}
        \hat{J}_i\hat{\tilde{G}}_{1,i,j}(E)
        \hat{J}_j\hat{\tilde{G}}_{2,j,i}(E)
    \right]
    \label{eq:trans_prob_lattice_1d}
\end{align}
with the current operator on the 1D N side
\begin{align}
    J
    =&
    \frac{e}{i\hbar}[\hat{x},H_0]
    =
    \sum_{i_1,i_2,\sigma_1,\sigma_2}c_{i_1,\sigma_1}^\dagger \check{J}_{i_1,\sigma_1,i_2,\sigma_2}c_{i_2,\sigma_2},
    \\
    \check{J}
    =&
    \frac{e\breve{t}}{i\hbar}(\delta_{i_1,i_2-1}-\delta_{i_1,i_2+1})\delta_{\sigma_1,\sigma_2},
    \\
    \hat{x}
    =&
    \sum_{j,\sigma} r_j c_{j,\sigma}^\dagger c_{j,\sigma},
    \\
    \hat{J}_j=&
    \begin{pmatrix}
        0&\check{J}_{j,j+1}
        \\
        \check{J}_{j+1,j}&0
    \end{pmatrix},
\end{align}
where $r_j$ is the spatial position of the $j$-th lattice site.
$\hat{\tilde{G}}_{1,i,j}(E)$ and $\hat{\tilde{G}}_{2,i,j}(E)$ are given by
\begin{align}
    \hat{\tilde{G}}_{1(2),i,j}(E)
    =&
    \begin{pmatrix}
        \tilde{G}_{1(2),i,j}(E) & \tilde{G}_{1(2),i,j+1}(E)
        \\
        \tilde{G}_{1(2),i+1,j}(E) & \tilde{G}_{1(2),i+1,j+1}(E)
    \end{pmatrix},
    \\
    \tilde{G}_{1,i,j}(E)
    =&
    \begin{cases}
        \frac{1}{2\pi i}
        \left[
        \check{G}^{\mathrm{A}}_{i,j}(E)-\check{G}^{\mathrm{R}}_{i,j}(E)
        \right]
        & i=j,\:i=j\pm 1,
        \\
        \frac{1}{\sqrt{2}\pi i}
        \left[
        \sin\theta \check{G}^{\mathrm{A}}_{i,j}(E)-\cos\theta \check{G}^{\mathrm{R}}_{i,j}(E)
        \right] & \mathrm{other\:cases},
    \end{cases}
    \\
    \tilde{G}_{2,i,j}(E)
    =&
    \begin{cases}
        \frac{1}{2\pi i}
        \left[
        \check{G}^{\mathrm{A}}_{i,j}(E)-\check{G}^{\mathrm{R}}_{i,j}(E)
        \right]
        & i=j,\:i=j\pm 1,
        \\
        \frac{1}{\sqrt{2}\pi i}
        \left[
        \cos\theta \check{G}^{\mathrm{A}}_{i,j}(E)-\sin\theta \check{G}^{\mathrm{R}}_{i,j}(E)
        \right] & \mathrm{other\:cases}.
    \end{cases}
\end{align}
Note that we introduce the same angle $\theta$ as in Eqs.~\eqref{eq:def_tilde1} and \eqref{eq:def_tilde2} in the continuum case.
In the main text, we use $\theta=\pi/2$ for 1D N (ladder)/2D S junctions since the numerical convergence is the fastest.

We denote the spatial averaging of Eq.~\eqref{eq:trans_prob_lattice_1d} as
\begin{align}
    \bar{\Gamma}_\mathrm{e}(E)
    =&
    \frac{1}{L^2}
    \sum_{-L\leq i,j\leq -1}
    \Gamma_\mathrm{e}(i,j,E).
    \label{eq:spatial_average_lattice_1d_app}
\end{align}

Here, let us show $\bar{\Gamma}_F^\mathrm{eo}(E)=0$ for the lattice model.
The BdG Hamiltonian $H$ has particle-hole symmetry:
\begin{align}
    C^{-1}HC=&-H,
    \\
    H=&
    \begin{pmatrix}
        h & \Delta
        \\
        \Delta^\dagger & -h^\mathrm{T}
    \end{pmatrix},
    \\
    C=&\hat{\tau}_1K
\end{align}
with complex conjugation operator $K$.
The Green function $G(z)={(z-H)}^{-1}$ ($z\in\mathbb{C}$) satisfies $G^\dagger(z)=G(z^*)$ and $C^{-1}G(z)C=-G(-z^*)$.
\begin{align}
    \hat{\tilde{G}}^\dagger_{1}(E)
    =&
    \hat{\tilde{G}}_{2}(E),
    \\
    {\left[\hat{\tilde{F}}_1^\mathrm{even(odd)}(E)\right]}^\dagger
    =&
    \hat{\tilde{F}}_2^\mathrm{even(odd)}(E),
    \\
    C^{-1}\hat{\tilde{G}}_{1(2)}(E)C
    =&
    \hat{\tilde{G}}_{1(2)}(-E),
    \\
    C^{-1}\hat{\tilde{F}}_{1}^{\mathrm{even(odd)}}(E)C
    =&
    -(+)\hat{\tilde{F}}_{2}^{\mathrm{even(odd)}}(E),
    \\
    C^{-1}JC
    =&
    -\frac{e}{i\hbar}C^{-1}[\hat{x},H_0]C
    =
    -J
\end{align}
with $H_0=\begin{pmatrix}
    h&0
    \\
    0&-h^\mathrm{T}
\end{pmatrix}$, 
$\hat{x}
    =
    \begin{pmatrix}
        x & 0
        \\
        0 & -x
    \end{pmatrix}$,
    $x=
    \mathrm{diag}
    (\ldots,r_{j-1},r_j,r_{j+1},r_{j+2},\ldots)$, 
    $C^{-1}\hat{x}C=-\hat{x}$,
$G^\mathrm{A}(E)=G(E-i\eta)$, and $G^\mathrm{R}(E)=G(E+i\eta)$.

The cross term between even and odd frequency pairing for the differential conductance satisfies
\begin{align}
    {[\Gamma_F^\mathrm{eo}(i,j,E)]}^*
    =&
    \pi\hbar
    \mathrm{Tr}
    {\left[
    P_\mathrm{e}
    \hat{J}_i \hat{\tilde{F}}^\mathrm{even}_{1,i,j}(E)
    \hat{J}_j \hat{\tilde{F}}^\mathrm{odd}_{2,j,i}(E)
    \right]}^\dagger
    +(\mathrm{even}\leftrightarrow\mathrm{odd})
    \nonumber\\
    =&
    \pi\hbar
    \mathrm{Tr}
    \left\{
        P_\mathrm{e}
        \hat{J}_i {\left[\hat{\tilde{F}}^\mathrm{odd}_{2,j,i}(E)\right]}^\dagger
        \hat{J}_j {\left[\hat{\tilde{F}}^\mathrm{even}_{1,i,j}(E)\right]}^\dagger
    \right\}
    +(\mathrm{even}\leftrightarrow\mathrm{odd})
    \nonumber\\
    =&
    \pi\hbar
    \mathrm{Tr}
    \left[
        P_\mathrm{e}
        \hat{J}_i \hat{\tilde{F}}^\mathrm{odd}_{1,i,j}(E)
        \hat{J}_j \hat{\tilde{F}}^\mathrm{even}_{2,j,i}(E)
    \right]
    +(\mathrm{even}\leftrightarrow\mathrm{odd})
    \nonumber\\
    =&
    \Gamma_F^\mathrm{eo}(i,j,E).
    \label{eq:app_gamma_F_eo_lattice1}
\end{align}
Here, we use the relation $P_\mathrm{e}\hat{J}_{i}=\hat{J}_{i}P_\mathrm{e}$.
Also, from particle-hole symmetry, $\Gamma_F^\mathrm{eo}(i,j,E)$ satisfies
\begin{align}
    {[\Gamma_F^\mathrm{eo}(i,j,E)]}^*
    =&
    \pi\hbar
    \mathrm{Tr}
    \left[
        C^{-1}
        P_\mathrm{e}
        CC^{-1}
        \hat{J}_i
        CC^{-1}
        \hat{\tilde{F}}^\mathrm{even}_{1,i,j}(E)
        CC^{-1}
        \hat{J}_j 
        CC^{-1}
        \hat{\tilde{F}}^\mathrm{odd}_{2,j,i}(E)
        C
    \right]
    +(\mathrm{even}\leftrightarrow\mathrm{odd})
    \nonumber\\
    =&
    -
    \pi\hbar
    \mathrm{Tr}
    \left[
        P_\mathrm{h}
        \hat{J}_i
        \hat{\tilde{F}}^\mathrm{even}_{2,i,j}(E)
        \hat{J}_j 
        \hat{\tilde{F}}^\mathrm{odd}_{1,j,i}(E)
    \right]
    +(\mathrm{even}\leftrightarrow\mathrm{odd})
    \nonumber\\
    =&
    -
    \Gamma_F^\mathrm{eo}(j,i,E)
    \label{eq:app_gamma_F_eo_lattice2}
\end{align}
with $P_\mathrm{h}=(\hat{\tau}_0-\hat{\tau}_3)/2$.
After averaging over $i$ and $j$ of Eqs.~\eqref{eq:app_gamma_F_eo_lattice1} and ~\eqref{eq:app_gamma_F_eo_lattice2}, we obtain $\bar{\Gamma}_F^\mathrm{eo}(E)=0$.

\subsection{\label{app_sub:Gamma_1d_lattice} $\bar{\Gamma}_\mathrm{e}(E)$ and odd-frequency pairing}
\begin{figure}[t]
    \centering
    \includegraphics[width=8.5cm]{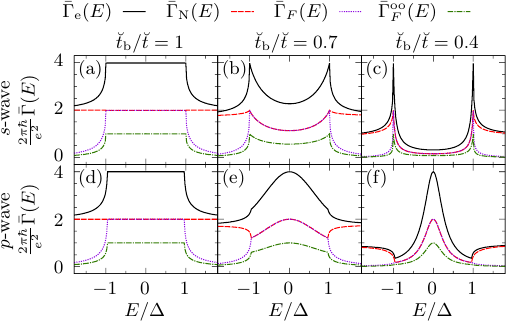}
    \caption{
       $\bar{\Gamma}_\mathrm{e}(E)$, $\bar{\Gamma}_\mathrm{N}(E)$, $\bar{\Gamma}_F(E)$, and $\bar{\Gamma}_F^\mathrm{oo}(E)$ are plotted as a function of $E$ for $\breve{t}_\mathrm{b}/\breve{t}=1$, $0.7$, and $0.4$ with $\Delta/\breve{t}=0.1$, $\mu/\breve{t}=-0.5$, $\eta/\breve{t}=10^{-7}$, and $L=500$. 
       (a)--(c) 1D N/1D $s$-wave junctions, and (d)--(f) 1D N/1D $p$-wave junctions.
    }
    \label{fig:Gamma_1d_lattice}
\end{figure}

The components of $\bar{\Gamma}_\mathrm{e}(E)$ for the $s$-wave junction are shown in Figs.~\ref{fig:Gamma_1d_lattice}(a)--(c).
The qualitative results are the same as those in the continuum model.
We attribute all differences to finite-size effects.
In Fig.~\ref{fig:Gamma_1d_lattice}(a), $\frac{2\pi\hbar}{e^2}\bar{\Gamma}_\mathrm{e}(E)$ at $E=0$ is almost $4$ (the maximum value). 
At $\mu=0$, we can analytically obtain the condition of $\frac{2\pi\hbar}{e^2}\bar{\Gamma}_\mathrm{e}(E=0)=4$ for the $s$-wave junction, which is given by
$\breve{t}_\mathrm{b}^8=\frac{\Delta \breve{t}^8\sqrt{\Delta^2+4\breve{t}^2}+2\breve{t}^{10}+\Delta^2 \breve{t}^8}{\Delta^2-\Delta\sqrt{\Delta^2+4\breve{t}^2}+2\breve{t}^2}$.
Figures~\ref{fig:Gamma_1d_lattice}(d)--(f) show $\bar{\Gamma}_\mathrm{e}(E)$ and its components for the $p$-wave junction.
At $E=0$, $\frac{2\pi\hbar}{e^2}\bar{\Gamma}_\mathrm{e}(E=0)=4$ holds due to the presence of the Majorana state independent of the value of $\breve{t}_\mathrm{b}$ since the chemical potential resides in the topological regime.

\begin{figure}[t]
    \centering
    \includegraphics[width=8.5cm]{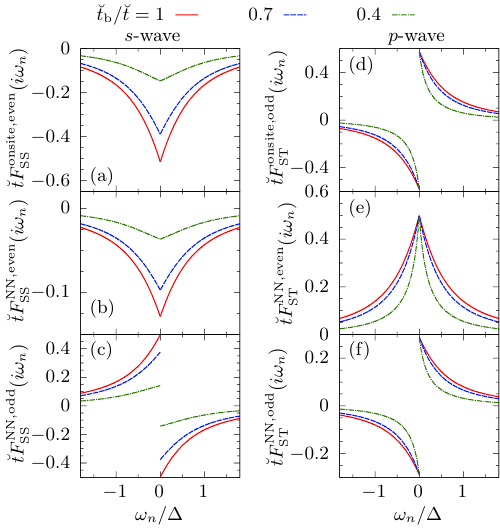}
    \caption{The anomalous Green function on the 1D N side is plotted as a function of $\omega_n$. (a)--(c) spin-singlet components of $s$-wave junctions, (d)--(e) spin-triplet components of $p$-wave junctions. 
    (a) Real part of onsite component, (b) real part of the NN even-frequency component, (c) imaginary part of the NN odd-frequency components. 
    (d) Real part of onsite component, (e) imaginary part of the NN even-frequency component, (f) real part of the NN odd-frequency components. 
    $\mathrm{Im}F_\mathrm{SS}^\mathrm{onsite,even}(i\omega_n)=\mathrm{Im}F_\mathrm{SS}^\mathrm{NN,even}(i\omega_n)=\mathrm{Re}F_\mathrm{SS}^\mathrm{NN,odd}(i\omega_n)=0$ for the $s$-wave junction.
    $\mathrm{Im}F_\mathrm{ST}^\mathrm{onsite,odd}(i\omega_n)=\mathrm{Re}F_\mathrm{ST}^\mathrm{NN,even}(i\omega_n)=\mathrm{Im}F_\mathrm{ST}^\mathrm{NN,odd}(i\omega_n)=0$ for the $p$-wave junction.
    $\Delta/\breve{t}=0.1$ and $\mu/\breve{t}=-0.5$.
    }
    \label{fig:F_1d_lattice}
\end{figure}

In Fig.~\ref{fig:F_1d_lattice}, the anomalous Green function as a function of the Matsubara frequency in 1D N is shown. 
The onsite and nearest neighbor (NN) spin-singlet (SS) and spin-triplet (ST) components are given by
\begin{align}
    F_\mathrm{SS(ST)}^\mathrm{onsite,even(odd)}(i\omega_n)
    =&
    \frac{1}{4}\left\{
    \left[F_{-1,-1,\uparrow,\downarrow}^{12}(i\omega_n)+\zeta F_{-1,-1,\downarrow,\uparrow}^{12}(i\omega_n)\right]
    \right.
    +\xi
    \left.
    \left[F_{-1,-1,\uparrow,\downarrow}^{12}(-i\omega_n)+\zeta F_{-1,-1,\downarrow,\uparrow}^{12}(-i\omega_n)\right]
    \right\},
    \label{eq:app_def_onsite_even_odd}
    \\
    F_\mathrm{SS(ST)}^{\mathrm{NN,even(odd)}}(i\omega_n)
    =&
    \frac{1}{4}\left\{
    \left[F_{-2,-1,\uparrow,\downarrow}^{12}(i\omega_n)+\zeta F_{-2,-1,\downarrow,\uparrow}^{12}(i\omega_n)\right]
    \right.
    +\xi
    \left.
    \left[F_{-2,-1,\uparrow,\downarrow}^{12}(-i\omega_n)+\zeta F_{-2,-1,\downarrow,\uparrow}^{12}(-i\omega_n)\right]
    \right\}
    \label{eq:app_def_nn_even_odd}
\end{align}
with $\zeta=-1$ for the SS case, $\zeta=1$ for the ST case, $\xi=1$ for the even frequency pair, and $\xi=-1$ for the odd frequency pair.

In Figs.~\ref{fig:F_1d_lattice}(a)--(c), anomalous Green functions for $s$-wave junctions are shown.
Figures.~\ref{fig:F_1d_lattice}(a) and (b) display the even frequency components.
Both of them have non-zero values.
Fig.~\ref{fig:F_1d_lattice}(c) displays the NN component of odd-frequency pairing.
Note that $\mathrm{Im}F^\mathrm{NN,odd}_\mathrm{SS}(i\omega_n)$ has jump at $\omega_n=0$.
This jump significantly contributes to $\bar{\Gamma}_F(E=0)$.
Figures~\ref{fig:F_1d_lattice}(d)--(f) show the anomalous Green functions for $p$-wave junctions.
Figures.~\ref{fig:F_1d_lattice}(d) and (f) are the corresponding odd-frequency components.
These components (with $\omega_n\rightarrow0$) contribute to $\bar{\Gamma}_F(E=0)$. 
Due to the presence of the Majorana state, the odd-frequency component of $F(i\omega_n\rightarrow\pm0)$ is independent of $\breve{t}_\mathrm{b}$.
Additionally, the even-frequency component of the anomalous Green function [Fig.~\ref{fig:F_1d_lattice}(e)] with $\omega_n\rightarrow0$ is independent of $\omega_n$.

\begin{figure}[t]
    \centering
    \includegraphics[width=17cm]{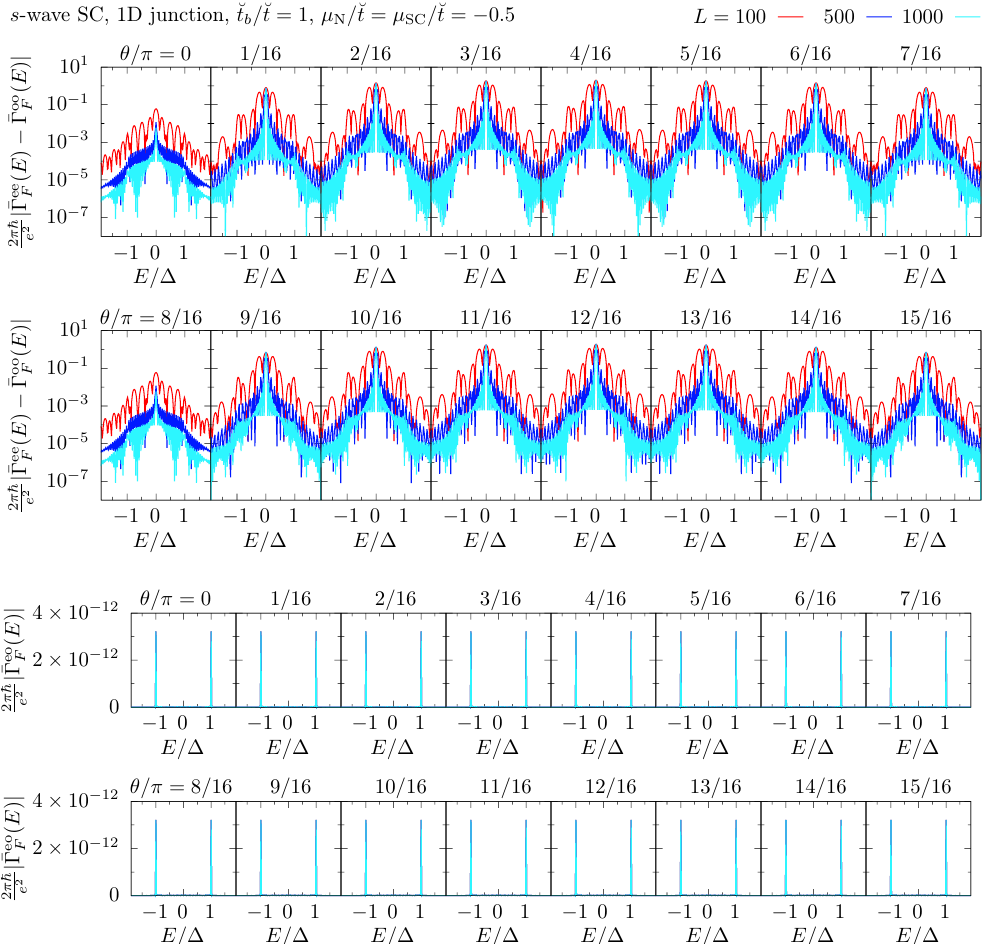}
    \caption{
    $L$ and $\theta$ dependence of $|\bar{\Gamma}_F^\mathrm{ee}(E)-\bar{\Gamma}_F^\mathrm{oo}(E)|$ and $|\bar{\Gamma}_F^\mathrm{eo}(E)|$ are plotted as a function of $E$ for 1D lattice $s$-wave junctions.
    $\Delta/\breve{t}=0.1$, $\breve{t}_\mathrm{b}/\breve{t}=1$, $\mu/\breve{t}=-0.5$, and $\eta/\breve{t}=10^{-7}$.
    }
    \label{fig:size_dep_s_1d}
\end{figure}

\begin{figure}[t]
    \centering
    \includegraphics[width=17cm]{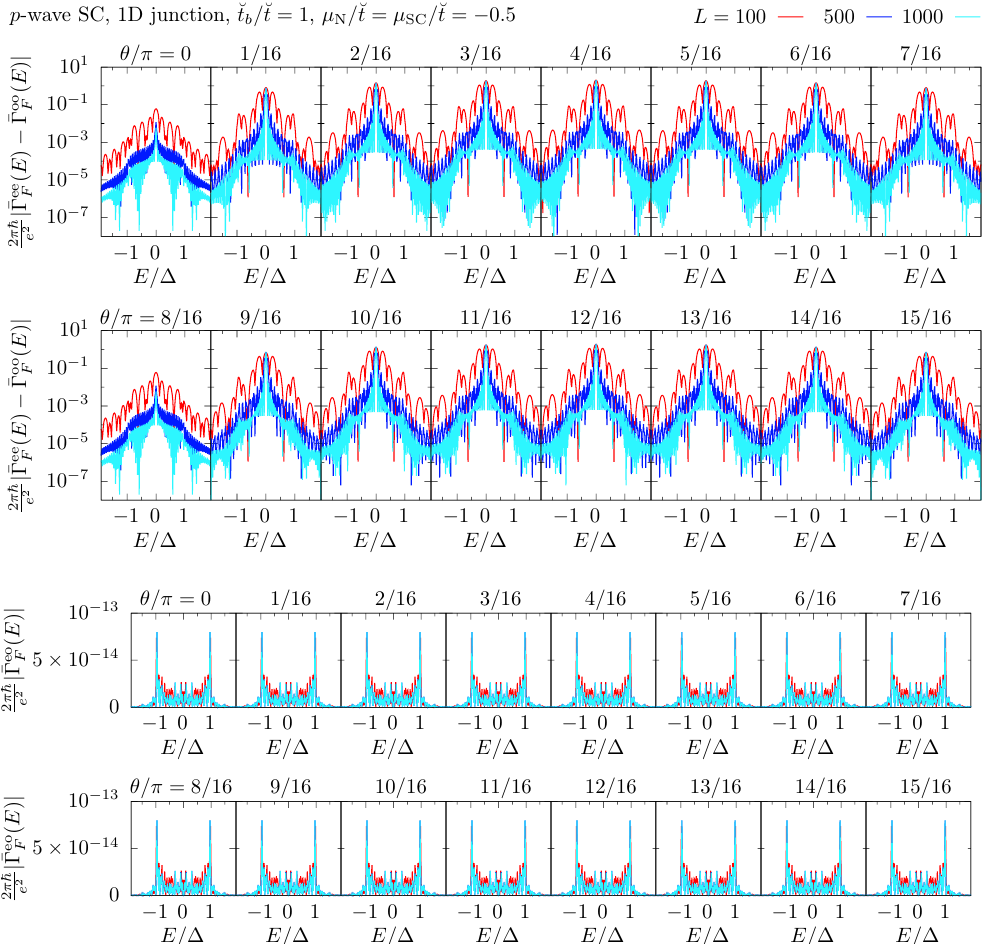}
    \caption{
    Size and $\theta$ dependence of $|\bar{\Gamma}_F^\mathrm{ee}(E)-\bar{\Gamma}_F^\mathrm{oo}(E)|$ and $|\bar{\Gamma}_F^\mathrm{eo}(E)|$ are plotted as a function of $E$ for 1D lattice $p$-wave junctions.
    $\Delta/\breve{t}=0.1$, $\breve{t}_\mathrm{b}/\breve{t}=1$, $\mu/\breve{t}=-0.5$, and $\eta/\breve{t}=10^{-7}$.
    }
    \label{fig:size_dep_p_1d}
\end{figure}

In Fig.~\ref{fig:size_dep_s_1d}, we show $\theta$ and $L$ dependence of $\bar{\Gamma}^{\mathrm{ee}}_{F}(E)-\bar{\Gamma}^{\mathrm{oo}}_{F}(E)$ and $\bar{\Gamma}^{\mathrm{eo}}_{F}(E)$ for $s$-wave junctions.
The numerical convergence of $\bar{\Gamma}^{\mathrm{ee}}_{F}(E)-\bar{\Gamma}^{\mathrm{oo}}_{F}(E)$ is the fastest for $\theta=0$ and $\pi/2$.
For the continuum model, in Eq.~\eqref{eq:Gamma_ee_oo_conti_theta_app}, the first term, which has longer spatial oscillation period compared with the second term, is zero for $\theta=0$ and $\pi/2$.
This indicates that at $\theta=0$ or $\pi/2$, the slowly varying term also vanishes in lattice models, which might be the reason that the numerical convergence is the fastest.
$\bar{\Gamma}_{F}^\mathrm{eo}(E)$ is zero within numerical errors.
For $p$-wave junctions shown in Figs.~\ref{fig:size_dep_p_1d}, the numerical convergence of $\bar{\Gamma}^{\mathrm{ee}}_{F}(E)-\bar{\Gamma}^{\mathrm{oo}}_{F}(E)$ is also the fastest for $\theta=0$ and $\pi/2$.
$\bar{\Gamma}^{\mathrm{eo}}_{F}(E)$ is also zero within numerical error.
We conclude that average values do not depend on $\theta$ for a sufficiently large averaging length $L$.

\subsection{\label{app_sub:x_dep_Gamma_1d_lattice} $i$ and $j$ dependence of $\Gamma_F^\mathrm{ee}(i,j,E)$ and $\Gamma_F^\mathrm{oo}(i,j,E)$}
\begin{figure}[t]
    \centering
    \includegraphics[width=12cm]{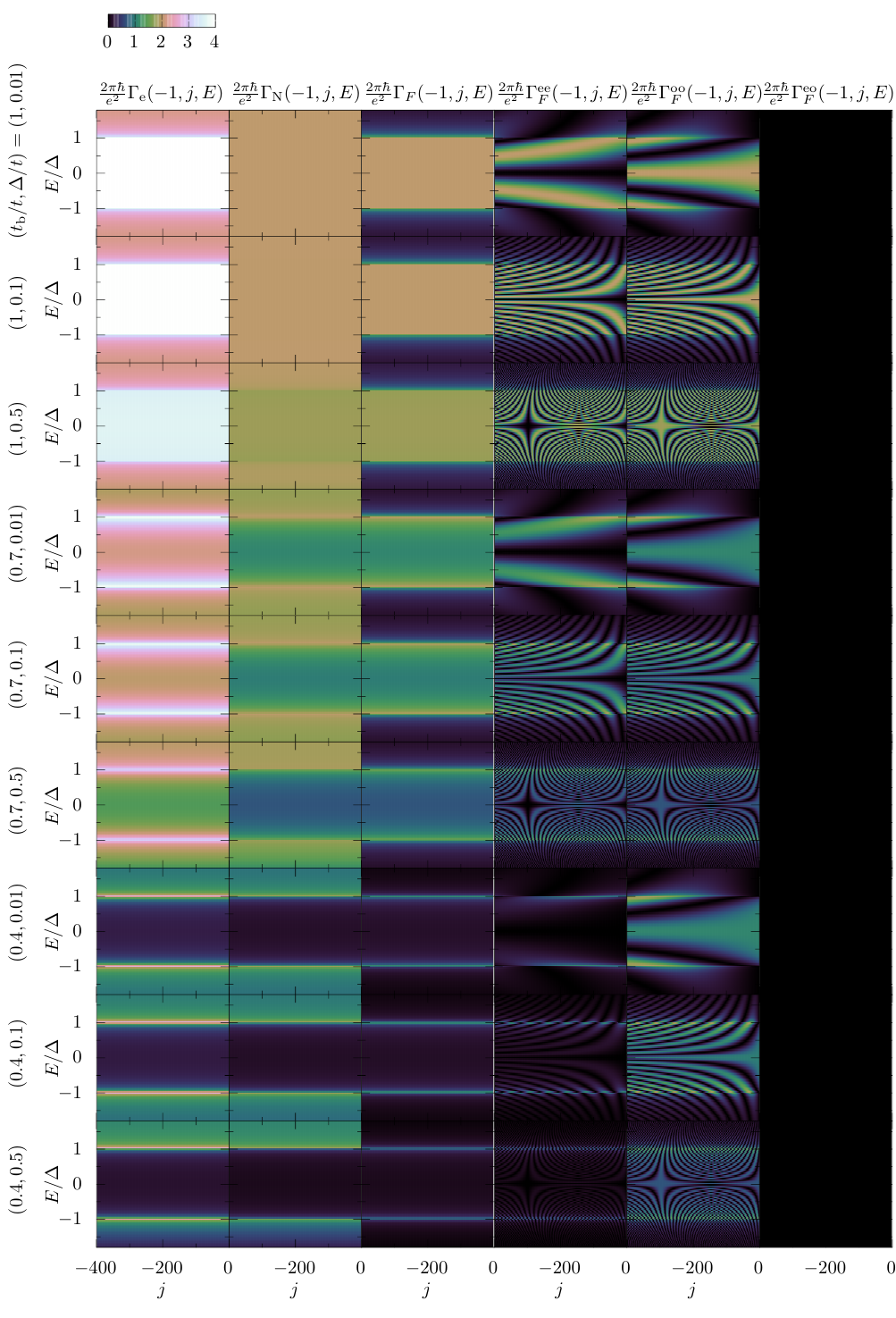}
    \caption{
    $\Gamma_\mathrm{e}(-1,j,E)$, $\Gamma_\mathrm{N}(-1,j,E)$, $\Gamma_F(-1,j,E)$, $\Gamma_F^\mathrm{ee}(-1,j,E)$, $\Gamma_F^\mathrm{oo}(-1,j,E)$, and $\Gamma_F^\mathrm{eo}(-1,j,E)$ are plotted as functions of $E$ and $j$ for 1D N/1D $s$-wave junction with $\mu/t=-1$, $(t_\mathrm{b}/t,\Delta/t)=(1,0.01)$, $(1,0.1)$, $(1,0.5)$, $(0.7,0.01)$, $(0.7,0.1)$, $(0.7,0.5)$, $(0.4,0.01)$, $(0.4,0.1)$, and $(0.4,0.5)$.
    }
    \label{fig:site_dep_s}
\end{figure}
\begin{figure}[t]
    \centering
    \includegraphics[width=12cm]{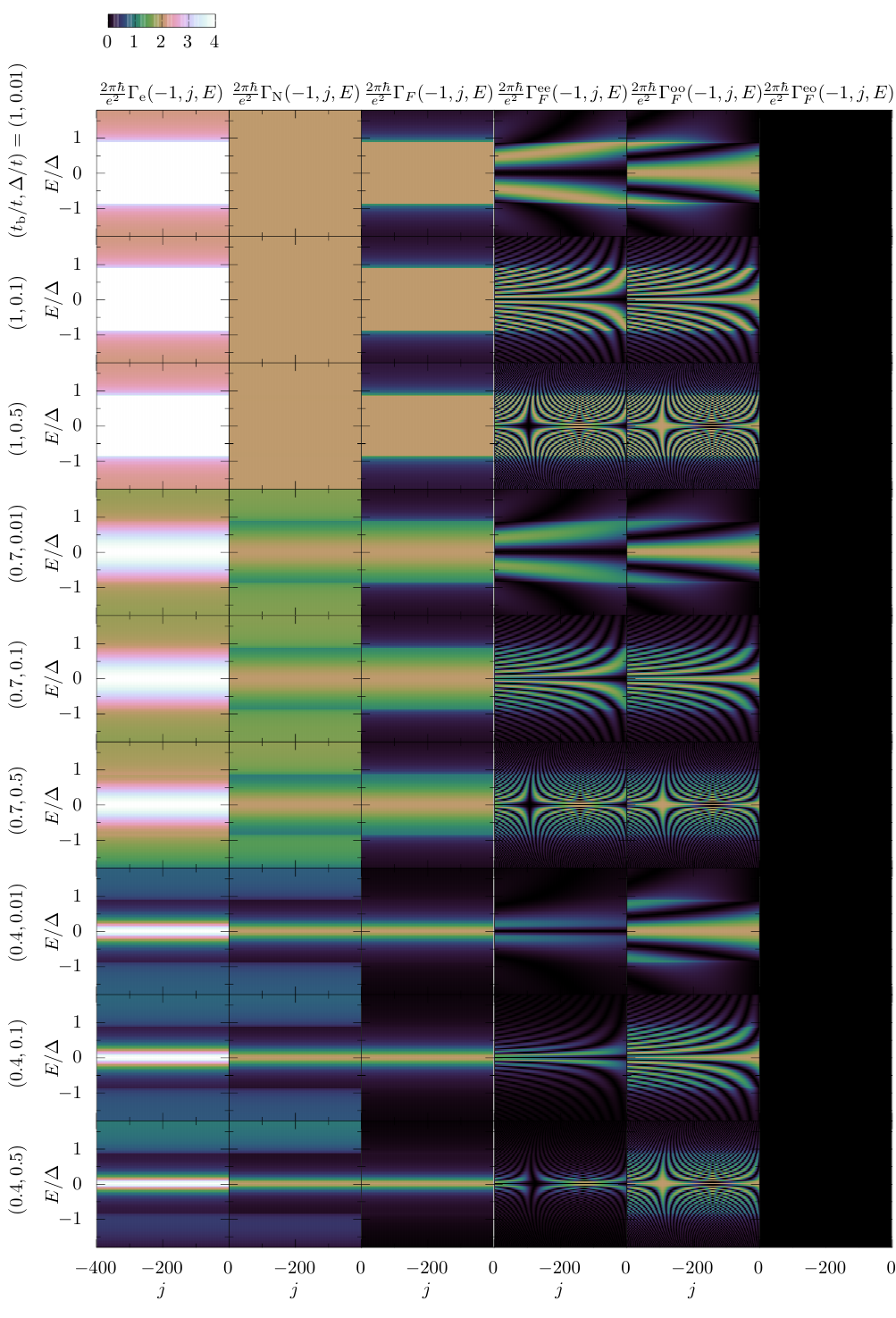}
    \caption{
    $\Gamma_\mathrm{e}(-1,j,E)$, $\Gamma_\mathrm{N}(-1,j,E)$, $\Gamma_F(-1,j,E)$, $\Gamma_F^\mathrm{ee}(-1,j,E)$, $\Gamma_F^\mathrm{oo}(-1,j,E)$, and $\Gamma_F^\mathrm{eo}(-1,j,E)$ are plotted as functions of $E$ and $j$ for 1D N/1D $p$-wave junction with $\mu/t=-1$, $(t_\mathrm{b}/t,\Delta/t)=(1,0.01)$, $(1,0.1)$, $(1,0.5)$, $(0.7,0.01)$, $(0.7,0.1)$, $(0.7,0.5)$, $(0.4,0.01)$, $(0.4,0.1)$, and $(0.4,0.5)$.
    }
    \label{fig:site_dep_p}
\end{figure}
We discuss the positional dependence of $\Gamma_F^\mathrm{ee,oo,eo}(i,j,E)$ for the 1D N/1D S junction.
In this section, we adopt $\theta=\pi/4$.
We fix $i=-1$ and study the $j$ dependence of the components of $\Gamma_\mathrm{e}(-1,j,E)$.

Figures~\ref{fig:site_dep_s} and \ref{fig:site_dep_p} display $\Gamma_\mathrm{e}(i=-1,j,E)$ and its components for the 1D N/1D $s$-wave and 1D N/1D $p$-wave junction, respectively.
As discussed in Sec.~\ref{sub:decomp_cond}, $\Gamma_\mathrm{e}(i,j,E)$, $\Gamma_\mathrm{N}(i,j,E)$, and $\Gamma_F(i,j,E)$ are independent of $i$ and $j$. 
In $s$-wave and $p$-wave junctions, $\Gamma_F^\mathrm{eo}(i,j,E)=0$ holds.
We see that $\Gamma_F^\mathrm{ee}(-1,j,E)$ and $\Gamma_F^\mathrm{oo}(-1,j,E)$ have a finite $j$ dependence.
From Figs.~\ref{fig:site_dep_s} and \ref{fig:site_dep_p}, we observe that the spatial dependence of $\Gamma_F^\mathrm{ee,oo}(-1,j,E)$ is smaller for smaller $\Delta$ ($\Delta/t=0.01$).
When $\Delta$ becomes larger ($\Delta/t=0.5$), $\Gamma_F^\mathrm{ee,oo}(-1,j,E)$ changes rapidly as a function of $j$.
These results indicate that the oscillation period is determined by the superconducting coherence length.

\begin{figure}[t]
    \centering
    \includegraphics[width=12cm]{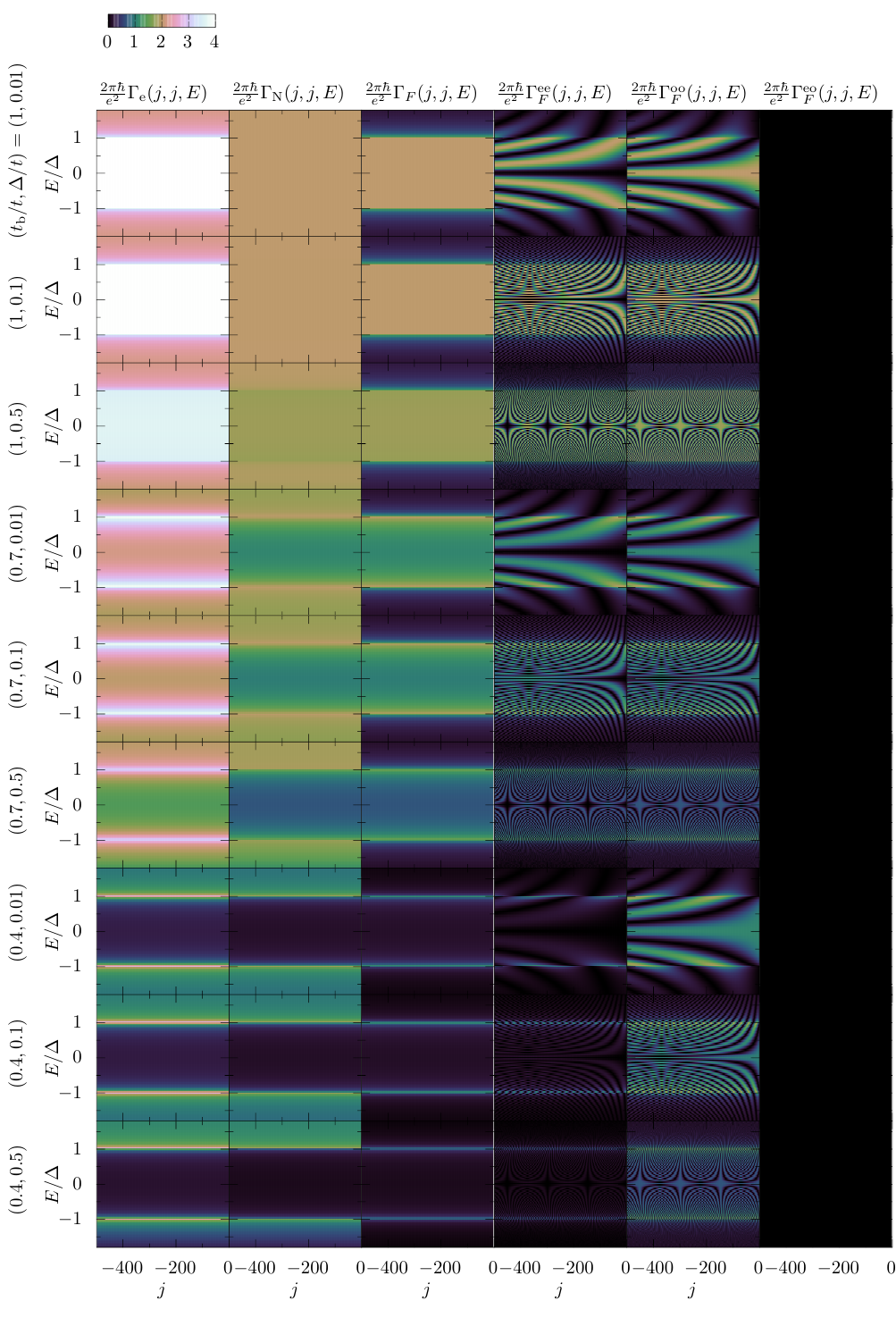}
    \caption{
    $\Gamma_\mathrm{e}(j,j,E)$, $\Gamma_\mathrm{N}(j,j,E)$, $\Gamma_F(j,j,E)$, $\Gamma_F^\mathrm{ee}(j,j,E)$, $\Gamma_F^\mathrm{oo}(j,j,E)$, and $\Gamma_F^\mathrm{eo}(j,j,E)$ are plotted as functions of $E$ and $j$ for 1D N/1D $s$-wave junction with $\mu/t=-1$, $(t_\mathrm{b}/t,\Delta/t)=(1,0.01)$, $(1,0.1)$, $(1,0.5)$, $(0.7,0.01)$, $(0.7,0.1)$, $(0.7,0.5)$, $(0.4,0.01)$, $(0.4,0.1)$, and $(0.4,0.5)$.
    }
    \label{fig:site_dep_s_same}
\end{figure}
\begin{figure}[t]
    \centering
    \includegraphics[width=12cm]{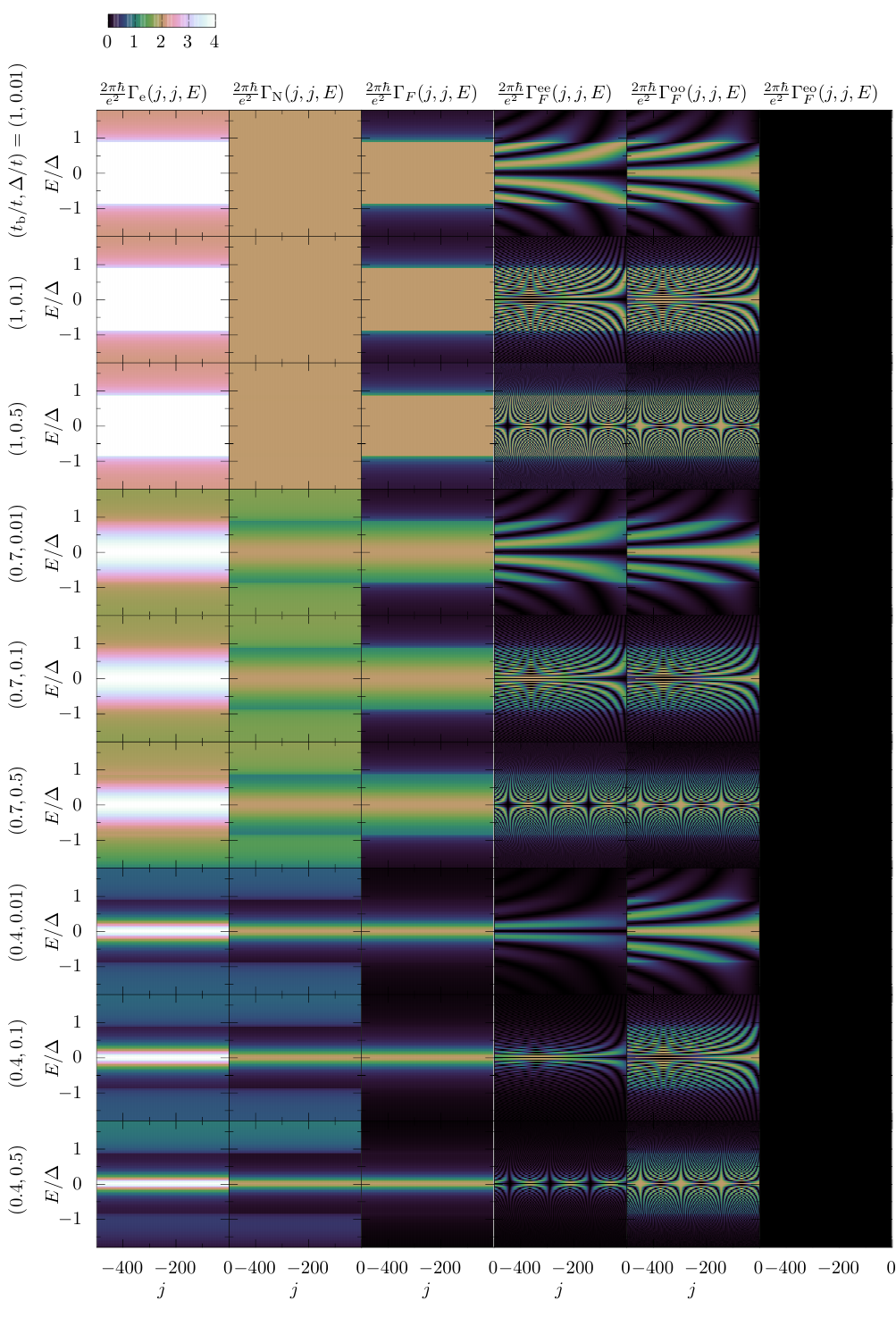}
    \caption{
    $\Gamma_\mathrm{e}(j,j,E)$, $\Gamma_\mathrm{N}(j,j,E)$, $\Gamma_F(j,j,E)$, $\Gamma_F^\mathrm{ee}(j,j,E)$, $\Gamma_F^\mathrm{oo}(j,j,E)$, and $\Gamma_F^\mathrm{eo}(j,j,E)$ are plotted as functions of $E$ and $j$ for 1D N/1D $p$-wave junction with $\mu/t=-1$, $(t_\mathrm{b}/t,\Delta/t)=(1,0.01)$, $(1,0.1)$, $(1,0.5)$, $(0.7,0.01)$, $(0.7,0.1)$, $(0.7,0.5)$, $(0.4,0.01)$, $(0.4,0.1)$, and $(0.4,0.5)$.
    }
    \label{fig:site_dep_p_same}
\end{figure}
Figures~\ref{fig:site_dep_s_same} and \ref{fig:site_dep_p_same} show $\Gamma_\mathrm{e}(j,j,E)$ and its components for the 1D N/1D $s$-wave and 1D N/1D $p$-wave junction, respectively.
In these figures, we can see that the qualitative behaviors are the same as Figs.~\ref{fig:site_dep_s} and \ref{fig:site_dep_p}, respectively.

\subsection{\label{sub:app_DN_junction}Robustness of $\bar{\Gamma}_F^\mathrm{ee}(E)=\bar{\Gamma}_F^\mathrm{oo}(E)$ against disorder}
\begin{figure}[htbp]
    \centering
    \includegraphics[width=10cm]{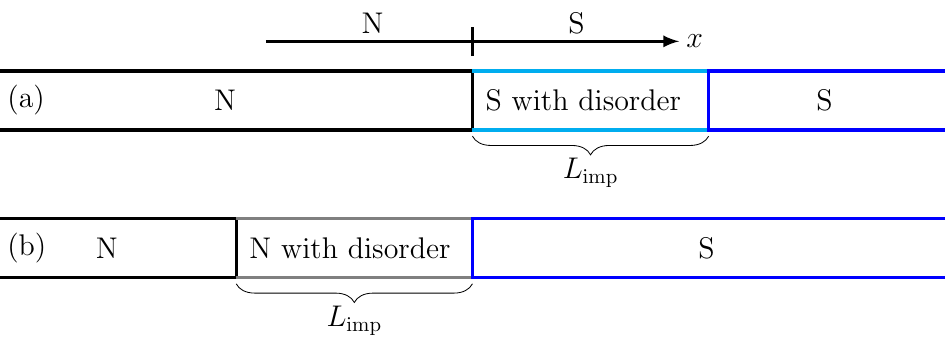}
    \caption{Schematic of junction with disorder in (a) the superconducting region and (b) the normal region.}
    \label{fig:app_schematic_junction_diffusive}
\end{figure}

Here, we show that $\bar{\Gamma}_F^\mathrm{ee}(E)=\bar{\Gamma}_F^\mathrm{oo}(E)$ holds in the presence of disorder by numerical calculation.
We consider two types of disordered 1D junctions as shown in Fig.~\ref{fig:app_schematic_junction_diffusive}. 
1) Normal metal/disordered S/S junction [Fig.~\ref{fig:app_schematic_junction_diffusive}(a)].
2) Normal metal/disordered metal/S junction [Fig.~\ref{fig:app_schematic_junction_diffusive}(b)].

The Hamiltonian for the normal metal/disordered S/S junction is given by
\begin{align}
    H
    =&
    H_\mathrm{kin}+H_\mathrm{onsite}+H_\Delta,
    \label{eq:app_H_diffusive}
    \\
    H_\mathrm{kin}
    =&
    -\breve{t}\sum_{j\neq 1,\sigma}
    \left(
        c_{j,\sigma}^\dagger c_{j+1,\sigma}+\mathrm{H.c.}
    \right)
    -\breve{t}_\mathrm{b}\sum_{\sigma}
    \left(
        c_{-1,\sigma}^\dagger c_{0,\sigma}+\mathrm{H.c.}
    \right),
    \\
    H_\mathrm{onsite}
    =&
    -\mu\sum_{j,\sigma}c_{j,\sigma}^\dagger c_{j,\sigma}
    +\sum_{0\le j<L_\mathrm{imp},\sigma}V_{j}c_{j,\sigma}^\dagger c_{j,\sigma},
    \\
    H_\mathrm{\Delta}
    =&
    \begin{cases}
        \Delta
        \sum_{0\le j<L_\mathrm{imp}}(1+\varepsilon_j)
        \left(
            e^{i\theta_j^\mathrm{imp}}
            c_{j,\uparrow}^\dagger 
            c_{j,\downarrow}^\dagger 
            +
            \mathrm{H.c.}
        \right)
        +
        \Delta
        \sum_{L_\mathrm{imp}\le j}
        \left(
            c_{j,\uparrow}^\dagger 
            c_{j,\downarrow}^\dagger 
            +
            \mathrm{H.c.}
        \right)
        & s\mathrm{-wave},
        \\
        \Delta
        \sum_{0\le j<L_\mathrm{imp}-1}(1+\varepsilon_j)
        \left(
            e^{i\theta_j^\mathrm{imp}}
            c_{j,\uparrow}^\dagger 
            c_{j+1,\downarrow}^\dagger 
            +
            \mathrm{H.c.}
        \right)
        +
        \Delta
        \sum_{L_\mathrm{imp}< j}
        \left(
            c_{j,\uparrow}^\dagger 
            c_{j+1,\downarrow}^\dagger 
            +
            \mathrm{H.c.}
        \right)
        & p\mathrm{-wave}.
    \end{cases}
\end{align}

\begin{figure}[t]
    \centering
    \includegraphics[width=15cm]{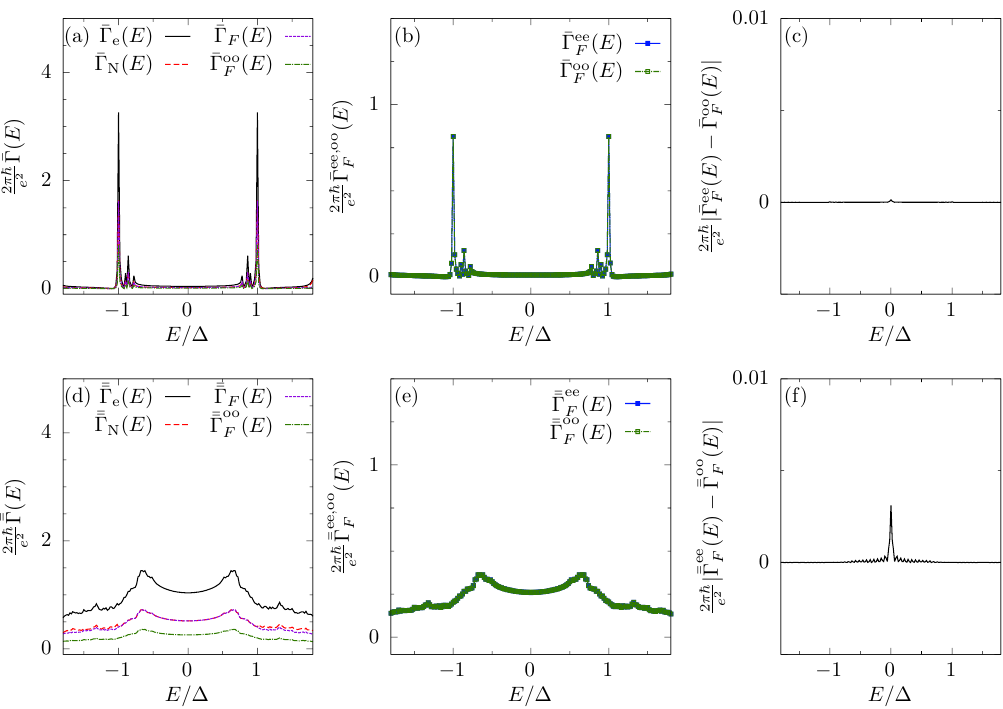}
    \caption{
        The differential conductance and its components for the normal metal/disordered $s$-wave S/$s$-wave S junction.
        (a) $\bar{\Gamma}_\mathrm{e}(E)$ and its components are plotted as a function of $E$.
        (b) $\bar{\Gamma}_F^\mathrm{ee}(E)$ and $\bar{\Gamma}_F^\mathrm{oo}(E)$ are plotted as a function of $E$.
        (c) The difference between $\bar{\Gamma}_F^\mathrm{ee}(E)$ and $\bar{\Gamma}_F^\mathrm{oo}(E)$ is plotted as a function of $E$.
        $\Delta/\breve{t}=0.1$, $\breve{t}_\mathrm{b}/\breve{t}=0.7$, $\mu/\breve{t}=-0.5$, $\eta/\breve{t}=10^{-8}$, $\theta_\mathrm{max}^\mathrm{imp}=\pi/2$, $L_\mathrm{imp}=100$, and the averaging area $L$ defined in Eq.~\eqref{eq:spatial_average_lattice_1d_app} is $L=500$.
        (d)--(f) Averaged results over $N_\mathrm{imp}=100$ impurity configurations.
    }
    \label{fig:app_imp_swave_s}
\end{figure}

In Fig.~\ref{fig:app_imp_swave_s}(a), $\bar{\Gamma}_\mathrm{e}(E)$ and its components are shown for the $s$-wave S junction given by Eq.~\eqref{eq:app_H_diffusive} with $(\theta_\mathrm{max}^\mathrm{imp}/\pi,\mu/\breve{t})=(0.5,-0.5)$. 
In Fig.~\ref{fig:app_imp_swave_s}(b), $\bar{\Gamma}_F^\mathrm{ee}(E)$ and $\bar{\Gamma}_F^\mathrm{oo}(E)$ are plotted.
We can confirm that $\bar{\Gamma}_F^\mathrm{ee}(E)=\bar{\Gamma}_F^\mathrm{oo}(E)$ holds within numerical accuracy.
Explicitly, we show the difference between $\bar{\Gamma}_F^\mathrm{ee}(E)$ and $\bar{\Gamma}_F^\mathrm{oo}(E)$ in Fig.~\ref{fig:app_imp_swave_s}(c).
The difference is small when $|E|/\Delta$ is not small. When $|E|$ is close to zero, the difference becomes larger, which we attribute to a finite size effect.
This behavior is the same as Fig.~\ref{fig:size_dep_s_1d}.
In Figs.~\ref{fig:app_imp_swave_s}(d--f), we show impurity configuration averaged values defined by
\begin{align}
    \bar{\bar{\Gamma}}_\mathrm{e}(E)
    =&
    \frac{1}{N_\mathrm{imp}}\sum_{l=1}^{N_\mathrm{imp}}\bar{\Gamma}_\mathrm{e}(E).
\end{align}
Here, we imply the impurity configuration average by the double over bar.
The components of $\bar{\bar{\Gamma}}_\mathrm{e}(E)$ are defined in the same manner.
The difference between $\bar{\bar{\Gamma}}_F^\mathrm{ee}(E)$ and $\bar{\bar{\Gamma}}_F^\mathrm{oo}(E)$ shown in Fig.~\ref{fig:app_imp_swave_s}(f) is larger than in Fig.~\ref{fig:app_imp_swave_s}(c) but is still small.
 
\begin{figure}[t]
    \centering
    \includegraphics[width=15cm]{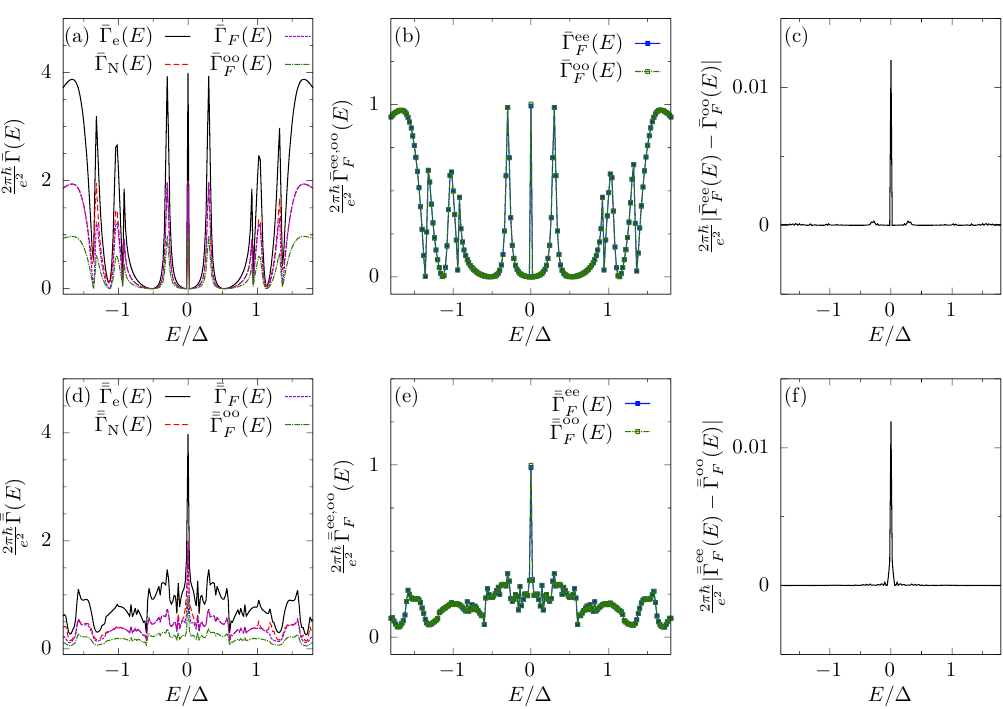}
    \caption{
        The differential conductance and its components for the normal metal/disordered $p$-wave S/$p$-wave S junction.
        (a) $\bar{\Gamma}_\mathrm{e}(E)$ and its components are plotted as a function of $E$.
        (b) $\bar{\Gamma}_F^\mathrm{ee}(E)$ and $\bar{\Gamma}_F^\mathrm{oo}(E)$ are plotted as a function of $E$.
        (c) The difference between $\bar{\Gamma}_F^\mathrm{ee}(E)$ and $\bar{\Gamma}_F^\mathrm{oo}(E)$ is plotted as a function of $E$.
        $\Delta/\breve{t}=0.1$, $\breve{t}_\mathrm{b}/\breve{t}=0.7$, $\mu/\breve{t}=-0.5$, $\eta/\breve{t}=10^{-8}$, $\theta_\mathrm{max}^\mathrm{imp}=\pi/2$, $L_\mathrm{imp}=100$, and the averaging area $L$ defined in Eq.~\eqref{eq:spatial_average_lattice_1d_app} is $L=500$.
        (d)--(f) Averaged results over $N_\mathrm{imp}=100$ impurity configurations.
    }
    \label{fig:app_imp_swave_p}
\end{figure}

$\bar{\Gamma}_\mathrm{e}(E)$ and its components for the normal metal/disordered S/S junction for the $p$-wave S is shown in Figs.~\ref{fig:app_imp_swave_p}(a) and (b) in the topologically nontrivial phase $(\theta_\mathrm{max}^\mathrm{imp}/\pi,\mu/\breve{t})=(0.5,-0.5)$.
The differential conductance at $E=0$ is quantized to $4\frac{e^2}{2\pi\hbar}$ since the disordered $p$-wave S has the same symmetry as the clean $p$-wave S, which is classified in class D in topological classification.
Figure~\ref{fig:app_imp_swave_p}(c) shows the difference between $\bar{\Gamma}_F^\mathrm{ee}(E)$ and $\bar{\Gamma}_F^\mathrm{oo}(E)$. 
Evidently, $\bar{\Gamma}_F^\mathrm{ee}(E)=\bar{\Gamma}_F^\mathrm{oo}(E)$ holds within numerical accuracy.
The qualitative behavior of $|\bar{\Gamma}_F^\mathrm{ee}(E)-\bar{\Gamma}_F^\mathrm{oo}(E)|$ is the same as for the $s$-wave junction i.e., the difference between $\bar{\Gamma}_F^\mathrm{ee}(E)$ and $\bar{\Gamma}_F^\mathrm{oo}(E)$ is small when $|E|/\Delta$ is large, and that is large when $E$ is close to zero.

In addition to the normal metal/disordered S/S-junction, we consider the normal metal/disordered metal/S-junction [Fig.~\ref{fig:app_schematic_junction_diffusive}(b)].
The Hamiltonian is given by
\begin{align}
    H
    =&
    H_\mathrm{kin}+H_\mathrm{onsite}+H_\Delta,
    \\
    H_\mathrm{kin}
    =&
    -
    \breve{t}\sum_{j\neq-1,\sigma}
    \left(
        c^\dagger_{j,\sigma}c_{j+1,\sigma}+\mathrm{H.c.}
    \right)
    -
    \breve{t}_\mathrm{b}\sum_{\sigma}
    \left(
        c_{-1,\sigma}^\dagger c_{0,\sigma} + \mathrm{H.c.}
    \right),
    \\
    H_\mathrm{onsite}
    =&
    -
    \mu
    \sum_{j,\sigma} c_{j,\sigma}^\dagger c_{j,\sigma}
    +
    \sum_{-L_\mathrm{imp}\le j< 0,\sigma}V_j c_{j,\sigma}^\dagger c_{j,\sigma},
    \\
    H_\Delta
    =&
    \begin{cases}
        \Delta
        \sum_{0\le j}
        \left(
            c_{j,\uparrow}^\dagger c_{j,\downarrow}^\dagger + \mathrm{H.c.}
        \right)
        &
        s\mathrm{-wave},
        \\
        \Delta
        \sum_{0\le j}
        \left(
            c_{j,\uparrow}^\dagger 
            c_{j+1,\downarrow}^\dagger 
            +
            \mathrm{H.c.}
        \right)
        & p\mathrm{-wave}.
    \end{cases}
\end{align}
with the random potential $V_j$ in the disordered metal region with $|V_j|/\breve{t}<0.1$.

\begin{figure}[htbp]
    \centering
    \includegraphics[width=15cm]{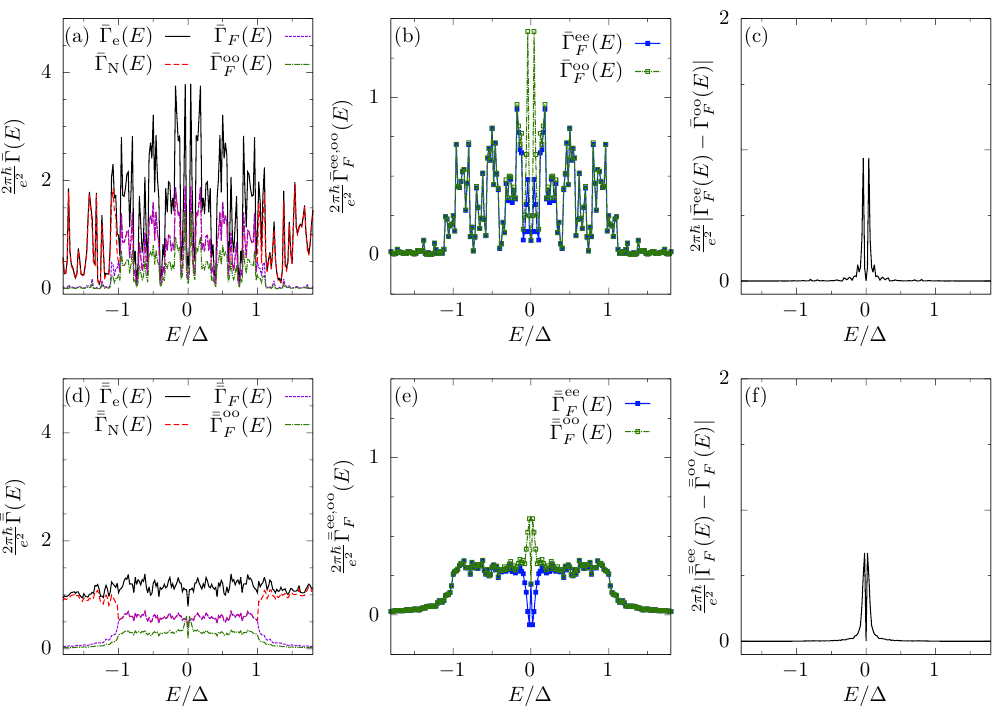}
    \caption{
        The differential conductance and its components for the normal metal/disordered metal/$s$-wave S junction.
        (a) $\bar{\Gamma}_\mathrm{e}(E)$ and its components are plotted as a function of $E$.
        (b) $\bar{\Gamma}_F^\mathrm{ee}(E)$ and $\bar{\Gamma}_F^\mathrm{oo}(E)$ are plotted as a function of $E$.
        (c) The difference between $\bar{\Gamma}_F^\mathrm{ee}(E)$ and $\bar{\Gamma}_F^\mathrm{oo}(E)$ is plotted as a function of $E$.
        $\Delta/\breve{t}=0.1$, $\breve{t}_\mathrm{b}/\breve{t}=1$, $\mu/\breve{t}=-0.5$, $\eta/\breve{t}=10^{-8}$, and $L_\mathrm{imp}=10^3$, and the averaging area $L$ defined in Eq.~\eqref{eq:spatial_average_lattice_1d_app} is $L=900$.
        (d)--(f) Averaged results over $N_\mathrm{imp}=100$ impurity configurations.
    }
    \label{fig:app_n_diffusive_s_ave}
\end{figure}

\begin{figure}[htbp]
    \centering
    \includegraphics[width=15cm]{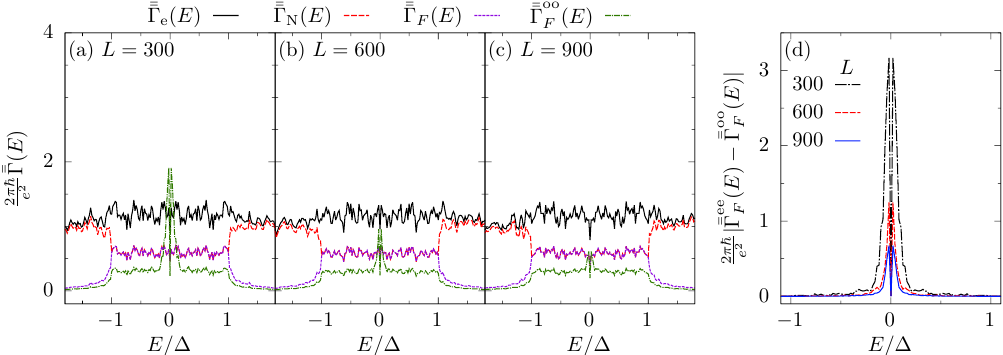}
    \caption{
        The impurity configuration averaged differential conductance and its components for the normal metal/disordered metal/$s$-wave S junction with several values of $L$.
        (a--c) $\bar{\bar{\Gamma}}_\mathrm{e}(E)$ and its components are plotted as a function of $E$ for (a) $L=300$, (b) $L=600$, and (c) $L=900$.
        (d) The difference between $\bar{\bar{\Gamma}}_F^\mathrm{ee}(E)$ and $\bar{\bar{\Gamma}}_F^\mathrm{oo}(E)$ for $L=300$, $600$, and $900$.
        $\Delta/\breve{t}=0.1$, $\breve{t}_\mathrm{b}/\breve{t}=1$, $\mu/\breve{t}=-0.5$, $\eta/\breve{t}=10^{-8}$, $L_\mathrm{imp}=10^3$, and $N_\mathrm{imp}=100$.
    }
    \label{fig:app_n_diffusive_s_ave_size}
\end{figure}

Figures~\ref{fig:app_n_diffusive_s_ave}(a) and (b) show the differential conductance and its components for the $s$-wave S junction.
The difference between $\bar{\Gamma}_F^\mathrm{ee}(E)$ and $\bar{\Gamma}_F^\mathrm{oo}(E)$ is larger than in Fig.~\ref{fig:app_imp_swave_s}(c).
The impurity configuration averaged values are plotted in Figs.~\ref{fig:app_n_diffusive_s_ave}(d--f).
The order of the magnitude the difference between $\bar{\bar{\Gamma}}_F^\mathrm{ee}(E)$ and $\bar{\bar{\Gamma}}_F^\mathrm{oo}(E)$ [Fig.~\ref{fig:app_n_diffusive_s_ave}(f)] is the same as the unaveraged one [Fig.~\ref{fig:app_n_diffusive_s_ave}(c)].
To confirm that the difference reduces with increasing system size,
we calculate $\bar{\bar{\Gamma}}_\mathrm{e}(E)$, its components, and $|\bar{\bar{\Gamma}}_F^\mathrm{ee}(E)-\bar{\bar{\Gamma}}_F^\mathrm{oo}(E)|$ for several values of $L$ as shown in Fig.~\ref{fig:app_n_diffusive_s_ave_size}.
From Fig.~\ref{fig:app_n_diffusive_s_ave_size}(d), we can see that the difference decreases as $L$ increases.
We expect that the difference approaches zero as $L\rightarrow\infty$.

\begin{figure}[htbp]
    \centering
    \includegraphics[width=15cm]{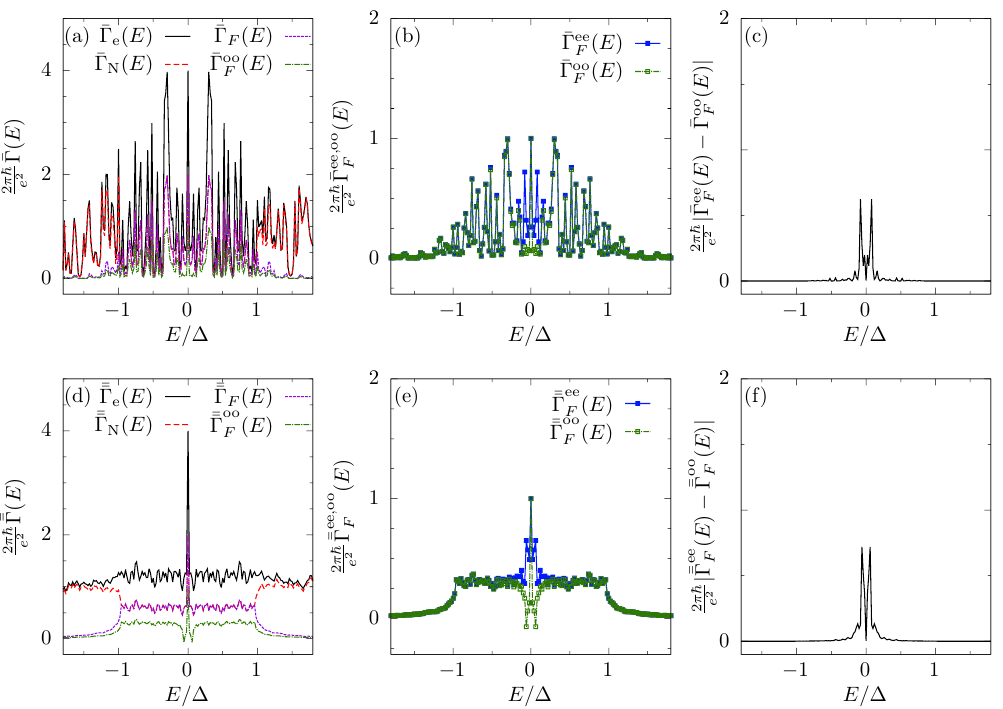}
    \caption{
        The differential conductance and its components for the normal metal/disordered metal/$p$-wave S junction.
        (a) $\bar{\Gamma}_\mathrm{e}(E)$ and its components are plotted as a function of $E$.
        (b) $\bar{\Gamma}_F^\mathrm{ee}(E)$ and $\bar{\Gamma}_F^\mathrm{oo}(E)$ are plotted as a function of $E$.
        (c) The difference between $\bar{\Gamma}_F^\mathrm{ee}(E)$ and $\bar{\Gamma}_F^\mathrm{oo}(E)$ is plotted as a function of $E$.
        $\Delta/\breve{t}=0.1$, $\breve{t}_\mathrm{b}/\breve{t}=1$, $\mu/\breve{t}=-0.5$, $\eta/\breve{t}=10^{-8}$, and $L_\mathrm{imp}=10^3$, and the averaging area $L$ defined in Eq.~\eqref{eq:spatial_average_lattice_1d_app} is $L=900$.
        (d)--(f) Averaged results over $N_\mathrm{imp}=100$ impurity configurations.
    }
    \label{fig:app_n_diffusive_p_ave}
\end{figure}

\begin{figure}[htbp]
    \centering
    \includegraphics[width=15cm]{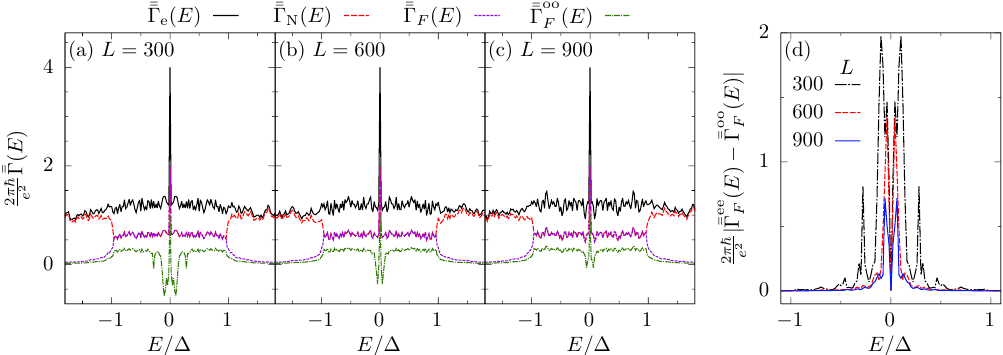}
    \caption{
        The impurity configuration averaged differential conductance and its components for the normal metal/disordered metal/$p$-wave S junction with several values of $L$.
        (a--c) $\bar{\bar{\Gamma}}_\mathrm{e}(E)$ and its components are plotted as a function of $E$ for (a) $L=300$, (b) $L=600$, and (c) $L=900$.
        (d) The difference between $\bar{\bar{\Gamma}}_F^\mathrm{ee}(E)$ and $\bar{\bar{\Gamma}}_F^\mathrm{oo}(E)$ for $L=300$, $600$, and $900$.
        $\Delta/\breve{t}=0.1$, $\breve{t}_\mathrm{b}/\breve{t}=1$, $\mu/\breve{t}=-0.5$, $\eta/\breve{t}=10^{-8}$, $L_\mathrm{imp}=10^3$, and $N_\mathrm{imp}=100$.
    }
    \label{fig:app_n_diffusive_p_ave_size}
\end{figure}

Figures~\ref{fig:app_n_diffusive_p_ave}(a) and (b) show $\bar{\Gamma}_\mathrm{e}(E)$ and its components for the normal metal/disordered metal/$p$-wave S junction.
The differential conductance at $E=0$ exhibits a quantized value, $\bar{\Gamma}_\mathrm{e}(E=0)=4\frac{e^2}{2\pi\hbar}$, since the chemical potential is in the topologically nontrivial phase ($\mu/\breve{t}=-0.5$).
As for the $s$-wave S junction, $\bar{\Gamma}_F^\mathrm{ee}(E)$ and $\bar{\Gamma}_F^\mathrm{oo}(E)$ are almost the same when $|E|$ is not too small. 
As $|E|$ becomes smaller, the difference becomes larger [Figs.~\ref{fig:app_n_diffusive_p_ave}(b) and (c)].
We show the disorder configuration averaged values in Figs.~\ref{fig:app_n_diffusive_p_ave}(d--f).
The order of the magnitude of the difference between $\bar{\bar{\Gamma}}_F^\mathrm{ee}(E)$ and $\bar{\bar{\Gamma}}_F^\mathrm{oo}(E)$ [Fig.~\ref{fig:app_n_diffusive_p_ave}(f)] is the same as the unaveraged one [Fig.~\ref{fig:app_n_diffusive_p_ave}(c)].
As shown in Fig.~\ref{fig:app_n_diffusive_p_ave_size}, the difference becomes smaller as $L$ increases.
We expect that the difference between $\bar{\bar{\Gamma}}_F^\mathrm{ee}(E)$ and $\bar{\bar{\Gamma}}_F^\mathrm{oo}(E)$ approaches zero as $L\rightarrow\infty$.

\subsection{\label{sub:1d_2d_lattilce_model}Hamiltonian for 1D N/2D S lattice model}
\begin{figure}[t]
    \centering
    \includegraphics[width=8.5cm]{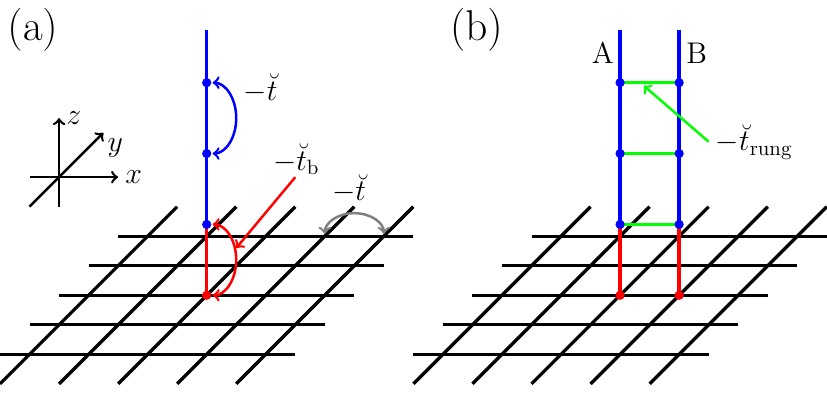}
    \caption{
    Schematic picture of 1D N/2D S junctions.
    (a) 1D N/2D S junction and (b) 1D N ladder/2D S junction.
    The 1D N is connected to $\mathbf{j}_0=(0,0)$ of the 2D S, and the 1D N ladder is connected to $\mathbf{j}_0$ and $\mathbf{j}_0+\hat{\mathrm{e}}_x$.
    }
    \label{fig:schematic_2d_1d_junction}
\end{figure}

The Hamiltonian for the 1D N (ladder)/2D S junction based on lattice models (Fig.~\ref{fig:schematic_2d_1d_junction}) is given by
\begin{align}
    H
    =&
    H_{\mathrm{1D}}^{\mathrm{(ladder)}}
    +
    H_{\mathrm{2D}}
    +
    H_{\mathrm{1D-2D}}^{\mathrm{(ladder)}},
    \\
    H_{\mathrm{1D}}
    =&
    -\breve{t}\sum_{j>0,\sigma}\left(c_{j,\sigma}^\dagger c_{j+1,\sigma}+\mathrm{H.c.}\right)
    -\mu_{\mathrm{N}}\sum_{j>0}n_{j},
    \\
    H_{\mathrm{1D}}^{\mathrm{ladder}}
    =&
    -\breve{t}\sum_{j>0,\nu,\sigma}\left(\bar{c}_{j,\nu,\sigma}^\dagger \bar{c}_{j+1,\nu,\sigma}+\mathrm{H.c.}\right)
    -\breve{t}_\mathrm{rung}\sum_{j>0,\sigma}\left(\bar{c}_{j,\mathrm{A},\sigma}^\dagger \bar{c}_{j,\mathrm{B},\sigma}+\mathrm{H.c.}\right)
    -\mu_{\mathrm{N}}\sum_{j>0}\bar{n}_{j},
    \\
    H_{\mathrm{2D}}
    =&
    -\breve{t}
    \sum_{\langle \mathbf{i},\mathbf{j}\rangle,\sigma}
    \left(
        b_{\mathbf{i},\sigma}^\dagger b_{\mathbf{j},\sigma}+\mathrm{H.c.}
    \right)
    -\mu_{\mathrm{S}}\sum_{\mathbf{j},\sigma}b_{\mathbf{j},\sigma}^\dagger b_{\mathbf{j},\sigma}
    +
    H_\Delta,
    \\
    H_\mathrm{1D-2D}
    =&
    -\breve{t}_\mathrm{b}\sum_{\sigma}
    \left(
        c_{1,\sigma}^\dagger b_{\mathbf{j}_{0},\sigma}+\mathrm{H.c.}
    \right),
    \label{eq:H_connect}
    \\
    H_\mathrm{1D-2D}^\mathrm{ladder}
    =&
    -\breve{t}_\mathrm{b}\sum_{\sigma}
    \left(
        \bar{c}_{1,\mathrm{A},\sigma}^\dagger b_{\mathbf{j}_{0},\sigma}
        +
        \bar{c}_{1,\mathrm{B},\sigma}^\dagger b_{\mathbf{j}_{0}+\hat{\mathrm{e}}_x,\sigma}
        +
        \mathrm{H.c.}
    \right)
    \label{eq:H_connect_ladder}
\end{align}
with $n_{j}=\sum_{\sigma}c_{j,\sigma}^\dagger c_{j,\sigma}$, $\bar{n}_{{j}}=\sum_{\sigma,\nu=\mathrm{A,B}}\bar{c}_{{j},\nu,\sigma}^\dagger \bar{c}_{{j},\nu,\sigma}$ and $\mathbf{j}_0$ in Eq.~\eqref{eq:H_connect} is $\mathbf{j}_0=(0,0)$ in 2D S\@.
Here, $c_{j,\sigma}$ ($c_{j,\sigma}^\dagger$) is an annihilation (creation) operator in 1D N with site $j$ and spin $\sigma$, $b_{\mathbf{j},\sigma}$ ($b_{\mathbf{j},\sigma}^\dagger$) is an annihilation (creation) operator in 2D S, and $\bar{c}_{\mathbf{j},\nu,\sigma}$ ($\bar{c}_{\mathbf{j},\nu,\sigma}^\dagger$) is an annihilation (creation) operator in 1D N ladder with the rung degrees of freedom $\nu=\mathrm{A,B}$.
$\breve{t}$ is a hopping integral in 1D N and 2D S, $\breve{t}_\mathrm{b}$ is a hopping integral between 1D N and 2D S, and $\mu_\mathrm{N}$ ($\mu_\mathrm{S}$) is a chemical potential in 1D N and 1D N ladder (2D S).
For simplicity we choose $\breve{t}_\mathrm{rung}=\breve{t}$.
We consider $s$-, $p_x(p_y)$-, and $d$-wave pair potentials, where $H_\Delta$ is given by $H_{\Delta}^s$, $H_{\Delta}^{p_x(p_y)}$, and $H_{\Delta}^d$, respectively:
\begin{align}
    H_{\Delta}^s
    =&
    \Delta
    \sum_{\mathbf{j}}
    \left(
        \tilde{c}_{\mathbf{j},\uparrow}^\dagger\tilde{c}_{\mathbf{j},\downarrow}^\dagger+\mathrm{H.c.}
    \right),
    \\
    H_{\Delta}^{p_{x}(p_y)}
    =&
    \frac{\Delta}{2}
    \sum_{\mathbf{j}}
    \left(
        \tilde{c}_{\mathbf{j},\uparrow}^\dagger\tilde{c}_{\mathbf{j}+\hat{\mathrm{e}}_{x(y)},\downarrow}^\dagger
        +
        \tilde{c}_{\mathbf{j},\downarrow}^\dagger\tilde{c}_{\mathbf{j}+\hat{\mathrm{e}}_{x(y)},\uparrow}^\dagger
        +
        \mathrm{H.c.}
    \right),
    \\
    H_{\Delta}^d
    =&
    \frac{\Delta}{4}
    \sum_{\mathbf{j}}
    \left(
        \tilde{c}_{\mathbf{j},\uparrow}^\dagger\tilde{c}_{\mathbf{j}+\hat{\mathrm{e}}_x,\downarrow}^\dagger
        -
        \tilde{c}_{\mathbf{j},\downarrow}^\dagger\tilde{c}_{\mathbf{j}+\hat{\mathrm{e}}_x,\uparrow}^\dagger
        -
        \tilde{c}_{\mathbf{j},\uparrow}^\dagger\tilde{c}_{\mathbf{j}+\hat{\mathrm{e}}_y,\downarrow}^\dagger
        +
        \tilde{c}_{\mathbf{j},\downarrow}^\dagger\tilde{c}_{\mathbf{j}+\hat{\mathrm{e}}_y,\uparrow}^\dagger
        +
        \mathrm{H.c.}
    \right),
\end{align}
with $\hat{\mathrm{e}}_x=(1,0)$ and $\hat{\mathrm{e}}_y=(0,1)$.
To calculate the Green function, we utilize the recursive Green function method~\cite{PhysRevB.55.5266}.
For 2D S, we impose periodic boundary conditions in $x$-direction with $L_x$ sites. 
We adopt $\theta=\pi/2$ in the following.

The spatial averaging is taken in 1D N (ladder) as
\begin{align}
    \bar{\Gamma}_\mathrm{e}(E)
    =&
    \frac{1}{L^2}
    \sum_{1\leq i,j\leq L}
    \Gamma_\mathrm{e}(i,j,E).
    \label{eq:spatial_average_lattice_1d_2d_app}
\end{align}

\subsection{\label{sec:F_1d_2d_lattice}Size dependence of $\bar{\Gamma}_\mathrm{e}(E)$ and its components and anomalous Green function in 1D N for 1D N/2D S junction}
\begin{figure}[t]
    \centering
    \includegraphics[width=17cm]{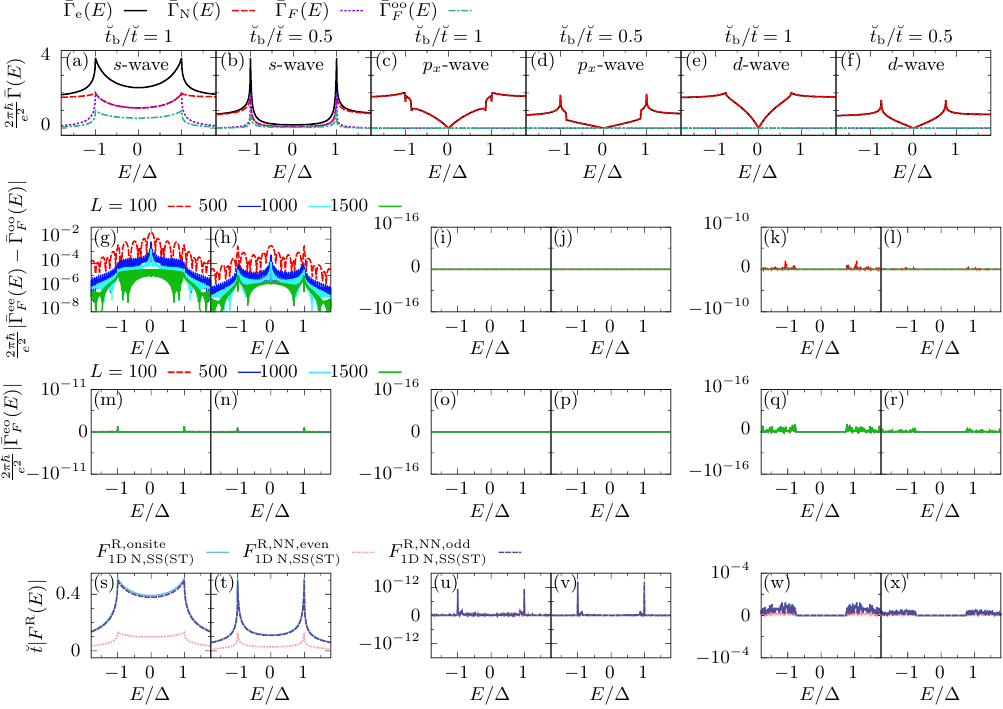}
    \caption{
    (a)-(f) $\bar{\Gamma}_\mathrm{e}(E)$ and its components are plotted as a function of $E$ with $L=500$.
    (g)--(l) $|\bar{\Gamma}_F^\mathrm{ee}(E)-\bar{\Gamma}_F^\mathrm{oo}(E)|$ is plotted as a function of $E$ for several values of $L$.
    (m)--(r) $|\bar{\Gamma}_F^\mathrm{eo}(E)|$ is plotted as a function of $E$ for several values of $L$.
    (s)--(x) absolute value of anomalous retarded Green function is plotted as a function of $E$.
    (a), (b), (g), (h), (m), (n), (s), and (t) $s$-wave junction.
    (c), (d), (i), (j), (o), (p), (u), and (v) $p_x$-wave junction.
    (e), (f), (k), (l), (q), (r), (w), and (x) $d$-wave junction.
    (a), (c), (e), (g), (i), (k), (m), (o), (q), (s), (u), and (w) $\breve{t}_\mathrm{b}/\breve{t}=1$ and
    (b), (d), (f), (h), (j), (l), (n), (p), (r), (t), (v), and (x) $\breve{t}_\mathrm{b}/\breve{t}=0.5$.
    $L_x=10^7$, $\Delta/\breve{t}=0.1$, and $\eta/\breve{t}=10^{-7}$.
    }
    \label{fig:size_1d_2d}
\end{figure}

\begin{figure}[t]
    \centering
    \includegraphics[width=3.5cm]{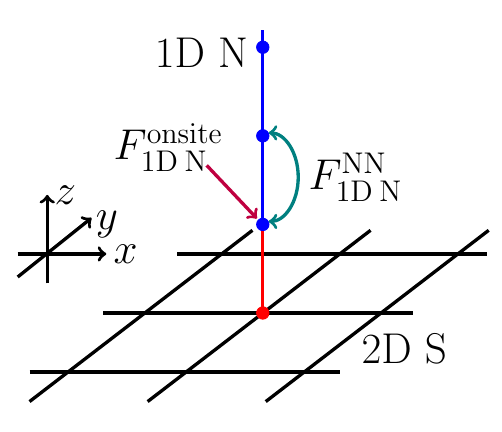}
    \caption{Schematic picture of $F_\mathrm{1D\:N}^{\mathrm{onsite}}$ and $F_\mathrm{1D\:N}^{\mathrm{NN}}$.
    }
    \label{fig:schematic_2d_1d_F}
\end{figure}

In Figs.~\ref{fig:size_1d_2d}(a)--(f), we show $\bar{\Gamma}_\mathrm{e}(E)$ and its components for $s$-, $p_x$- and $d$-wave junctions.
Figures~\ref{fig:size_1d_2d}(a), (c), and (e) are also shown in the main text.
For the $s$-wave junction, we show $\bar{\Gamma}_\mathrm{e}(E)$ and its components with $\breve{t}_\mathrm{b}/\breve{t}=0.5$ in Fig.~\ref{fig:size_1d_2d}(b).
Similar to continuum and lattice 1D N/1D S $s$-wave SC junctions, as $\breve{t}_\mathrm{b}/\breve{t}$ decreases, the shape of $\bar{\Gamma}_\mathrm{e}(E)$ approaches the U-shaped density of states.
For $p_x$-wave and $d$-wave junctions, independent of the value of $\breve{t}_\mathrm{b}$, Andreev reflection is zero: $\bar{\Gamma}_F(E)=0$.
The difference between $\bar{\Gamma}_F^\mathrm{ee}(E)$ and $\bar{\Gamma}_F^\mathrm{oo}(E)$ are plotted in Figs.~\ref{fig:size_1d_2d}(g)--(l).
This difference for $s$-wave junctions [Figs.~\ref{fig:size_1d_2d}(g) and (h)] becomes smaller for increasing $L$.
For $p_x$- and $d$-wave junctions, $\bar{\Gamma}_F^\mathrm{ee}(E)-\bar{\Gamma}_F^\mathrm{oo}(E)$ is zero within numerical errors [Figs.~\ref{fig:size_1d_2d}(i)--(l)].
$\bar{\Gamma}_F^\mathrm{eo}(E)$ is plotted in Figs.~\ref{fig:size_1d_2d}(m)--(r).
It is zero for all cases.
The penetrated anomalous even and odd-frequency pairings are shown in Figs.~\ref{fig:size_1d_2d}(s)--(x) (see also Fig.~\ref{fig:schematic_2d_1d_F}).
The onsite and nearest neighbor (NN) pairings in 1D N are defined as
\begin{align}
    F_\mathrm{1D\:N,SS(ST)}^\mathrm{onsite,even(odd)}(i\omega_n)
    =&
    \frac{1}{4}\left\{
    \left[F_{1,1,\uparrow,\downarrow}(i\omega_n)+\zeta F_{1,1,\downarrow,\uparrow}(i\omega_n)\right]
    \right.
    +
    \xi
    \left.
    \left[F_{1,1,\uparrow,\downarrow}(-i\omega_n)+\zeta F_{1,1,\downarrow,\uparrow}(-i\omega_n)\right]
    \right\},
    \label{eq:def_F_2d_1d_onsite_1d}
    \\
    F_\mathrm{1D\:N,SS(ST)}^\mathrm{NN,even(odd)}(i\omega_n)
    =&
    \frac{1}{4}\left\{
    \left[F_{1,2,\uparrow,\downarrow}(i\omega_n)+\zeta F_{1,2,\downarrow,\uparrow}(i\omega_n)\right]
    \right.
    +
    \xi
    \left.
    \left[F_{1,2,\uparrow,\downarrow}(-i\omega_n)+\zeta F_{1,2,\downarrow,\uparrow}(-i\omega_n)\right]
    \right\},
    \label{eq:def_F_2d_1d_nn_1d}
\end{align}
with $\zeta=-1$ for the SS case, $\zeta=1$ for the ST case, $\xi=1(-1)$ for even (odd) frequency pairing.
By analytic continuation, $i\omega_n\rightarrow E+i\eta$, we obtain the retarded anomalous Green function shown in Figs.~\ref{fig:size_1d_2d}(s)--(x).
As discussed in the main text, even and odd-frequency pairings only penetrate into 1D N for $s$-wave junctions.
For $p_x$- and $d$-wave junctions, even and odd-frequency pairings do not penetrate into 1D N within numerical errors.

\subsection{\label{sec:F_1d_2d_lattice}NN anomalous Green function in 2D S for 1D N/2D S junction on lattice}
\begin{figure}[t]
    \centering
    \includegraphics[width=10.0cm]{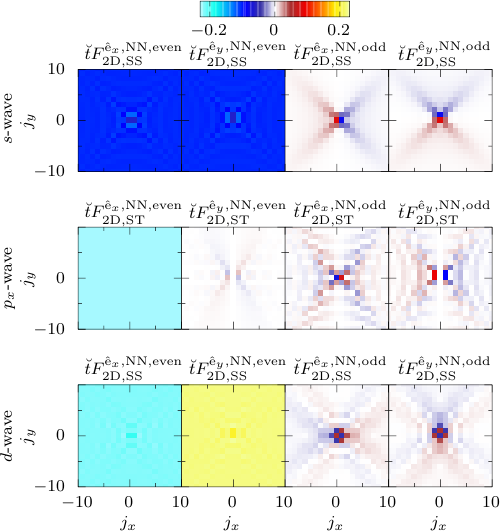}
    \caption{
        NN components of the anomalous Green function in 2D S are plotted as functions of $j_x$ and $j_y$ at $\omega_n/\Delta=0.1$ for the 1D N/2D S junction.
        Here, $\mu_\mathrm{N}/\breve{t}=-0.5$, $\mu_\mathrm{S}/\breve{t}=-1$, $\breve{t}_\mathrm{b}/\breve{t}=1$, $\Delta/\breve{t}=0.1$, and $\mathbf{j}_0=(0,0)$ is connected to the 1D N side.
        For $s$-wave and $d$-wave junctions, the spin-singlet component is plotted.
        For $p_x$-wave S junctions, the spin-triplet component is plotted.
        For $s$-wave and $d$-wave junctions with even-frequency components, the real part is plotted.
        For the corresponding odd-frequency components, the imaginary part is plotted.
        For $p_x$-wave junctions with even-frequency components, the imaginary part is plotted.
        For the corresponding odd-frequency components, the real part is plotted.
        The counterparts are numerically zero.
    }
    \label{fig:F2d_1d_2d_nn}
\end{figure}

We calculate the nearest neighbor (NN) component of the anomalous Green function in 2D S shown in Fig.~\ref{fig:schematic_2d_1d_F}.
It can be written as

\begin{align}
    F_\mathrm{2D,SS(ST)}^{\hat{\mathrm{e}}_{x(y)},\mathrm{NN,even(odd)}}(\mathbf{j},i\omega_n)
    =&
    \frac{1}{4}
    \left\{
       \left[
          F_{\mathbf{j}-\hat{\mathrm{e}}_{x(y)},\mathbf{j},\uparrow,\downarrow}(i\omega_n)
          +
          \zeta F_{\mathbf{j}-\hat{\mathrm{e}}_{x(y)},\mathbf{j},\downarrow,\uparrow}(i\omega_n)
          \right]
       \right.
       \nonumber\\
       &
       \left.
       +
       \xi
       \left[
          F_{\mathbf{j}-\hat{\mathrm{e}}_{x(y)},\mathbf{j},\uparrow,\downarrow}(-i\omega_n)
          +
          \zeta F_{\mathbf{j}-\hat{\mathrm{e}}_{x(y)},\mathbf{j},\downarrow,\uparrow}(-i\omega_n)
          \right]
       \right\}
    \label{eq:def_F_2d_1d_nn_2d}
\end{align}
with $\zeta=-1$ for the SS case, $\zeta=1$ for the ST case, $\xi=1(-1)$ for the even (odd) frequency pairing.
The NN components of the anomalous Green function in 2D S are shown in Fig.~\ref{fig:F2d_1d_2d_nn}.
Let us define $F^\mathrm{NN,even(odd)}_{s,d}$ for $s$ and $d$-wave junctions as
\begin{align}
    F^\mathrm{NN,even(odd)}_{\mathrm{SS}}
    =&
    F_\mathrm{2D,SS}^{\hat{\mathrm{e}}_{x},\mathrm{NN,even(odd)}}(\mathbf{j}_0,i\omega_n)
    +(-)
    F_\mathrm{2D,SS}^{\hat{\mathrm{e}}_{x},\mathrm{NN,even(odd)}}(\mathbf{j}_0+\hat{\mathrm{e}}_x,i\omega_n)
    \nonumber\\
    &
    +
    F_\mathrm{2D,SS}^{\hat{\mathrm{e}}_{y},\mathrm{NN,even(odd)}}(\mathbf{j}_0,i\omega_n)
    +(-)
    F_\mathrm{2D,SS}^{\hat{\mathrm{e}}_{y},\mathrm{NN,even(odd)}}(\mathbf{j}_0+\hat{\mathrm{e}}_y,i\omega_n),
    \label{eq:criteria_s_d_app}
\end{align}
and $F^\mathrm{NN,even(odd)}_{p_x}$ for the $p_x$-wave junction as
\begin{align}
    F^\mathrm{NN,even(odd)}_{\mathrm{ST}}
    =&
    F_\mathrm{2D,ST}^{\hat{\mathrm{e}}_{x},\mathrm{NN,even(odd)}}(\mathbf{j}_0,i\omega_n)
    -(+)
    F_\mathrm{2D,ST}^{\hat{\mathrm{e}}_{x},\mathrm{NN,even(odd)}}(\mathbf{j}_0+\hat{\mathrm{e}}_x,i\omega_n)
    \nonumber\\
    &
    +
    F_\mathrm{2D,ST}^{\hat{\mathrm{e}}_{y},\mathrm{NN,even(odd)}}(\mathbf{j}_0,i\omega_n)
    -(+)
    F_\mathrm{2D,ST}^{\hat{\mathrm{e}}_{y},\mathrm{NN,even(odd)}}(\mathbf{j}_0+\hat{\mathrm{e}}_y,i\omega_n).
    \label{eq:criteria_p_app}
\end{align}
If $F^\mathrm{NN,even(odd)}_{s,p_x,d}$ is not zero, then, the NN component of the even (odd) frequency pairing can penetrate into 1D N\@.

\begin{figure}[t]
    \centering
    \includegraphics[width=12.0cm]{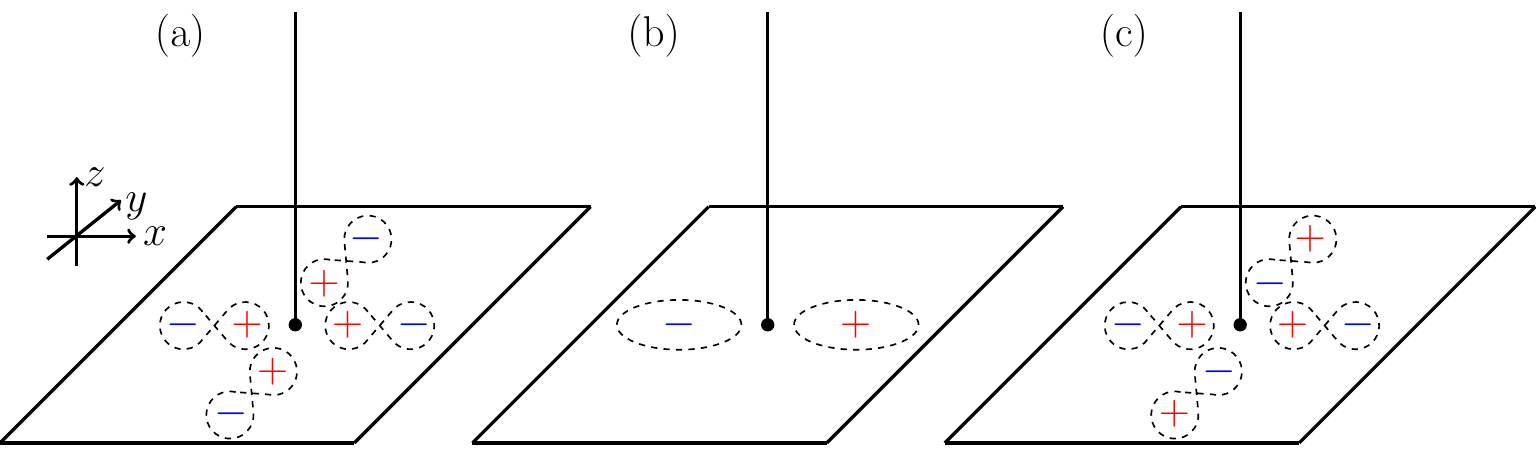}
    \caption{
    Schematic illustration of (a) $F_{\mathrm{2D,SS}}^{\hat{\mathrm{e}}_{x(y)},\mathrm{NN,odd}}(\mathbf{j},i\omega_n)$ for $s$-wave junction,
    (b) $F_{\mathrm{2D,ST}}^{\hat{\mathrm{e}}_{x(y)},\mathrm{NN,odd}}(\mathbf{j},i\omega_n)$ for $p_x$-wave junction, and
    (c) $F_{\mathrm{2D,SS}}^{\hat{\mathrm{e}}_{x(y)},\mathrm{NN,odd}}(\mathbf{j},i\omega_n)$ for $d$-wave junction.
    }
    \label{fig:F_2dS_schematic_app}
\end{figure}

For $s$-wave junctions, $F_\mathrm{2D,SS}^{\hat{\mathrm{e}}_{x},\mathrm{NN,even}}(\mathbf{j},i\omega_n)$ and $F_\mathrm{2D,SS}^{\hat{\mathrm{e}}_{y},\mathrm{NN,even}}(\mathbf{j},i\omega_n)$ have $s$-wave spatial structure.
Hence, they have the same sign for any $\mathbf{j}$ since these pairings exist in bulk.
Then, $F^\mathrm{NN,even}_{\mathrm{SS}}\neq0$ holds, and NN even-frequency pairing penetrates into 1D N\@.
$F_\mathrm{2D,SS}^{\hat{\mathrm{e}}_{x(y)},\mathrm{NN,odd}}(\mathbf{j},i\omega_n)$ has $p_{x(y)}$-wave spatial structure.
The sign of $F_\mathrm{2D,SS}^{\hat{\mathrm{e}}_{x(y)},\mathrm{NN,odd}}(\mathbf{j},i\omega_n)$ is opposite for $j_{x(y)}\leq0$ and $j_{x(y)}\geq1$ since this pairing is induced by 1D N.
Consequently, $F^\mathrm{NN,odd}_{\mathrm{SS}}\neq0$ holds [see Fig.~\ref{fig:F_2dS_schematic_app}(a)].

For $p_x$-wave junctions, $F_\mathrm{2D,ST}^{\hat{\mathrm{e}}_{x},\mathrm{NN,even}}(\mathbf{j},i\omega_n)$ is related to the bulk pair potential. 
Thus, it has a large amplitude.
$F_\mathrm{2D,ST}^{\hat{\mathrm{e}}_{x},\mathrm{NN,even}}(\mathbf{j},i\omega_n)$ has a uniform sign and $F_\mathrm{2D,ST}^{\hat{\mathrm{e}}_{y},\mathrm{NN,even}}(\mathbf{j},i\omega_n)$ is zero at $j_x=0$.
Consequently, $F^\mathrm{NN,even}_{\mathrm{ST}}=0$ holds. 
$F_\mathrm{2D,ST}^{\hat{\mathrm{e}}_{x},\mathrm{NN,odd}}(\mathbf{j},i\omega_n)$ and
$F_\mathrm{2D,ST}^{\hat{\mathrm{e}}_{y},\mathrm{NN,odd}}(\mathbf{j},i\omega_n)$ are induced by 1D N\@. 
In $x$-direction, $F_\mathrm{2D,ST}^{\hat{\mathrm{e}}_{x},\mathrm{NN,odd}}(\mathbf{j},i\omega_n)$ changes its sign as shown in Fig.~\ref{fig:F_2dS_schematic_app}(b) and the NN odd-frequency components cancel each other.
In $y$-direction, $F_\mathrm{2D,ST}^{\hat{\mathrm{e}}_{y},\mathrm{NN,odd}}(\mathbf{j},i\omega_n)$ is zero at $j_x=0$ and cannot penetrate into 1D N\@.
Totally, $F^\mathrm{NN,odd}_{\mathrm{ST}}=0$ holds.

For $d$-wave junctions, $F_\mathrm{2D,SS}^{\hat{\mathrm{e}}_{x},\mathrm{NN,even}}(\mathbf{j},i\omega_n)$ and $F_\mathrm{2D,SS}^{\hat{\mathrm{e}}_{y},\mathrm{NN,even}}(\mathbf{j},i\omega_n)$ have the opposite sign due to $d$-wave symmetry.
The $x$-direction and $y$-direction components cancel each other and $F^\mathrm{NN,even}_{\mathrm{SS}}=0$ holds.
$F_\mathrm{2D,SS}^{\hat{\mathrm{e}}_{x},\mathrm{NN,odd}}(\mathbf{j},i\omega_n)$ and
$F_\mathrm{2D,SS}^{\hat{\mathrm{e}}_{y},\mathrm{NN,odd}}(\mathbf{j},i\omega_n)$ are induced by 1D N\@.
As illustrated in Fig.~\ref{fig:F_2dS_schematic_app}(c), $x$- and $y$-directional components cancel each other, and $F^\mathrm{NN,odd}_{\mathrm{SS}}=0$ holds.


\subsection{\label{sec:tb_dep_Gamma_2d_1d}$\breve{t}_\mathrm{b}$ dependence of $\bar{\Gamma}_\mathrm{e}(E=0)$}
\begin{figure}[t]
    \centering
    \includegraphics[width=17cm]{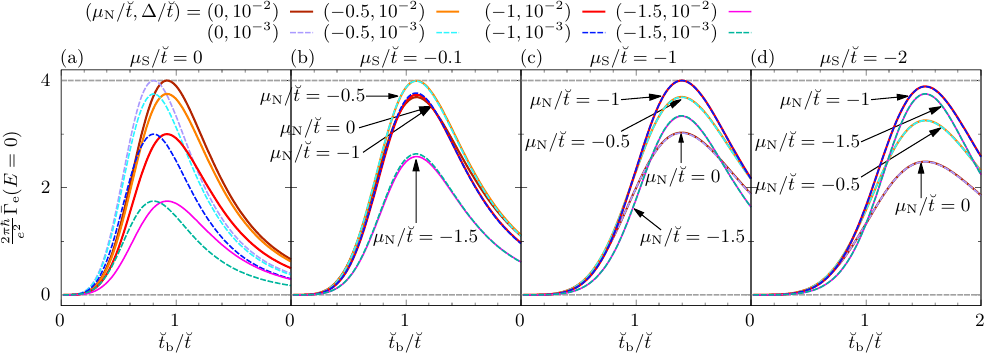}
    \caption{
        $\bar{\Gamma}_\mathrm{e}(E=0)$ is plotted as a function of $\breve{t}_\mathrm{b}$ for the 1D N/2D $s$-wave S junction with $L_x=2\times10^6$, and $\eta/\breve{t}=10^{-7}$ for several values of $\mu_\mathrm{N}$ and $\Delta$.
        (a) $\mu_\mathrm{S}/\breve{t}=0$,
        (b) $\mu_\mathrm{S}/\breve{t}=-0.1$,
        (c) $\mu_\mathrm{S}/\breve{t}=-1$, and
        (d) $\mu_\mathrm{S}/\breve{t}=-2$.
    }
    \label{fig:tb_dep}
\end{figure}

We now illustrate the $\breve{t}_\mathrm{b}$ dependence of $\bar{\Gamma}_\mathrm{e}(E=0)$ for 1D N/2D $s$-wave junctions in Fig.~\ref{fig:tb_dep}.
For $\mu_\mathrm{S}=0$ [Fig.~\ref{fig:tb_dep}(a)], the maximum value of $\bar{\Gamma}_\mathrm{e}(E=0)$ is obtained at approximately $\breve{t}_\mathrm{b}/\breve{t}=0.9$ for $\Delta/\breve{t}=10^{-2}$ and $\breve{t}_\mathrm{b}/\breve{t}=0.8$ for $\Delta/\breve{t}=10^{-3}$.
For $\mu_\mathrm{S}=-0.1$, $-1$, and $-2$ [Fig.~\ref{fig:tb_dep}(b), (c), and (d), respectively], the maximum value of $\bar{\Gamma}_\mathrm{e}(E=0)$ is obtained at approximately $\breve{t}_\mathrm{b}/\breve{t}=1.1$, $1.4$, and $1.5$, respectively.
$\breve{t}_\mathrm{b}$ that gives the maximum value of $\bar{\Gamma}_\mathrm{e}(E=0)$ is almost the same for $\Delta/\breve{t}=10^{-2}$ and $10^{-3}$ for $\mu_\mathrm{S}=-0.1$, $-1$, and $-2$, .

\subsection{\label{sec:tb_dep_Gamma_2d_1d_d_wave}$\bar{\Gamma}_\mathrm{e}(E)$ for 1D N ladder/$d$-wave S junction close to $\mu=0$}
\begin{figure}[t]
    \centering
    \includegraphics[width=8.5cm]{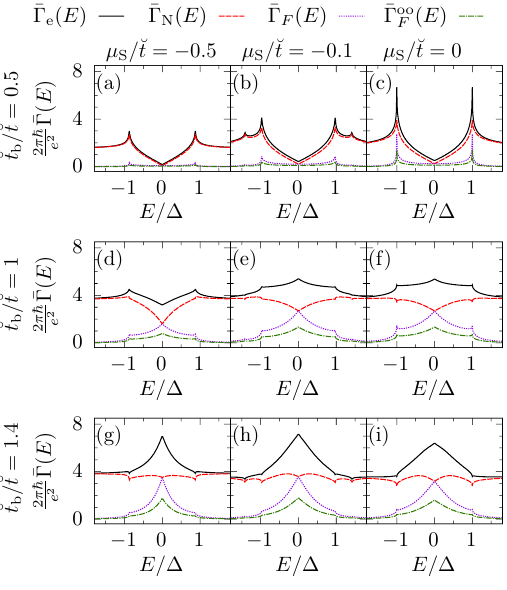}
    \caption{
    The components of $\bar{\Gamma}_\mathrm{e}(E)$ for the 1D N ladder/$d$-wave S junction is plotted as a function of $E$ for several values of $\mu_\mathrm{S}$ and $\breve{t}_\mathrm{b}$ with $\Delta/\breve{t}=0.1$ and $\mu_\mathrm{N}/\breve{t}=-0.5$.
    $(\mu_\mathrm{S}/\breve{t},\breve{t}_\mathrm{b}/\breve{t})$ is $(-0.5,0.5)$ for (a), $(-0.1,0.5)$ for (b), $(0,0.5)$ for (c), $(-0.5,1)$ for (d), $(-0.1,1)$ for (e), $(0,1)$ for (f), $(-0.5,1.4)$ for (g), $(-0.1,1.4)$ for (h), and $(0,1.4)$ for (i).
    $L=500$, $L_x=2\times10^6$ and $\eta/\breve{t}=10^{-7}$.
    }
    \label{fig:Gamma_d_wave_ladder}
\end{figure}

We show $\mu_\mathrm{S}$ and $\breve{t}_\mathrm{b}$ dependence of $\bar{\Gamma}_\mathrm{e}(E)$ for the 1D N ladder/2D $d$-wave junction in Fig.~\ref{fig:Gamma_d_wave_ladder}.
We calculate $\bar{\Gamma}_\mathrm{e}(E)$ for $\mu_\mathrm{S}/\breve{t}=-0.5$, $-0.1$, and $0$.
In Figs.~\ref{fig:Gamma_d_wave_ladder}(e)--(i), 
$\bar{\Gamma}_\mathrm{e}(E)$ shows zero energy peaks. 
These peaks originate from the peak structure of $\bar{\Gamma}_F(E)$.
The 2D square lattice model has a von Hove singularity at $\mu_\mathrm{S}=0$.
Hence, the zero energy peak of $\bar{\Gamma}_\mathrm{e}(E)$ might come from the von Hove singularity.

\subsection{\label{sec:odd_2d_SC_ladder_in_1dN}Size dependence of $\bar{\Gamma}_\mathrm{e}(E)$ and its components and anomalous Green function in 1D N for 1D N ladder/2D S junctions}
\begin{figure}[t]
    \centering
    \includegraphics[width=12cm]{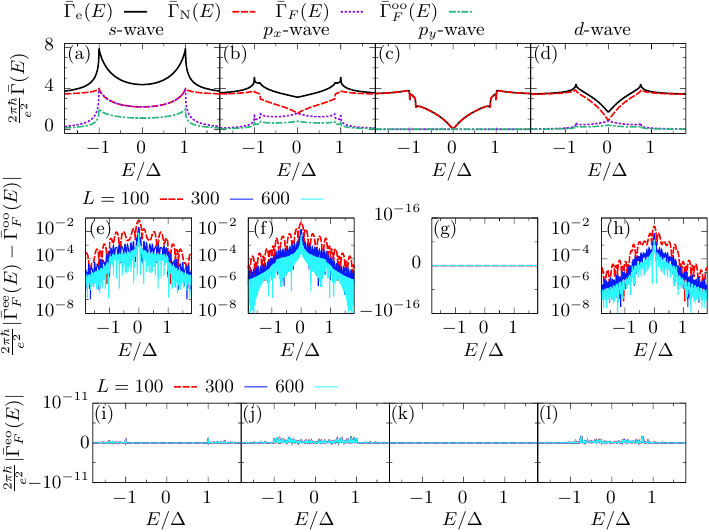}
    \caption{
    (a)--(d) $\bar{\Gamma}_\mathrm{e}(E)$ and its components are plotted as a function of $E$.
    (e)--(h) $|\bar{\Gamma}^\mathrm{ee}_{F}(E)-\bar{\Gamma}^\mathrm{oo}_{F}(E)|$ is plotted as a function of $E$ for several values of $L$.
    (i)--(l) $|\bar{\Gamma}^\mathrm{eo}_{F}(E)|$ is plotted as a function of $E$ for several values of $L$.
    (a), (e), and (i) $s$-wave, 
    (b), (f), and (j) $p_x$-wave, 
    (c), (g), and (k) $p_y$-wave, and
    (d), (h), and (l) $d$-wave junction.
    $\breve{t}_\mathrm{b}/\breve{t}=1$, $\Delta/\breve{t}=0.1$, $L_x=2\times10^6$, and $\eta/\breve{t}=10^{-7}$.
    }
    \label{fig:size_ladder_app}
\end{figure}
In Figs.~\ref{fig:size_ladder_app}(a)--(d), $\bar{\Gamma}_\mathrm{e}(E)$ and its components for 1D N ladder/2D S junctions are plotted [Figs.~\ref{fig:size_ladder_app}(b)--(d) are also shown in the main text].
The maximum value of $\bar{\Gamma}_\mathrm{e}(E)$ is eight since there are two conducting channels in 1D N ladder.
The $s$-wave result [Fig.~\ref{fig:size_ladder_app}(a)] is qualitatively the same as that for 1D N/2D S junction.

We show the difference between $\bar{\Gamma}^\mathrm{ee}_{F}(E)$ and $\bar{\Gamma}^\mathrm{oo}_{F}(E)$ in Figs.~\ref{fig:size_ladder_app}(e)--(h).
For $s$-, $p_x$-, and $d$-wave junctions [Figs.~\ref{fig:size_ladder_app}(e), (f), and (h), respectively], the difference becomes smaller as $L$ increases.
For $p_y$-wave junctions, the difference is zero within numerical errors [Fig.~\ref{fig:size_ladder_app}(g)].
$\bar{\Gamma}^\mathrm{eo}_{F}(E)$ is shown in Figs.~\ref{fig:size_ladder_app}(i)--(l).
In all cases, it is zero within numerical errors.

\begin{figure}[t]
    \centering
    \includegraphics[width=10.0cm]{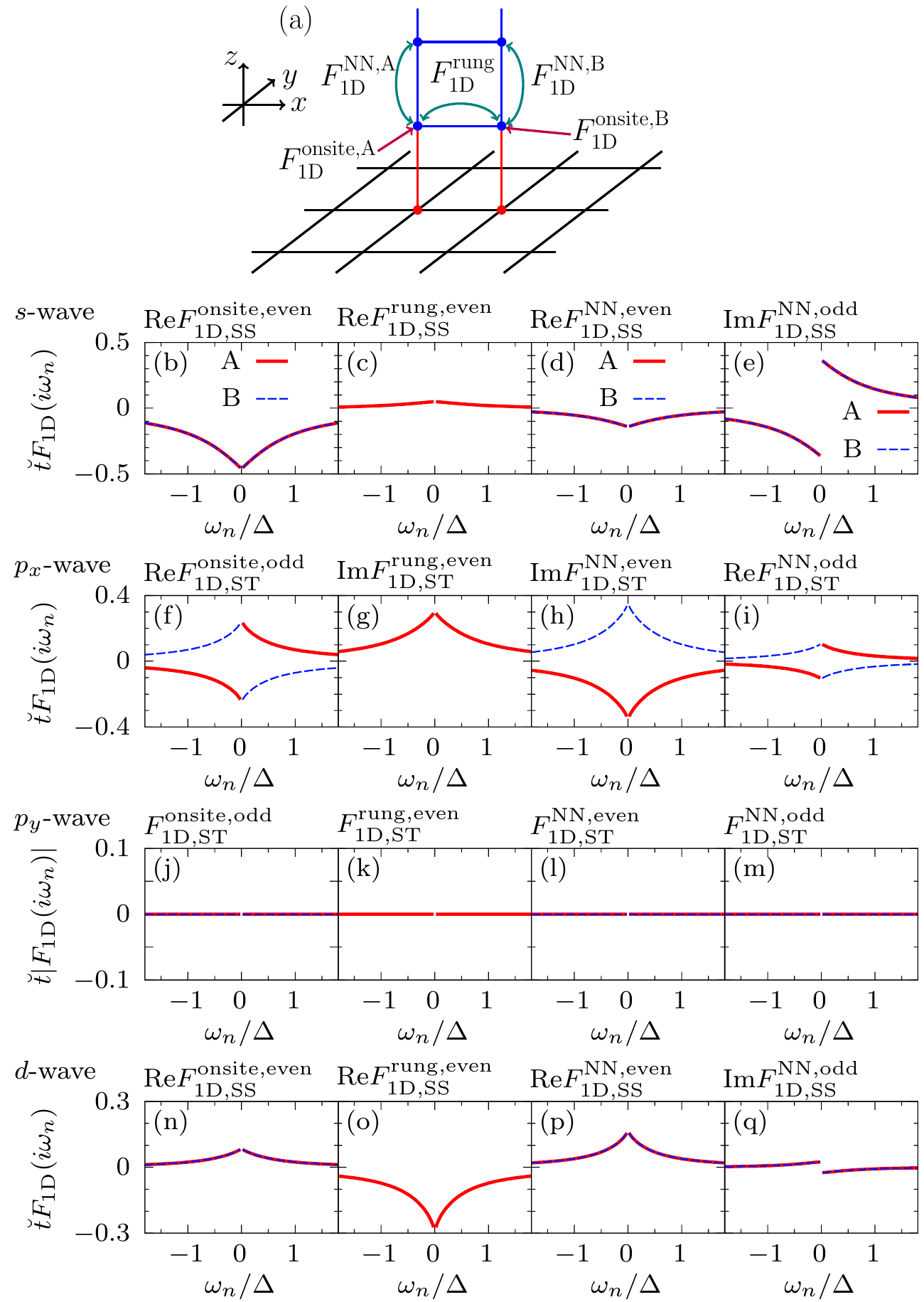}
    \caption{
    (a) Schematic picture of anomalous Green functions.
    (b)--(q) The anomalous Green function is plotted as a function of $\omega_n$ with $\breve{t}_\mathrm{b}/\breve{t}=1$, $\Delta/\breve{t}=0.1$, $\mu_\mathrm{N}/\breve{t}=-0.5$, $\mu_\mathrm{S}/\breve{t}=-1$, and $L_x=10^{6}$.
    $F_\mathrm{1D}^\mathrm{rung,odd}$ is zero for all cases and is not plotted.
    }
    \label{fig:F1d_ladder_app}
\end{figure}
In Fig.~\ref{fig:F1d_ladder_app}, the odd-frequency pairing in the 1D N ladder for the $s$-, $p_x$-, $p_y$- and $d$-wave junctions are shown.
For $s$-, $p_x$-, and $d$-wave junctions, even and odd-frequency pairings penetrate into the 1D N ladder.
However, for $p_y$-wave junction, they do not penetrate like for 1D N/2D $p_x$-wave junctions.

\subsection{\label{sec:odd_2d_SC_ladder_in_2dS}Even and odd-frequency pairings in 2D S for 1D N ladder/2D S junctions}
\begin{figure}[t]
    \centering
    \includegraphics[width=10.5cm]{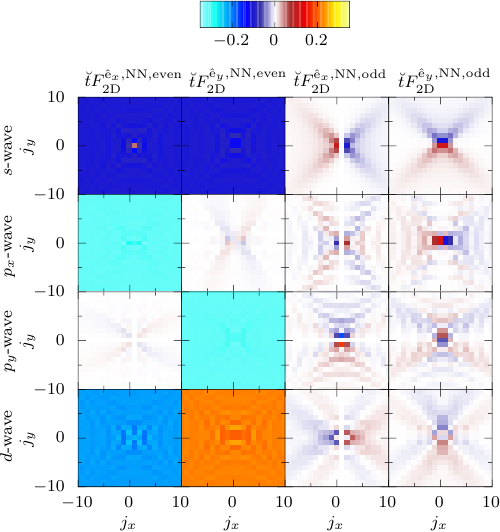}
    \caption{
        The NN components of the anomalous Green function is plotted as functions of $j_x$ and $j_y$ at $\omega_n/\Delta=0.1$ for the 1D N ladder/2D S junction.
        Here, $(j_x,j_y)=(0,0)$ and $(1,0)$ are connected to the 1D N ladder.
        For the $s$-wave, $p_y$-wave and $d$-wave S junctions with even-frequency components, the real part is plotted, and for the odd-frequency components, the imaginary part is plotted.
        For the $p_x$-wave S junction with even-frequency components, the imaginary part is plotted, and for the odd-frequency components, the real part is plotted.
        The counterparts are numerically zero.
    }
    \label{fig:F2d_1d_2d_ladder_nn_app}
\end{figure}

In Fig.~\ref{fig:F2d_1d_2d_ladder_nn_app}, NN components of anomalous Green functions in 2D S are plotted.
We can employ Eq.~\eqref{eq:criteria_s_d_app} for $s$-wave and $d$-wave junctions, and Eq.~\eqref{eq:criteria_p_app} for $p_x$-wave and $p_y$-wave junctions for $\mathbf{j}=\mathbf{j}_0$.
For $s$-wave junctions, we confirm $F_\mathrm{SS}^{\mathrm{NN,even}}\neq0$ and $F_\mathrm{SS}^{\mathrm{NN,odd}}\neq0$.
Hence, NN even and odd-frequency pairing penetrate into the 1D N ladder.
We can discuss the same properties for another 1D N ladder point ($\mathbf{j}=\mathbf{j}_0+\hat{
\mathrm{e}}_x$).
For $p_x$-wave junctions, NN even-frequency components in $y$-direction do not cancel each other and penetrate into 1D N ladder.
Likewise, NN odd-frequency pairings do not cancel each other and penetrate into the 1D N ladder.
For $p_y$-wave junction, NN even and odd-frequency pairings are qualitatively the same as 1D N/2D $p_x$-wave junction.
Then, NN even and odd-frequency contributions cancel each other and do not penetrate into the 1D N ladder.
For $d$-wave junctions, although even and odd-frequency pairings cancel each other in $x$ and $y$-direction in 1D N/2D $d$-wave junction, this symmetry is broken by the 1D N ladder.
Hence, these two directional components do not cancel each other.
Then, NN even and odd-frequency pairings penetrate into the 1D N ladder.

\section{\label{app:green_lattice}Recursive Green function method}
We now explain the recursive Green function method for 1D N/2D S junctions. 
The extension to 1D N ladder/2D S junctions is straightforward.
To derive the Green function in 2D S, we consider periodic boundary conditions in $x$-direction and infinite system size in $y$-direction.
The bulk Green function with momentum $k_x$ is obtained by the surface Green function of left [$\check{G}^{\mathrm{2D},\dashv}_{k_x,i_y,j_y}(z)$] and right [$\check{G}^{\mathrm{2D},\vdash}_{k_x,i_y,j_y}(z)$] semi-infinite systems: 
\begin{align}
    \check{G}_{k_x,i_y,j_y}^\mathrm{2D}(z)
    =&
    {\left\{{\left[\check{G}^{\mathrm{2D},\dashv}(z)\right]}^{-1}-\check{t}_\mathrm{2D}\check{G}^{\mathrm{2D},\vdash}(z)\check{t}_\mathrm{2D}^\dagger\right\}}^{-1},
    \label{eq:connect}
\end{align}
with $\check{t}_\mathrm{2D}=-\breve{t}\hat{\sigma}_0\hat{\tau}_3$ for the $s$-wave and $p_x$-wave superconductors, 
$\check{t}_\mathrm{2D}=-\breve{t}\hat{\sigma}_0\hat{\tau}_3+\Delta\hat{\sigma}_1\hat{\tau}_2/2$ for the $p_y$-wave case, and
$\check{t}_\mathrm{2D}=-\breve{t}\hat{\sigma}_0\hat{\tau}_3+\Delta\hat{\sigma}_2\hat{\tau}_2/4$ for the $d$-wave case.
Here, $z\in\mathbb{C}$ is a complex frequency, where $z=i\omega_n$ for the Matsubara frequency representation, and $z=E-(+)i\eta$ for the advanced (retarded) Green function.
Then, the real space representation of the Green function is
\begin{align}
    \check{G}_{\mathbf{i},\mathbf{j}}^\mathrm{2D}(z)
    =&
    \frac{1}{L_x}
    \sum_{k_x}
    \check{G}_{k_x,i_y,j_y}^\mathrm{2D}(z)e^{ik_x(i_x-j_x)}.
\end{align}
The Green function in 1D N is obtained by the surface Green function in the 1D N $\check{G}^{\mathrm{1D},\dashv}(z)$ and $\check{G}_{\mathbf{j}_0,\mathbf{j}_0}^\mathrm{2D}(z)$:
\begin{align}
    \check{G}_{1,1}^{\mathrm{1D}}(z)
    ={\left\{{\left[\check{G}^{\mathrm{1D},\dashv}(z)\right]}^{-1}-\check{t}_\mathrm{b}\check{G}_{\mathbf{j}_0,\mathbf{j}_0}^{\mathrm{2D}}(z)\check{t}_\mathrm{b}^\dagger\right\}}^{-1}
    \label{eq:app_G11}
\end{align}
with $\check{t}_\mathrm{b}=-\breve{t}_\mathrm{b}\hat{\tau}_3$.
To calculate the differential conductance, Green functions in 1D N are needed.
For instance, $\check{G}_{1,1}^{\mathrm{1D}}(z)$, $\check{G}_{1,2}^{\mathrm{1D}}(z)$, $\check{G}_{2,1}^{\mathrm{1D}}(z)$, and $\check{G}_{2,2}^{\mathrm{1D}}(z)$ are given by
\begin{align}
    \check{G}_{2,2}^\mathrm{1D}(z)
    =&
    {\left[(z+\mu_\mathrm{N}\hat{\tau}_3)-\check{t}_\mathrm{b}\check{G}^{\mathrm{2D}}_{\mathbf{j}_0,\mathbf{j}_0}(z)\check{t}_\mathrm{b}^\dagger\right]}^{-1},
    \label{eq:app_G22}
    \\
    \check{G}_{1,2}^\mathrm{1D}(z)
    =&
    \check{G}^\mathrm{1D,\dashv}(z)\check{t}_\mathrm{1D}\check{G}_{1,1}^{\mathrm{1D}}(z),
    \label{eq:app_G12}
    \\
    \check{G}_{2,1}^\mathrm{1D}(z)
    =&
    \check{G}_{1,1}^{\mathrm{1D}}(z) \check{t}_\mathrm{1D}^\dagger \check{G}^\mathrm{1D,\dashv}(z)
    \label{eq:app_G21}
\end{align}
with $\check{t}_\mathrm{1D}=-\breve{t}\hat{\tau}_3$.
In the same manner, we can obtain the matrix components of the Green function in 1D N ladder and 2D S.
To calculate the Green function in the 2D S, we calculate $\check{G}_{k_x,i_y,j_y}^{\mathrm{2D}}(z)$ [Eq.~\eqref{eq:connect}] by the form
\begin{align}
    \left[1-\check{G}^{\mathrm{2D},\dashv}(z)\check{t}_\mathrm{2D}\check{G}^{\mathrm{2D},\vdash}(z)\check{t}_\mathrm{2D}^\dagger\right]
    \check{G}_{k_x,i_y,j_y}^\mathrm{2D}(z)
    =&
    \check{G}^{\mathrm{2D},\dashv}(z).
\end{align}
Then, we do not have to calculate the inverse of matrices.


\end{document}